\documentclass[12pt]{amsart}
\usepackage{amssymb}
\usepackage{amsbsy}
\usepackage{amscd}

\usepackage[mathscr]{eucal}
\usepackage{nicefrac}

\usepackage{a4wide}
\usepackage[pdftex]{pict2e}
\usepackage{curve2e}
\makeatletter
\renewcommand{\thesubsection}{\thesection(\@roman\c@subsection)}
\makeatother
\newenvironment{aenume}{%
  \begin{enumerate}%
  }{\end{enumerate}}
\newcounter{number}
\newenvironment{enume_add}{%
  \begin{enumerate}%
  \setcounter{enumi}{\thenumber}
  \refstepcounter{number}
  }
  {
    \setcounter{number}{\value{enumi}}
\end{enumerate}}

\usepackage{refcount}

\usepackage{verbatim}
\usepackage{version}
\usepackage{color}
\newenvironment{NB}{
\color{red}{\bf NB}. \footnotesize
}{}
\newenvironment{NB2}{
\color{blue}{\bf NB}. \footnotesize
}{}
\excludeversion{NB}
\excludeversion{NB2}
\newtheorem{Theorem}[equation]{Theorem}

\newtheorem{Lemma}[equation]{Lemma}

\theoremstyle{definition}
\newtheorem{Definition}[equation]{Definition}

\theoremstyle{remark}
\newtheorem{Remark}[equation]{Remark}
\newtheorem{Remarks}[equation]{Remarks}

\numberwithin{equation}{section}

\newcommand{\secref}[1]{\S\ref{#1}}
\newcommand{\lemref}[1]{Lemma~\ref{#1}}

\newcommand{\subsecref}[1]{\S\ref{#1}}

\newcommand{\remref}[1]{Remark~\ref{#1}}

\newcommand{\lsp}[2]{{\mskip-.3mu}{}^{#1}\mskip-1mu{#2}}
\newcommand{\defeq}{\overset{\operatorname{\scriptstyle def.}}{=}}
\newcommand{\CC}{{\mathbb C}}
\newcommand{\ZZ}{{\mathbb Z}}

\newcommand{\RR}{{\mathbb R}}
\newcommand{\proj}{{\mathbb P}}

\newcommand{\SL}{\operatorname{\rm SL}}
\newcommand{\SU}{\operatorname{\rm SU}}
\newcommand{\GL}{\operatorname{GL}}

\newcommand{\U}{\operatorname{\rm U}}

\newcommand{\SO}{\operatorname{\rm SO}}
\newcommand{\grpSp}{\operatorname{\rm Sp}}
\newcommand{\grpO}{\operatorname{\rm O}}
\newcommand{\algsl}{\operatorname{\mathfrak{sl}}} 
\newcommand{\su}{\operatorname{\mathfrak{su}}}

\newcommand{\so}{\operatorname{\mathfrak{so}}}
\newcommand{\algsp}{\operatorname{\mathfrak{sp}}}
\newcommand{\g}{{\mathfrak g}}

\newcommand{\End}{\operatorname{End}}
\newcommand{\Hom}{\operatorname{Hom}}

\newcommand{\Ker}{\operatorname{Ker}}

\newcommand{\Ima}{\operatorname{Im}}

\newcommand{\rank}{\operatorname{rank}}
\newcommand{\ind}{\mathop{\text{\rm ind}}\nolimits}
\newcommand{\tr}{\operatorname{tr}}

\newcommand{\DB}{\overline{\partial}}

\newcommand{\ve}{\varepsilon}
\renewcommand{\MR}[1]{}
\newcommand{\Wedge}{{\textstyle \bigwedge}}

\newcommand{\HH}{\mathbb H}
\newcommand{\dslash}{/\!\!/}
\newcommand{\vin}[1]{\operatorname{i}(#1)} 
\newcommand{\vout}[1]{\operatorname{o}(#1)} 
\newcommand{\bM}{\mathbf M}
\newcommand{\bN}{\mathbf N}
\newcommand{\CS}{\mathrm{CS}}
\newcommand{\LL}{\mathbb L}
\newcommand{\wt}{\operatorname{wt}}

\newcommand{\shfO}{\mathcal O}
\newcommand{\tslash}{/\!\!/\!\!/}
\newcommand{\tslabar}{\mathbin{
\setbox0=\hbox{/\!\!/\!\!/}\rule[0.4\ht0]{\wd0}{.3\dp0}\kern-\wd0\box0}}
\newcommand{\Hyp}{\operatorname{Hyp}}
\newcommand{\bmu}{\boldsymbol\mu}

\newcommand{\la}{\lambda}
\newcommand{\aff}{\mathrm{aff}}

\newcommand{\Gr}{\mathrm{Gr}}

\usepackage{url}

\usepackage[
bookmarks=true,
colorlinks=true]{hyperref}

\begin{document}

\title[Coulomb branches of $3d$ $\mathcal N=4$ gauge theories, I]
{Towards a mathematical definition of Coulomb branches of
  $3$-dimensional $\mathcal N=4$ gauge theories, I
}
\author[H.~Nakajima]{Hiraku Nakajima}
\address{Research Institute for Mathematical Sciences,
Kyoto University, Kyoto 606-8502,
Japan}
\email{nakajima@kurims.kyoto-u.ac.jp}

\subjclass[2000]{}
\begin{abstract}
    Consider the $3$-dimensional $\mathcal N=4$ supersymmetric gauge
    theory associated with a compact Lie group $G$ and its
    quaternionic representation $\bM$. Physicists study its Coulomb
    branch, which is a noncompact hyper-K\"ahler manifold, such as
    instanton moduli spaces on $\RR^4$, $\SU(2)$-monopole moduli
    spaces on $\RR^3$, etc.
    In this paper and its sequel, we propose a mathematical definition
    of the coordinate ring of the Coulomb branch, using the vanishing
    cycle cohomology group of a certain moduli space for a gauged
    $\sigma$-model on the $2$-sphere associated with $(G,\bM)$. In
    this first part, we check that the cohomology group has the correct
    graded dimensions expected from the monopole formula proposed by
    Cremonesi, Hanany and Zaffaroni \cite{Cremonesi:2013lqa}.
    A ring structure (on the cohomology of a modified moduli space)
    will be introduced in the sequel of this paper.
\end{abstract}

\maketitle

\setcounter{tocdepth}{2}

\section{Introduction}

\subsection{\texorpdfstring{$3$}{3}-dimensional \texorpdfstring{$\mathcal N=4$}{N=4} SUSY gauge theories}

Let $\bM$ be a quaternionic representation (also called a {\it
  pseudoreal\/} representation) of a compact Lie group $G$.
Let us consider the $4$-dimensional gauge theory with $\mathcal N=2$
supersymmetry associated with $(G,\bM)$. It has been studied by
physicists for many years.
It is closely related to pure mathematics, because the correlation
function of its topological twist ought to give the Donaldson
invariant \cite{MR1066174} of $4$-manifolds for $(G,\bM) =
(\SU(2),0)$, as proposed by Witten \cite{MR953828}. Thus physics and
mathematics influence each other in this class of gauge theories.
For example, Seiberg-Witten's ansatz \cite{MR1293681} led to a
discovery of a new invariant, namely the Seiberg-Witten invariant. It
is associated with $(G,\bM) = (\U(1),\HH)$.
Nekrasov partition function \cite{Nekrasov} gave a mathematically
rigorous footing on Seiberg-Witten's ansatz, and hence has been
studied by both physicists and mathematicians.

In this paper, we consider the $3$-dimensional gauge theory with
$\mathcal N=4$ supersymmetry, obtained from the $4d$ theory, by
considering on $(\text{$3$-manifold})\times S^1_R$ and taking $R\to
0$.
We denote the $3d$ gauge theory by $\Hyp(\bM)\tslabar G$ following
\cite{Tach-review}, though it is used for the $4d$ gauge theory
originally.
One usually studies only asymptotically conformal or free theories in
$4d$, while we do not have such restriction in $3d$.
Also some aspects are easier, simplified and clarified in $3d$, and
our hope is that to use understanding in $3d$ to study problems in
$4d$.

\begin{NB}
    There are two kinds of fields in this gauge theory. One set
    consists of a $G$-connection $A$ on a $G$-bundle $P$, three
    sections $\phi_1$, $\phi_2$, $\phi_3$ of $P\times_G\g$ together
    with their fermionic partners. They are called vector
    multiplets. Another set .... It is not helpful, so let us not to
    explain.
\end{NB}%

We do not review what is $\Hyp(\bM)\tslabar G$. It is a quantum field
theory in dimension $3$, and is not rigorously constructed
mathematically.

Instead we will take the following strategy: physicists associate
various mathematical objects to $\Hyp(\bM)\tslabar G$ and study their
properties. They ask mathematicians to construct those objects,
instead of $\Hyp(\bM)\tslabar G$ itself, in mathematically rigorous
ways so that expected properties can be checked.

This strategy is posed explicitly in \cite{MR2985331} for a particular
object, which is very close to what we consider here. (See
\subsecref{subsec:MT} for detail on the precise relation.) Many
physically oriented mathematical works in recent years are more or
less take this strategy anyway.

\subsection{Topological twist}
An example of mathematical objects is a topological invariant defined
so that it ought to be a correlation function of a topologically
twisted version of $\Hyp(\bM)\tslabar G$.
For the $4$-dimensional $\mathcal N=2$ SUSY pure $\SU(2)$-theory
(i.e., $(G,\bM) = (\SU(2),0)$), Witten claimed that the correlation
function gives Donaldson invariants, as mentioned above.
In \cite{MR1094734} Atiyah-Jeffrey understood the correlation function
heuristically as the Euler class of an infinite rank vector bundle of
an infinite dimensional space with a natural section $s$ in the
Mathai-Quillen formalism. The zero set of $s$ is the moduli space of
$\SU(2)$ anti-self-dual connections for $(G,\bM) = (\SU(2),0)$, hence
Witten's claim has a natural explanation as a standard result in
differential topology applied formally to infinite dimension.
It was also observed that the $3$-dimensional story explains an
Taubes' approach to the Casson invariant \cite{MR1037415}, at least
for homological $3$-spheres.
And $3$ and $4$-dimensional stories are nicely combined to Floer's
instanton homology group \cite{MR956166} and its relation to Donaldson
invariants in the framework of a $(3+1)$-dimensional topological
quantum field theory (TQFT) \cite{MR1883043}.
Another example of a similar spirit is the Seiberg-Witten invariant in
dimensions $3/4$ (see e.g., \cite{MR1306021} for $4d$ and
\cite{MR1418579} for $3d$). This is the case $(G,\bM) = (\U(1),\HH)$.

Atiyah-Jeffrey's discussion is heuristic. In particular, it is not
clear how to deal with singularities of the zero set of $s$ in
general.
We also point out that a new difficulty, besides singularities of
$\operatorname{Zero}(s)$, failure of compactness occurs in
general. See \subsecref{sec:relat-topol-invar} and
\subsecref{sec:roles-higgs} below.
Therefore it is still an open problem to define topological invariants
rigorously for more general $\Hyp(\bM)\tslabar G$. It is beyond the
scope of this paper. But we will use a naive or heuristic analysis of
would-be topological invariants to help our understanding of
$\Hyp(\bM)\tslabar G$.
We hope the study in this paper and its sequel \cite{BFN} might be
relevant to attack the problem of the definition of topological
invariants.

\subsection{Coulomb branch}\label{sec:coulomb-branch}

Instead of giving definitions of topological invariants, we consider
the so-called Coulomb branch $\mathcal M_C$ of $\Hyp(\bM)\tslabar G$.

Physically it is defined as a specific branch of the space of {\it
  vacua}, where the potential function takes its minimum. However it
is only a classical description, and receives a quantum correction due
to the integration of massive fields. At the end, there is no
definition of $\mathcal M_C$, which mathematicians could understand in
the literature up to now. However it is a physicists consensus that
the Coulomb branch 
is a
\begin{NB}
conical
\end{NB}%
hyper-K\"ahler manifold with an $\SU(2)$-action rotating
hyper-K\"ahler structures $I$, $J$, $K$ \cite{MR1490862}.
Physicists also found many hyper-K\"ahler manifolds as Coulomb
branches of various gauge theories, such as toric hyper-K\"ahler
manifolds, moduli spaces of monopoles on $\RR^3$, instantons on
$\RR^4$ and ALE spaces, etc. (Reviewed below.)

\begin{NB}
    The goal of this paper and its sequel \cite{BFN} is to propose a
    mathematically rigorous definition of $\mathcal M_C$ as an affine
    scheme, more precisely, the definition of its coordinate ring
    $\CC[\mathcal M_C]$. See \subsecref{sec:prop-defin-texorpdfs}
    below for the statement.
We also give several evidences to believe our definition is correct.
\end{NB}

\subsection{A relation to topological invariants}\label{sec:relat-topol-invar}

The Coulomb branch $\mathcal M_C$ might play some role in the study of
would-be topological invariants for $\Hyp(\bM)\tslabar G$. As
topological invariants are currently constructed only $(G,\bM) =
(\SU(2),0)$ and $(G,\bM)=(\U(1),\HH)$, we could touch this aspect
superficially. Nevertheless we believe that it is a good starting
point.

First consider $\Hyp(0)\tslabar\SU(2)$.
Let us start with the $4$-dimensional case. A pseudo physical review
for mathematicians was given in \cite[\S1]{MR2095899}, hence let us
directly go to the conclusion.
The space of vacua is parametrized by a complex parameter $u$, and
hence called the {\it $u$-plane} (Seiberg-Witten ansatz).
We have a family of elliptic curves $E_u$ parametrized by $u$, where
$E_u$ degenerates to rational curves at $u=\pm 2\Lambda^2$. The
so-called prepotential of the gauge theory is recovered from the
period integral of $E_u$. From the gauge theoretic view point, $u$ is
understood as a `regularized' integration of a certain equivariant
differential form over the framed moduli space of $\SU(2)$-instantons
on $\RR^4$. More rigorously we define the integration by a
regularization cooperating $T^2$-action on $\RR^4$ (Nekrasov's
$\Omega$-background).
The prepotential above determines the equivariant variable $a$ as a
function of $u$, and hence $u$ as an inverse function of $a$.

Witten explained that the Donaldson invariant is given by a $u$-plane
integral, and the contribution at the singularities $u=\pm 2\Lambda^2$
is given by the Seiberg-Witten invariant \cite{MR1306021}. See also \cite{MR1605636} for a further development.
This picture was mathematically justified, in a slightly modified way, 
for projective surfaces in \cite{GNY1,GNY3}.

Now we switch to the $3$-dimensional case. As is mentioned above, we
first consider $\RR^3\times S^1_R$ and take the limit $R\to
0$. Seiberg-Witten determines the space of vacua for large $R$ as the
{\it total space\/} of the family $E_u$ for $u\in\CC$
\cite{MR1490862}. When we make $R\to 0$, points in $E_u$ are removed,
and get the moduli space of charge $2$ centered monopoles on
$\RR^3$. This is a $4$-dimensional hyper-K\"ahler manifold studied
intensively by Atiyah-Hitchin \cite{MR934202}. This is the Coulomb
branch $\mathcal M_C$ of the $3$-dimensional gauge theory
$\Hyp(0)\tslabar \SU(2)$.
It means that the $3$-dimensional gauge theory $\Hyp(0)\tslabar
\SU(2)$ reduces at low energies to a sigma-model whose target is
$\mathcal M_C$.

This picture gives us the following consequence for topological
invariants.
The partition function for the twisted $\Hyp(0)\tslabar\SU(2)$ is the
Casson-Walker-Lescop invariant as above.
On the other hand, the partition function for the sigma-model with
target $\mathcal M_C$ is the topological invariant constructed by
Rozansky-Witten \cite{MR1481135}.
Then as an analog of Seiberg-Witten = Donaldson in $4$-dimension, it
is expected that the Casson-Walker-Lescop invariant coincides with the
Rozansky-Witten invariant.

The Rozansky-Witten invariant for $\mathcal M_C$ is a finite type
invariant \cite{MR1373813,MR1604883,MR1736483} of order $3$, which is
unique up to constant multiple. Therefore the coincidence of two
invariants is not a big surprise. 
But it at least gives an expectation of a generalization to
$G=\SU(r)$. The Coulomb branch of $\Hyp(0)\tslabar \SU(r)$ is the
moduli space of charge $r$ centered monopoles on $\RR^3$
\cite{MR1443803}. Neither the $\SU(r)$-Casson-Walker-Lescop invariant
and the Rozansky-Witten invariant associated with $\mathcal M_C$ are
not mathematically rigorously defined yet, but it is natural to expect
that they coincide once they would be defined.
For the former, singularities of moduli spaces must be treated
appropriately. See \cite{MR1016041,MR1678493,MR1879803,MR1933787}, for
example, studying this problem.
The convergence of an integral must be proved for Rozansky-Witten
invariants, as the monopole moduli spaces are noncompact.

Let us continue examples from \cite{MR1490862}. Let $G=\U(1)$,
$\bM=\HH^N = \CC^N\oplus(\CC^*)^N$, the direct sum of $N$ copies of
the vector representation plus its dual.
The Coulomb branch $\mathcal M_C$ is the multi-Taub-NUT space, which
is $\CC^2/(\ZZ/N\ZZ)$ as a complex variety when $N > 0$, and is
$\RR^3\times S^1 = \CC\times\CC^*$ when $N=0$.

The case $N=0$ is trivial, so let us exclude it. There is a big
distinction between $N=1$ and $N>1$ cases, where we have a singularity
at the origin or not. For $N=1$, the partition function gives the
$3$-dimensional Seiberg-Witten invariant. On the other hand, $\mathcal
M_C$ is the Taub-NUT space, the associated Rozansky-Witten invariant
is again finite type of order $3$, hence should be equal to the
Casson-Walker-Lescop invariant up to multiple again. Marino-Moore
\cite{MR1697354}, Blau-Thompson \cite{MR1898369} argued that the
Casson-Walker-Lescop invariant is equal to the $3$-dimensional
Seiberg-Witten invariant, more precisely, the (regularized) sum over
all $\operatorname{Spin}^c$ classes, as proved earlier by Meng-Taubes
\cite{MR1418579} for $b_1 > 0$. In the $b_1=0$ case, the definition of
the Seiberg-Witten invariant is more subtle, and the claim was shown
later by Marcolli-Wang \cite{MR1919892}.

For $N>1$, the Coulomb branch $\mathcal M_C$ has a singularity at the
origin, whose contribution to the Rozansky-Witten invariant needs to
be clarified. In the Seiberg-Witten side, compactness of moduli spaces
fails, as we will review in \subsecref{sec:roles-higgs}. Therefore it
is not yet clear how to define the invariant. The singularity comes
from the Higgs branch $\mathcal M_H$, explained later. Therefore it is
natural to expect that the Rozansky-Witten invariant for $\mathcal
M_H$ enters the picture.

Let us note that $\mathcal M_C$ for $\Hyp(0)\tslabar\SU(2)$ can be
defined as a limit of the total space of $E_u$, which is rigorously
recovered from Nekrasov's partition function. This method could
probably apply to if $(G,\bM)$ gives an asymptotically conformal or
free theory in dimension $4$, but not in general. For example, it is
not clear how to do for $(\U(1),\HH)$.

\subsection{\texorpdfstring{$(2+1)$}{2+1}-dimensional TQFT}\label{sec:2+1-tqft}

The Rozansky-Witten invariant associated with a hyper-K\"ahler
manifold $M$ is expected to fit in the framework of a
$(2+1)$-dimensional TQFT. For a $2$-manifold $\Sigma$, one associates
a quantum Hilbert space $\mathcal H_\Sigma$, and an invariant of a
$3$-manifold $X$ with boundary $\Sigma$ takes value in $\mathcal
H_\Sigma$. Then the gluing axiom is satisfied.

In \cite[\S5]{MR1481135}, it is proposed that
\begin{equation*}
    \mathcal H_{\Sigma_g} = (-1)^{1+g}
    \bigoplus_q H^q(M, (\Wedge^* V)^{\otimes g}),
\end{equation*}
where $\Sigma_g$ is a $2$-manifold of genus $g$, $V$ is the natural
$\grpSp(\dim_\HH M)$-bundle over $M$, and the sign $(-1)^{1+g}$ is
introduced so that $\mathcal H_{\Sigma_g}$ has a correct
$\ZZ/2$-graded vector space structure. Rozansky-Witten wrote that it
is hazardous to apply this definition for noncompact $M$, like our
Coulomb branch $\mathcal M_C$.

Nonetheless consider the case $g=0$ assuming $\mathcal M_C$ is affine:
\begin{equation*}
    \mathcal H_{S^2} = \bigoplus_q (-1)^{1+q} H^q(\mathcal M_C,\mathcal O)
    = - \CC[\mathcal M_C].
\end{equation*}

Let us give a nontrivial check for this hazardous assertion. By the
gluing axiom, the invariant of $S^2\times S^1$ is equal to the
dimension of $\mathcal H_{S^2}$. On the other hand, from the knowledge
of the Casson-Walker-Lescop invariant for $S^2\times S^1$, it should
be equal to $1/12$ if $\mathcal M_C$ is the Taub-NUT space or the
Atiyah-Hitchin manifold (see \cite[(4.4), (5.15)]{MR1481135} and
\cite[(2.3)]{MR1898369}. Therefore we should have $\dim \CC[\mathcal
M_C] = -1/12$. For the Taub-NUT space, which is isomorphic to $\CC^2$
as an affine variety, this is true after $\zeta$-regularization:
\begin{equation}\label{eq:31}
    \left.\frac1{(1-t)^2}\right|_{t=1}
    \begin{NB}
    = \left.\frac{d}{dt} \frac1{(1-t)}\right|_{t=1}
    (\text{  or just expand})
    \end{NB}%
    = \left.\sum_{n=1}^\infty n t^{n-1}\right|_{t=1}
    = \zeta(-1) = -\frac1{12}.
\end{equation}
The expression $1/(1-t)^2$ is coming from the natural grading on
$\CC[\mathcal M_C] = \CC[x,y]$ as $\deg x=\deg y=1$. See
\subsecref{sec:examples} for more detail.
A closely related observation on $-1/12$ can be found at
\cite[(6.16)]{MR1697354}, \cite[\S2.3]{MR1898369}.
It is interesting to look for a similar explanation for the
Atiyah-Hitchin manifold, as well as a deeper understanding of this
regularization process.
\begin{NB}
    The AH manifold is $x + x^2 v = y^2$ with $\deg x=2$, $\deg y= 1$,
    $\deg v = -2$.
\end{NB}

On the other hand, the Casson and Seiberg-Witten invariants are also
expected to fit in the TQFT framework. The quantum Hilbert spaces
$\mathcal H_\Sigma$ are the cohomology group of moduli spaces of flat
$\SU(2)$-bundles and solutions of the (anti-)vortex equation over
$\Sigma$ respectively \cite{MR1001453,MR1734402} (see also
\cite{MR3160606}). This cannot be literally true for the Casson case,
as there is no nontrivial flat $\SU(2)$-bundle on $\Sigma = S^2$. In
the Seiberg-Witten case, $3d$ invariants depend on a choice of
perturbation of the equation, and its dependence must be understood
via the wall-crossing formula.
We need to choose the corresponding perturbation of the (anti-)vortex
equation to have a nonempty moduli space in $2d$. For a `positive'
(resp.\ `negative') perturbation, moduli spaces for the vortex (resp.\
anti-vortex) equation are nonempty, and symmetric products of
$\Sigma$. For $\Sigma = S^2$, the $(n-1)$th symmetric product
$S^{n-1}\Sigma$ is $\proj^{n-1}$, and its cohomology is
$n$-dimensional. It is compatible with the above naive computation
\eqref{eq:31}, as $n$ appears in the middle.

If Casson-Walker-Lescop = Rozansky-Witten would be true as
$(2+1)$-TQFT's, we conclude that
\begin{equation*}
    \CC[\text{$\mathcal M_C$ of $\Hyp(0)\tslabar\SU(2)$}]
    \overset{\operatorname{\scriptstyle ?}}{=}  H^*(\text{moduli spaces of
      flat bundles on $S^2$}),
\end{equation*}
and similarly for $\CC[\text{$\mathcal M_C$ of
  $\Hyp(\HH)\tslabar\U(1)$}]$ and the cohomology group of moduli
spaces of solutions of the (anti-)vortex equation on $\Sigma =
S^2$. This could {\it not\/} be true for flat bundles as we have
remarked above. For the (anti-)vortex equation we do not have an
immediate contradiction, but it is too strong to be true, as the
Coulomb branch seems to be independent of the choice of perturbation.

\subsection{Monopole formula}\label{subsec:monopole-formula-1}

As we have explained just above, it seems difficult to use our current
understanding of TQFT to determine $\mathcal M_C$ in a mathematically
rigorous way.

A work, more tractable to mathematicians, has been done recently by
Cremonesi, Hanany and Zaffaroni \cite{Cremonesi:2013lqa}. They write
down a combinatorial expression, which gives the Hilbert series of the
Coulomb branch $\mathcal M_C$. It is called {\it the monopole
  formula}.

The monopole formula is a formal Laurent power series
\begin{equation}\label{eq:11}
    H_{G,\bM}(t) \defeq \sum_{\la\in Y/W} t^{2\Delta(\la)} P_G(t;\la),
\end{equation}
where $Y$ is the coweight lattice of $G$, $W$ is the Weyl group, and
$\Delta(\la)$, $P_G(t;\la)$ are certain an integer and a rational
function in $t$ respectively.
We postpone a detailed discussion of the monopole formula to
\subsecref{sec:formula}. Let us give a brief comment here:
\eqref{eq:11} is a combinatorial expression, and mathematically makes
sense contrary to the case of $\mathcal M_C$.

It is worthwhile to keep a physical origin of the monopole formula in
mind. The monopole formula counts {\it monopole operators\/}, which
are defined by fields having point singularities
\cite{MR1955002}. Taking a radial coordinate system around a singular
point, the singularity is modeled on a connection on $S^2$, whose
topological charge is given by a coweight of $G$. This is the reason
why the coweight $\la$ appears in \eqref{eq:11}. We do not review
physical origins of expressions $\Delta(\la)$, $P_G(t;\la)$. See
\cite[\S2]{Cremonesi:2013lqa} and the references therein.

It is also clear that monopole operators belong to the quantum Hilbert
space $\mathcal H_{S^2}$ for $S^2$. They form a ring (called a {\it
  chiral ring\/} in physics literature), by considering the
topological quantum field theory associated with $S^3$ with three
punctures.

Combining with our heuristic consideration in the TQFT framework, we
will take the following strategy to find a definition of $\mathcal
M_C$. We will start with the cohomology group of a moduli space, and
look for its modification so that its Poincar\'e polynomial reproduces
\eqref{eq:11}. Then we will study properties of the proposed Coulomb
branch, whether they are compatible with physical expectations. We
will propose such modification in this paper and its sequel
\cite{BFN}.

Let us give a remark. In \eqref{eq:11} we need to assume
$2\Delta(\la)\ge 1$ for any $\la\neq 0$, a `good' or `ugly' theory in
the sense of \cite{MR2610576}, hence no negative powers of $t$
appear. Then \eqref{eq:11} makes sense as a formal power series. This
assumption fails for example the pure theory $(G,\bM) = (\SU(2),0)$,
though $\mathcal M_C$ still exists and is the monopole moduli space as
we discussed above.
Our proposed definition, though motivated by \eqref{eq:11}, will make
sense without this assumption.

\subsection{Higgs branch, \texorpdfstring{$3d$}{3d} mirror symmetry and symplectic duality}\label{sec:3d-mirror-symmetry}

There is a construction of a
\begin{NB}
conical    
\end{NB}%
hyper-K\"ahler manifold from $G$ and $\bM$, which mathematicians can
understand. It is the hyper-K\"ahler quotient construction
\cite{HKLR}. (See \subsecref{sec:hyper-kahl-quot} below for detail.)
However the Coulomb branch $\mathcal M_C$ is {\it not\/} the
hyper-K\"ahler quotient of $\bM$ by $G$. In the above examples with
$\bM = 0$, the hyper-K\"ahler quotient $\bM\tslash G$ is just $\{0\}$
for any $G$. (Or $\emptyset$ if we consider only {\it free\/} orbits.)
For $(G,\bM)=(\U(1), \HH^N)$, the hyper-K\"ahler quotient $\bM\tslash
G$ is the closure of the minimal nilpotent orbit in $\algsl(N,\CC)$,
or the cotangent bundle of $\CC\proj^{N-1}$ if the real part of the
level of the hyper-K\"ahler moment map is nonzero.
In fact, the hyper-K\"ahler quotient of $\bM$ by $G$ arises as the
{\it Higgs branch\/} $\mathcal M_H$, which is yet another mathematical
object associated with the gauge theory $\Hyp(\bM)\tslabar G$. 

For this class of 3-dimensional supersymmetric theories, it has been
noticed that two theories often appear in pairs.
It is called the {\it mirror symmetry\/} in 3-dimensional theories, as
it is similar to more famous mirror symmetry between two Calabi-Yau's.
The first set of examples was found by Intriligator and Seiberg
\cite{MR1413696}. In fact, the above is one of their examples, where
the mirror theory is the gauge theory for $G=\U(1)^N/\U(1)$ with $\bM
= \HH^N$ associated with the affine quiver of type
$A^{(1)}_{N-1}$. (See \subsecref{sec:quiver-type} below.)

When two theories $A$, $B$ form a mirror pair, their Higgs and Coulomb
branches are swapped:
\begin{equation*}
    \mathcal M^A_C = \mathcal M^B_H, \quad
    \mathcal M^A_H = \mathcal M^B_C.
\end{equation*}
This is indeed the case for our example. The hyper-K\"ahler quotient
of $\HH^N$ by $\U(1)^N/\U(1)$ is $\CC^2/(\ZZ/N\ZZ)$, Kronheimer's
construction of ALE spaces for type $A$ \cite{Kr}.

\begin{Remark}
    In order to the above equalities to be literally true, we need to
    replace the multi-Taub-NUT spaces by the corresponding ALE
    spaces. (The multi-Taub-NUT metric has a parameter
    $g_{\mathrm{cl}}$, and it becomes the ALE space when
    $g_{\mathrm{cl}}\to \infty$.) This is because the mirror symmetry
    is a duality in infrared. See \cite[\S3.1]{MR1413696} for a
    physical explanation why this is necessary.
    As we are only interested in the complex structure of $\mathcal
    M_C$, this process makes no change for us. We ignore this point
    hereafter.
\end{Remark}

Therefore the monopole formula \eqref{eq:11} for the theory $A$
computes the Hilbert series of the Higgs branch $\mathcal M^B_H$ of
the mirror theory $B$.
\begin{NB}
Since it has been known that many 
\begin{NB2}
conical
\end{NB2}%
hyper-K\"ahler manifolds are constructed by hyper-K\"ahler
constructions (see \secref{sec:hyper-kahl-quot} below), it is
interesting to look for the mirror theory whose monopole formula
computes the Hilbert series.
\end{NB}%
And there are lots of examples of mirror pairs, and we have a
systematic explanation via branes, dualities, $M$-theory, etc. See
e.g.\ \cite{MR1454291,MR1452322,MR1451054,MR1454292} and also \S\S\ref{sec:type-A-quiver}, \secref{sec:more_mirror}.

Unfortunately these techniques have no mathematically rigorous
foundation, and hence there is no definition of the mirror, which
mathematicians can understand. Moreover, the mirror of a gauge theory
$\Hyp(\bM)\tslabar G$ may not be of a form $\Hyp(\bM')\tslabar G'$ for
some $\bM'$ and $G'$ in general, as we will explain in
\subsecref{sec:mirror-could-be}.
Thus it seems that the mirror symmetry is more difficult to work with,
and we will not use it to look for the mathematical definition of
$\mathcal M_C^A$. In turn, we could hope that our proposed definition
of $\mathcal M^A_C$ will shed some light on the nature of the $3d$
mirror symmetry.

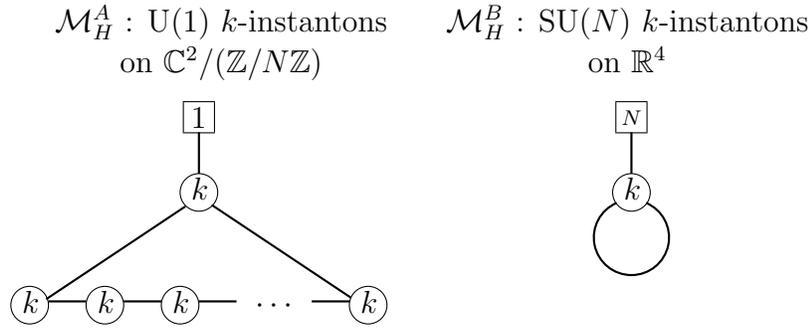
\begin{figure}[htbp]
    \centering
\setlength{\unitlength}{1mm}
\begin{picture}(105,40)
    \put(3,32){
      \shortstack{$\mathcal M^A_H$ : $\U(1)$ $k$-instantons \\ on $\CC^2/(\ZZ/N\ZZ)$}}
    \multiput(1,1)(10,0){3}{\circle{5}}
    \multiput(0,0)(10,0){3}{$k$}
    \multiput(3.5,1.5)(10,0){3}{\thicklines\line(1,0){5}}
    \put(31,0){$\cdots$}
    \put(38.5,1.5){\thicklines\line(1,0){5}}
    \put(46,1){\circle{5}}
    \put(45,0){$k$}
    \put(23.5,16){\circle{5}}
    \put(22.5,15){$k$}
    \put(3.2,1.8){\thicklines\line(3,2){18.5}}
    \put(43.7,2){\thicklines\line(-3,2){18.5}}
    \put(21.5,24){\framebox(4,4){$1$}}
    \put(23.5,18.5){\thicklines\line(0,1){5.5}}
%
    \put(55,32){
      \shortstack{$\mathcal M^B_H$ : $\SU(N)$ $k$-instantons \\on $\RR^4$}}
    \put(81,16){\circle{5}}
    \put(80,15){$k$}
    \put(79,24){\framebox(4,4){$\scriptstyle N$}}
    \put(81,18.5){\thicklines\line(0,1){5.5}}
    \put(81,10){\thicklines\arc[-247,67]{5}}
\end{picture}
\caption{An example of a mirror pair}
    \label{fig:mirror}
\end{figure}

Let us continue an example of a mirror pair.
The pair in Figure~\ref{fig:mirror}, both of quiver types, is known to
be mirror each other \cite{MR1454291}.
Higgs branches are the $k^{\mathrm{th}}$ symmetric power of
$\CC^2/(\ZZ/N\ZZ)$ and the framed moduli space space of $\SU(N)$
$k$-instantons on $\RR^4$, given by the ADHM description
respectively. This includes the above example as $k=1$ case.
The former can be considered as the framed moduli space of $\U(1)$ $k$-instantons on $\CC^2/(\ZZ/N\ZZ)$.
Their Hilbert series has been computed. For $\mathcal M^A_H$, it can
be written in terms of the Hilbert series of $\CC^2/(\ZZ/N\ZZ)$ as it
is a symmetric product. For $\mathcal M^B_H$, the Hilbert series is
the $K$-theoretic Nekrasov partition function (or $5$-dimensional
partition function in physics literature), and can be computed via
fixed point localization or the recursion by the blowup equation,
e.g., see \cite{MR2183121}.

We should emphasize that it is not obvious to see why the monopole
formula reproduces those results, as \eqref{eq:11} looks very
different from the known expression. For $\mathcal M^B_C$, this is
possible after some combinatorial tricks. See \secref{sec:app}.
(It was checked in \cite{Cremonesi:2013lqa} for small $k$.)
But it is not clear how to check for $\mathcal M^A_C$.

Let us mention that Braden et al.\ \cite{2014arXiv1407.0964B} expect
that the mirror symmetry is related to the symplectic duality, which
states an equivalence between categories attached to symplectic
resolutions of two different conical hyper-K\"ahler manifolds
$\mathcal M^A_H = \mathcal M^B_C$ and $\mathcal M^A_C = \mathcal
M^B_H$. The definition of categories and the dual pair require also
that both symplectic resolutions have torus action with {\it finite\/}
fixed points.
Since $\mathcal M^A_H$, $\mathcal M^A_C$ do not have symplectic
resolutions nor torus action with finite fixed points in general, the
symplectic duality deals with much more restrictive situations than
ones considered here.
\begin{NB}
    Also it is not clear, at least to the author, that these
    categories are related to our main object, the critical cohomology
    of the space of connections and sections of bundles over $\proj^1$
    as we will introduce below.
\end{NB}%
If $\mathcal M^A_H$ satisfies these two conditions, it is natural to
expect the same is true for $\mathcal M^A_C$, as twos are interchanged
under the mirror symmetry, as will be explained in
\subsecref{subsec:linebundle}. Note that it is usually easy to check
these conditions for $\mathcal M^A_H$, and we have lots of examples,
say quiver varieties of type $A$ or affine type $A$, toric
hyper-K\"aher manifolds.

No general recipe to construct a symplectic duality pair was given in
\cite{2014arXiv1407.0964B}. Since we will propose a definition of
$\mathcal M^A_C$ in this paper, this defect will be fixed. Moreover we
hope that we could give a better understanding on the symplectic
duality, and gain a possibility to generalize it to more general cases
when two conditions above are not satisfied.
We have interesting sets of examples, where only one of two conditions
is satisfied.

\subsection{Hikita conjecture}

Recently Hikita \cite{2015arXiv150102430H} proposes a remarkable
conjecture. Suppose that $G$ is a product of general linear groups,
such as quiver gauge theories \subsecref{sec:quiver-type} and abelian
cases \subsecref{sec:abelian-theory}. Using perturbation of the moment
map equation, we can modify the hyper-K\"ahler quotient $\bM\tslash G$
to $\bmu^{-1}(\zeta)/G$. In many cases, it is a smooth manifold, and assume that this happens.

On the Coulomb branch side, we have an action of a torus $T$ on
$\mathcal M_C$, where $T$ is the Pontryagin dual of $\pi_1(G)$. See
(\ref{item:group}) in \subsecref{sec:expect-properties} below. Let
$\mathcal M_C^T$ be the fixed point subscheme.
Hikita conjectures that there exists a ring isomorphism
\begin{equation*}
    \CC[\mathcal M_C^T] \overset{\operatorname{\scriptstyle ?}}{\cong} H^*(\bmu^{-1}(\zeta)/G),
\end{equation*}
and checks for several nontrivial cases. This conjecture obviously has
a similar flavor with our study, though a precise relation is not
clear yet. The author is currently considering what happens if we
consider the equivariant quantum cohomology group of
$\bmu^{-1}(\zeta)/G$. It seems the quantized Coulomb branch defined in
\cite{BFN} plays a role.

\subsection{A proposal of a definition of \texorpdfstring{$\mathcal M_C$}{MC}}\label{sec:prop-defin-texorpdfs}

We take the cohomology group of moduli spaces of solutions of the
generalized vortex equation for the gauged nonlinear $\sigma$-model on
$S^2$ as a starting point for a definition of $\mathcal M_C$, as
discussed in \subsecref{sec:2+1-tqft}. The equation will be discussed
in detail in \secref{sec:topological-twist}. See
\eqref{eq:30}. However we need to modify the definition, as we cannot
obtain a reasonable answer for $\Hyp(0)\tslabar\SU(2)$, as we have
already remarked.

Our proposal here is the following modifications:
\begin{enumerate}
      \item Drop the last equation in \eqref{eq:30},
which is related to the stability condition via the Hitchin-Kobayashi
correspondence.
  \item Consider the cohomology group with coefficients in the sheaf
of a vanishing cycle.
\end{enumerate}
Thus we propose
\begin{equation*}
    \CC[\mathcal M_C] \overset{\operatorname{\scriptstyle ?}}{=}
    H^{*+\dim\mathcal F-\dim\mathcal G_\CC(P)}_{c,\mathcal G_\CC(P)}
    (\left\{ (A,\Phi) \middle|
      \begin{aligned}[m]
          & (\DB+A)\Phi = 0\\
        & \bmu_\CC(\Phi) = 0
      \end{aligned}
    \right\}, 
    \varphi_{\CS}(\CC_{\mathcal F}))^*,
\end{equation*}
where $\DB+A$ is a partial connection on a $G_\CC$-bundle $P$ on $S^2
= \proj^1$, $\Phi$ is a section of an associated vector bundle twisted
by $\shfO_{\proj^1}(-1)$, and $\bmu_\CC$ is the complex moment
map. And $\mathcal F$ is the space of all $(A,\Phi)$ imposing no
equations, $\mathcal G_\CC(P)$ the complex gauge group, and
$\varphi_{\CS}$ is the vanishing cycle functor associated with the
generalized Chern-Simons functional $\CS$ defined on $\mathcal F$. See
\secref{sec:motiv-DT-invariants} for more detail.

The moduli space above can be {\it loosely\/} regarded as the space
parametrizing twisted holomorphic maps from $\proj^1$ to the Higgs
branch $\mathcal M_H$. It is literally true if we replace $\mathcal
M_H = \bmu_\CC^{-1}(0)\dslash G_\CC$ by the quotient stack
$[\bmu_\CC^{-1}(0)/G_\CC]$.
\begin{NB}
    Here twisted maps mean sections of the bundle
    $\bmu_\CC^{-1}(0)/G_\CC\times_{\CC^\times}(\CC^2\setminus\{0\})$
    over $\proj^1$, where $\CC^\times$ acts on $\Phi$ by
    multiplication with weight $1$.
\end{NB}%

The vanishing cycle functor $\varphi_{\CS}$ with respect to the
generalized Chern-Simons functional $\CS$ is strongly motivated by the
one appearing in the theory of Donaldson-Thomas invariants for
Calabi-Yau 3-categories.

The whole paper is devoted to explain why these modifications are
natural. We see that (1) is inevitable even at this stage:
\eqref{eq:30} is just $F_A = 0$ for $\Hyp(0)\tslabar\SU(2)$. We cannot
think of any reasonable modification, which gives us a non-trivial
solution for $S^2$.
For $\Hyp(\HH)\tslabar \U(1)$, moduli spaces depend on the choice of a
stability condition, though the Coulomb branch should not. This
problem apparently is related to the dependence of invariants of
perturbation, mentioned at the end of \subsecref{sec:2+1-tqft}. It has
a similar flavor with the problem arising the definition of $\SU(r)$
Casson invariants, mentioned in~\subsecref{sec:relat-topol-invar}.
Thus forgetting the equation and considering all connections seem the
only reasonable candidate for the modification.

The definition of the multiplication, when $\bM$ is of cotangent type
as explained below, will be postponed to \cite{BFN}. The goal of
\cite{BFN} will be to propose a definition of $\mathcal M_C$ as an
affine scheme, i.e., the definition of its coordinate ring as a
commutative ring. There remain lots to be done, in particular, we have
no idea how to define a hyper-K\"ahler metric on $\mathcal M_C$ at
this moment, though we could construct a natural noncommutative
deformation (or quantization) of $\mathcal M_C$.

Also we are very far from checking our proposal reproduces various
known examples already mentioned above, except $\bM=0$ and toric
cases. Nevertheless we will reproduce the monopole formula
\eqref{eq:11}, so we believe that our proposal passes the first check
that it is a correct mathematical definition of Coulomb branches.

The paper is organized as follows. 
In \secref{sec:exampl-hyper-kahl} we review the hyper-K\"ahler
quotient construction and examples of hyper-K\"ahler manifolds arising
in this way. 
In \secref{sec:more_mirror} we give examples where the mirror of a gauge theory is not a gauge theory though it is still a reasonable theory.
In \secref{sec:monopole-formula} we review the monopole formula and
its various properties. In particular, we start to list expected
properties of the Coulomb branch suggested from the monopole formula.
In \secref{sec:flavor-symmetry} we review the monopole formula when a
gauge theory has an additional flavor symmetry. We add a few
properties to the list.
Up to here, all materials are review of earlier works.

In \secref{sec:topological-twist} we consider a generalized
Seiberg-Witten equation associated with a gauge theory
$\Hyp(\bM)\tslabar G$ and study the compactness property of the moduli
space. We also study the dimension reduction of the equation to write
down the generalized vortex equation, which will lead us to the
proposed definition of the Coulomb branch.
In \secref{sec:motiv-DT-invariants} we observe that the complex part
of the reduced equation on a Riemann surface arises the Euler-Lagrange
equation of an analog of the holomorphic Chern-Simons functional, and
hence it is natural to consider the analog of Donaldson-Thomas
invariants, or more precisely the cohomology of the vanishing
cycle. We then formally apply results on the vanishing cycle, known in
the finite dimensional situation, to our case to reduce the equation
further.
In \secref{sec:comp-P1} we compute the dimension of the cohomology
group and check that it reproduces the monopole formula.
\secref{sec:c=cc} is a detour where we find that a few earlier works
are nicely fit with the framework of \secref{sec:motiv-DT-invariants} when the curve is the complex line $\CC$.

In \secref{sec:classical} we give further examples of hyper-K\"ahler
quotients related to instantons for classical groups.
In \secref{sec:app} we give a computation of the monopole formula in a
particular example when the Coulomb branch is a symmetric product of
a surface.

\subsection*{Acknowledgment}

This study of Coulomb branches was started after Amihay Hanany's talk
at Warwick EPSRC Symposium: McKay correspondence, Orbifolds, Quivers,
September 2014. The author is grateful to him for his talk as well as
detailed discussion on the monopole formula. The author's knowledge on
Coulomb branches is based on Edward Witten's three lectures at Newton
Institute, Cambridge in November 1996. It was great pleasure for the
author to report this work 18 years later at the workshop in
celebration of the 30th Kyoto Prize awarded to Professor Witten.
He also thanks
Alexander Braverman,
Tudor Dimofte,
Michael Finkelberg,
Tatsuyuki Hikita,
Justin Hilburn,
Kentaro Hori,
Anton Kapustin,
Yoshiyuki Kimura,
Alexei Oblomkov,
Andrei Okounkov,
Bal\'azs Szendr\"oi,
and Yuji Tachikawa
for discussion and comments on the subject. Parts of this paper were
written while the author was visiting MSRI, Higher School of Economics, Institute Mittag-Leffler. He is grateful to their hospitality.
Last but not least, the author would like to express his hearty thanks to
the late Dr.\ Kentaro Nagao.
Many techniques used in this paper originate Nagao's works.

This research is supported by JSPS Kakenhi Grant Numbers 
22244003, 
23224002, 
23340005, 
24224001, 
25220701. 

\section{Examples of hyper-K\"ahler quotients}\label{sec:exampl-hyper-kahl}

In view of the $3d$ mirror symmetry, it is natural to expect that the
Higgs branch $\mathcal M_H$ and the Coulomb branch $\mathcal M_C$
share similar properties. Therefore it is important to have examples
of gauge theories whose Higgs branches (i.e., hyper-K\"ahler quotients
of linear spaces) are well-understood. In this section we prepare
notation and basics of hyper-K\"ahler quotients in the first three
subsections, and then we review two important classes of gauge
theories, quiver gauge theories and abelian theories. Further examples
will be given in \secref{sec:classical}.

Here we consider hyper-K\"ahler quotients only in finite dimension. If
we allow infinite dimensional ones, we have more examples, such as
instanton moduli spaces for arbitrary gauge groups, solutions of
Nahm's equation, etc. However it is not clear how to consider the
corresponding Coulomb branches, in particular, the monopole formula
introduced in \secref{sec:monopole-formula} below.

\subsection{Hyper-K\"ahler quotients of linear spaces}\label{sec:hyper-kahl-quot}

Let $G$ be a compact Lie group with the Lie algebra $\g$. Let $\bM$ be
its quaternionic representation. 
Let $I$, $J$, $K$ denote multiplication by $i$, $j$, $k$,
considered as linear operators on $\bM$.
A quaternionic representation of $G$ is a representation such that
the $G$-action commutes with $I$, $J$, $K$.
\begin{NB}
Viewing $\bM = \mathbb H^N$ as row vectors, we consider $G$ as a
subgroup of $\GL(N,\mathbb H)$, acting on $\bM$ by the right
multiplication. It is $\mathbb H$-linear with respect to the {\it
  left\/} multiplication of $\mathbb H$.
\end{NB}%
We suppose $\bM$ has a $G$-invariant inner product $(\ ,\ )$ which is
hermitian with respect to all $I$,$J$,$K$. Therefore $\bM$ is a
hyper-K\"ahler manifold with a $G$-action preserving the
hyper-K\"ahler structure.
We have the hyper-K\"ahler moment map $\bmu\colon \bM\to \mathfrak g^*\otimes\RR^3$, vanishing at the origin:
\begin{equation*}
    \langle \xi, \bmu(\phi) \rangle = 
    \frac{1}2 ((I\xi\phi,\phi), (J\xi\phi,\phi), (K\xi\phi,\phi)),
\end{equation*}
where $\phi\in \bM$, $\xi\in\mathfrak g$, and $\langle\ ,\ \rangle$ is
the pairing between $\mathfrak g$ and its dual $\mathfrak g^*$. A
hyper-K\"ahler moment map is, by definition, (a) $G$-equivariant, and
(b) satisfying
\begin{equation*}
    \langle\xi,d\bmu_\phi(\dot\phi)\rangle 
    = (\omega_I(\xi^*, \dot\phi),\omega_J(\xi^*, \dot\phi),
    \omega_K(\xi^*, \dot\phi)),
\end{equation*}
where $\dot\phi$ is a tangent vector, $\xi^*$ is the vector field
generated by $\xi$, and $\omega_I$, $\omega_J$, $\omega_K$ are
K\"ahler forms associated with three complex structures $I$, $J$, $K$
and the inner product. It is direct to see that two properties are
satisfied in the above formula.

\begin{NB}
    Suppose $(G,\bM) = (\U(1),\HH)$.  Let $\phi = z_1 + z_2 j$. Here
    $G=\U(1)$ acts by the right multiplication: $\lambda\cdot \phi =
    \phi\lambda^{-1} = \lambda^{-1} z_1 + \lambda z_2 j$. Therefore
    $\xi\phi = -\xi z_1 + \xi z_2 j$. Then
    \begin{equation*}
        \left\{
        \begin{aligned}[m]
            & \langle \xi,\bmu_I(\phi)\rangle = -\frac{i}2\left(
              |z_1|^2 - |z_2|^2
            \right)\xi,
            \\
            & \langle \xi,\bmu_J(\phi)\rangle + i\langle\xi,\bmu_K(\phi)\rangle
            = -z_1z_2\xi.
        \end{aligned}\right.
    \end{equation*}
    If we identify $\g^*$ with $\g$ by the inner product
    $(\bullet,\circ) = \operatorname{Re}(\bullet\overline{\circ})$, we
    have
    \begin{equation*}
        \bmu_I(\phi) = \frac{i}2\left(
              |z_1|^2 - |z_2|^2
            \right),\quad
            \bmu_J(\phi)+i\bmu_K(\phi) = z_1 z_2.
    \end{equation*}
    We use an $i \su(2)$-expression
    \begin{equation*}
        \begin{split}
        & \bmu_I(\phi)
        \begin{bmatrix}
            -i & 0 \\ 0 & i
        \end{bmatrix}
        + \bmu_J(\phi)
        \begin{bmatrix}
            0 & -1 \\ 1 & 0
        \end{bmatrix}
        + \bmu_K(\phi)
        \begin{bmatrix}
            0 & i \\ i & 0
        \end{bmatrix}
\\
        =\; &
        \begin{bmatrix}
            \frac12(|z_1|^2 - |z_2|^2) & \overline{z_1 z_2}
            \\
            z_1z_2 & -\frac12(|z_1|^2 - |z_2|^2)
        \end{bmatrix}
        =
        \left(\begin{bmatrix}
            \overline{z_1} \\ z_2
        \end{bmatrix}
        \begin{bmatrix}
            z_1 & \overline{z_2}
        \end{bmatrix}\right)_0,
        \end{split}
    \end{equation*}
    which is a standard notation in the Seiberg-Witten theory.
\end{NB}

In the following, we only use the underlying complex symplectic
structure. Let us give another formulation. Let $G_\CC$ be the
complexification of $G$. Let $\bM$ be its complex representation, which
has a complex symplectic form $\omega_\CC$ preserved by $G_\CC$. We
have the complex moment map $\bmu_\CC\colon \bM\to \mathfrak g_\CC$
\begin{equation*}
    \langle\xi,\bmu_\CC(\phi)\rangle = \frac12 \omega_\CC(\xi\phi,\phi),
\end{equation*}
where $\xi\in\mathfrak g_\CC$. This is the complex part of the
hyper-K\"ahler moment map under the identification $\RR^3\cong
\RR\oplus\CC$.

We consider the hyper-K\"ahler quotient
\begin{equation*}
    \bM\tslash G \defeq
    \bmu^{-1}(0)/G \cong \bmu_\CC^{-1}(0)\dslash G_\CC,
\end{equation*}
where $\dslash G_\CC$ is the affine algebro-geometric invariant theory
quotient, and the second isomorphism follows from a result of
Kempf-Ness (see e.g., \cite[Th.~3.12]{Lecture}).
Let $\bmu^{-1}(0)^{\text{reg}}$ be the (possibly empty) open subset of
$\bmu^{-1}(0)$, consisting of points with trivial stabilizers. Then $G$
acts freely on $\bmu^{-1}(0)^{\text{reg}}$, and the quotient
$\bmu^{-1}(0)^{\text{reg}}/G$ is a smooth hyper-K\"ahler manifold.
The Higgs branch of the 3d $N=4$ SUSY gauge theory $\Hyp(\bM)\tslabar
G$ is $\bmu^{-1}(0)^{\text{reg}}/G$ or its closure in $\bM\tslash G$
depending on the situation.
We only consider $\Hyp(\bM)\tslabar G$ hereafter, but we actually mean
$\bmu^{-1}(0)^{\text{reg}}/G$ when we talk the hyper-K\"ahler structure
on it.

The hyper-K\"ahler quotient $\bM\tslash G$ has a natural
$\SU(2)=\grpSp(1)$-action induced from the $\HH$-module structure of
$\bM$. It commutes with the $G$-action, and hence descends to the
quotient. It rotates the hyper-K\"ahler structure.
In the complex symplectic notation, its restriction to $\U(1)$-action
is induced by the scalar multiplication on $\bM$.

\subsection{Cotangent type}

There is a class of quaternionic representations, which we call {\it
  cotangent type}. Let $\bN$ be a complex representation of $G$. We
put a $G$-invariant hermitian inner product on $\bN$.
Then $\bM = \bN\oplus \bN^*$ is a quaternionic representation of $G$.
We define $J$ by $J(x,y) = (-y^\dagger, x^\dagger)$ for $x\in\bN$,
$y\in\bN^*$, where $x^\dagger\in\bN^*$ is defined by
\(
   \langle x^\dagger, n\rangle = (n,x)
\)
for $n\in \bN$, and $y^\dagger$ is defined so that $(x^\dagger)^\dagger
= x$ for all $x\in\bN$.
Then $J$ is skew-linear as the hermitian
inner product is skew-linear in the second variable. We have $J^2 =
-\operatorname{id}$ from the definition.

\begin{NB}
    An easiest example of {\it non-cotangent type\/} $\bM$ is
    $G=\grpSp(1)$ with $\bM = \HH$. We regard $\HH$ as an $\HH$-module
    by the left multiplication, where $G$ acts by the right
    multiplication. They commute. For this example, it is claimed that
    the Coulomb branch does not exist \cite{MR1490862}. I do not
    understand a mathematical explanation of this claim.
\end{NB}

In the (complex) symplectic formulation, we start with a complex
representation $\bN$ of $G_\CC$, and take $\bM = \bN\oplus \bN^*$ with a
natural symplectic structure. Then $\bM$ is naturally a representation
of $G_\CC$, and has a symplectic form preserved by $G_\CC$.

\subsection{Complete intersection}\label{sec:compl-inters}

Although $\bmu_\CC^{-1}(0)\dslash G_\CC$ makes sense as an affine
scheme without any further condition, we do not expect they behave
well in general. We propose to assume
\begin{itemize}
      \item $\bmu_\CC^{-1}(0)$ is a complete intersection in $\bM$.
\end{itemize}
More precisely it means as follows. We construct a Koszul complex from
$\bmu_\CC$:
\begin{equation*}
    0\to \Wedge^{\dim\mathfrak g_\CC}\mathfrak g_\CC\otimes\shfO_{\bM}
    \to\cdots\to \Wedge^2 \mathfrak g_\CC\otimes\shfO_{\bM} 
    \to \mathfrak \g_\CC\otimes\shfO_{\bM} \to \shfO_{\bM}
    \to \shfO_{\bmu_\CC^{-1}(0)}\to 0.
\end{equation*}
Our assumption says that this is exact. Under this assumption, the
coordinate ring of $\bmu_\CC^{-1}(0)$ is given by
\begin{equation*}
    \CC[\bmu_\CC^{-1}(0)] = \sum_{i=0}^{\dim\mathfrak g_\CC}
    (-1)^i \Wedge^i \mathfrak g_\CC\otimes \CC[\bM].
\end{equation*}
This is an equality of virtual $\CC^*\times G_\CC$-modules. Taking the
$G_\CC$-invariant part, we get the Hilbert series of
$\CC[\bmu_\CC^{-1}(0)\dslash G_\CC]$. This is the definition used in
the $K$-theoretic Nekrasov partition function for instantons of
classical groups. See \cite{MR2104883}.

We expect that this complete intersection assumption is related to the
`good' or `ugly' condition, appearing in the monopole formula. But it
is a superficial observation as we are discussing the Higgs branch
now, while the monopole formula is about the Coulomb branch.

\subsection{Quiver gauge theory}\label{sec:quiver-type}

Let $Q = (Q_0,Q_1)$ be a quiver, where $Q_0$ is the set of vertices and
$Q_1$ is the set of arrows.
We can construct a quaternionic representation of $G = G_V =
\prod_{i\in Q_0}\U(V_i)$ associated with $Q_0$-graded representations
$V = \bigoplus V_i$, $W = \bigoplus W_i$. It is a cotangent type, and
given in the complex symplectic description by
\begin{equation*}
    \begin{split}
    & \bN = \bigoplus_{h\in Q_1} \Hom(V_{\vout{h}}, V_{\vin{h}})
    \oplus\bigoplus_{i\in Q_0} \Hom(W_i,V_i),
\\
    & \bM = 
    \begin{aligned}[t]
    \bigoplus_{h\in Q_1} \Hom(V_{\vout{h}}, & V_{\vin{h}})
    \oplus \Hom(V_{\vin{h}},V_{\vout{h}})
    \\
    & \oplus\bigoplus_{i\in Q_0} \Hom(W_i,V_i)\oplus\Hom(V_i,W_i).
    \end{aligned}
    \end{split}
\end{equation*}
Here $\vout{h}$ and $\vin{h}$ are the outgoing and incoming vertices
of the oriented edge $h\in Q_1$ respectively.

If $Q$ is the Jordan quiver, the hyper-K\"ahler quotient
$\bmu^{-1}(0)/G$ is the ADHM description of the framed moduli space of
$\SU(W)$-instantons on $\RR^4$, or more precisely its Uhlenbeck
partial compactification (see e.g., \cite[Ch.3]{Lecture} and the
reference therein).
More generally, hyper-K\"ahler quotient $\bM\tslash G =
\bmu_\CC^{-1}(0)\dslash G_{\CC}$ is the quiver variety (with complex
and stability parameters $0$), introduced in \cite{Na-quiver}.

For a quiver gauge theory, $Z(\mathfrak g^*) = \{
\zeta_\RR\in\mathfrak g^* \mid \text{$\operatorname{Ad}^*_g(\zeta_\RR)
  = \zeta_\RR$ for any $g\in G$}\}$ is non\-trivial, and is isomorphic
to $\bigoplus_{i\in Q_0} \RR \sqrt{-1}\tr_{V_i}$, where
$\tr_{V_i}\colon\mathfrak u(V_i)\to\sqrt{-1}\RR$ is the trace.
Then we can form a perturbed hyper-K\"ahler quotient $\bmu^{-1}(
\zeta)/G$ for $\zeta\in\RR^3\otimes Z(\mathfrak g^*)$.
If we decompose $\zeta$ as
$(\zeta_\RR,\zeta_\CC)$ according to $\RR^3 = \RR \oplus \CC$, we have
an algebro-geometric description
\begin{equation*}
    \bmu^{-1}(\zeta)/G
    \cong \bmu_\CC^{-1}(\zeta_\CC)
    \dslash\!\raisebox{-4pt}{$\scriptstyle\zeta_\RR$} G_\CC,
\end{equation*}
where $\dslash\!\raisebox{-4pt}{$\scriptstyle\zeta_\RR$}$ is the GIT
quotient with respect to the $\zeta_\RR$-stability. (See
\cite[\S3]{Na-quiver}.) If $\zeta$ is generic, $\bmu^{-1}(\zeta)/G$ is
a smooth hyper-K\"ahler manifold whose metric is complete. Moreover,
we have a projective morphism
\begin{equation*}
    \pi\colon\bmu_\CC^{-1}(\zeta_\CC)
    \dslash\!\raisebox{-4pt}{$\scriptstyle\zeta_\RR$} G_\CC
    \to \bmu_\CC^{-1}(\zeta_\CC)\dslash G_\CC = \bmu^{-1}(0,\zeta_\CC)/G.
\end{equation*}

In an algebro-geometric approach to hyper-K\"ahler quotients, it is
more natural to replace $\zeta_\RR$ by a corresponding element for the
group $G$, i.e., a character $\chi\colon G\to\U(1)$, where they are
related by $\zeta_\RR = d\chi$.
The character $\chi$ defines a $G_\CC$-equivariant structure on the
trivial line bundle over $\bmu_\CC^{-1}(\zeta_\CC)$. We introduce the
stability condition and form a GIT quotient with a natural projective
morphism to $\bmu_\CC^{-1}(\zeta_\CC)\dslash G$:
\begin{equation}\label{eq:37}
    \pi\colon \bmu_\CC^{-1}(\zeta_\CC)
    \dslash\!\raisebox{-4pt}{$\scriptstyle\chi$} G_\CC
    \to \bmu_\CC^{-1}(\zeta_\CC)\dslash G_\CC 
\end{equation}
It is equipped with a relatively ample line bundle $\mathcal
L_{\chi}$ in a natural way. See \cite[\S3]{Lecture} for detail.

In Figure~\ref{fig:mirror} we follow physicists convention. The
underlying graph of the quiver is circled vertices and edges
connecting them. (An orientation of the quiver is not relevant, and
usually omitted.) Dimensions of $V_i$ are put in circled vertices,
while dimensions of $W_i$ are put in the boxed vertices, connected to
the corresponding circled vertices. It is more or less the same as the
original convention in \cite{KN}.

When $W=0$, the scalar $\U(1)$ acts trivially, so we replace $G$ by
$\prod_{i\in Q_0}\U(V_i)/\U(1)$. This is the case for $\bM$ used by
Kronheimer \cite{Kr}, mentioned in Introduction.

For a quiver gauge theory, good and ugly conditions were analyzed in
\cite[\S2.4, \S5.4]{MR2610576}. It is conjectured, for example, that a
quiver gauge theory of finite type is good or ugly if and only if
\begin{equation}\label{eq:17}
    \dim W_i - \sum_j (2\delta_{ij} - a_{ij})\dim V_j \ge -1
\end{equation}
for any $i\in Q_0$.\footnote{The conjecture is stated only for
  goodness, and the right hand side is replaced by $0$ in
  \cite{MR2610576}.} Here $a_{ij}$ is the number of edges (regardless
of orientation) between $i$ and $j$ if $i\neq j$, and its twice if $i=j$.

For quiver varieties $\bmu_\CC^{-1}(0)\dslash G_\CC$ with $W=0$,
Crawley-Boevey \cite[Th.1.1]{CB} gave a combinatorial condition for
the complete intersection assumption above. It can be modified to
cover the $W\neq 0$ case using the trick \cite[the end of
Introduction]{CB}. To state the result, let us prepare notation.
Let $\mathbf C = (2\delta_{ij} - a_{ij})$ be the Cartan matrix. We
denote the dimension vectors $(\dim V_i)_{i\in Q_0}$, $(\dim
W_i)_{i\in Q_0}$ by $\mathbf v$, $\mathbf w$ respectively. A {\it
  root\/} is an element of $\ZZ^{Q_0}$ obtained from the coordinate
vector at a loopfree vertex or $\pm$ an element of the fundamental
region by applying a sequence of reflections at loopfree vertices.
If there are no loops, this notion coincides with the usual notion of
roots of the corresponding Kac-Moody Lie algebra by
\cite[Th.~5.4]{Kac}.

Then $\mu_\CC^{-1}(0)$ is a complete intersection if and only if the
following is true:
\begin{itemize}
      \item ${}^t \mathbf v (2\mathbf w - \mathbf C \mathbf v) \ge
    {}^t\mathbf v^0 (2\mathbf w - \mathbf C \mathbf v^0) + \sum_k (2 -
    {}^t \beta^{(k)}\mathbf C\beta^{(k)})$ for any decomposition
    $\mathbf v = \mathbf v^0 + \sum_k \beta^{(k)}$ such that $\mathbf
    w - \mathbf v^0$ is a weight of an irreducible highest weight
    module $V(\mathbf w)$ of the highest weight $\mathbf w$, and
    $\beta^{(k)}$ is a positive root.
\end{itemize}
(cf.\ \cite[Th.2.15(2)]{Na-branching} for a closely related
condition.)
The dominance condition \eqref{eq:17} is a necessary condition, from
the decomposition $\mathbf v = \mathbf v^0 + \alpha_i$, but not a
sufficient if there is $\beta^{(k)}$ with $\lsp{t}\beta^{(k)}\mathbf
C\beta^{(k)}< 2$, i.e., an imaginary root. Anyway these two conditions
are closely related. This is the reason why we expect the complete
intersection assumption and the `good or ugly' condition are the same.

\subsection{Type \texorpdfstring{$A$}{A} quiver, nilpotent orbits and
  affine Grassmannian}\label{sec:type-A-quiver}

As a special class of quiver gauge theories, type A (or linear) quiver
gauge theories are important. The hyper-K\"ahler quotient, in other
words, Higgs branch was identified with $\mathcal O_\mu\cap \mathcal
S_\lambda$, where $\mathcal O_\mu$ is a nilpotent orbit and $\mathcal
S_\lambda$ is Slodowy slice to another orbit $\mathcal O_\lambda$ of
type A (see \cite[\S8]{Na-quiver}). Here we assume
$\bmu^{-1}(0)^{\text{reg}}\neq\emptyset$, and two partitions
$\lambda$, $\mu$ are defined from dimensions of $V$, $W$ by an
explicit formula.\footnote{There is typo in
  \cite[\S8]{Na-quiver}. $\mu$ must be replaced by its transpose.}
Conversely any $\mathcal O_\mu\cap\mathcal S_\lambda$ for type $A$ is
described as a hyper-K\"ahler quotient.

Let us recall how the identification is constructed. The starting
point is Kronheimer's realization \cite{MR1072915} of $\mathcal
O_\mu\cap \mathcal S_\lambda$ as moduli spaces of $\SU(2)$-equivariant
instantons on $\RR^4$. This construction works for any compact Lie
groups. For type $A$, we apply the ADHM transform to these instantons.
Suppose that an instanton corresponds to $(B_1,B_2,a,b)\in \bM =
\Hom(V,V)^{\oplus 2}\oplus\Hom(W,V)\oplus \Hom(V,W)$, the data for the
Jordan quiver.
If the original instanton is $\SU(2)$-equivariant, $V$, $W$ are
representations of $\SU(2)$, and $(B_1,B_2)$, $a$, $b$ are
$\SU(2)$-linear. Here we mean the pair $(B_1,B_2)$ is
$\SU(2)$-equivariant, when it is considered as a homomorphism in
$\Hom(V,V\otimes \rho_2)$, where $\rho_2$ is the vector representation
of $\SU(2)$.
Let $\rho_i$ be the $i$-dimensional irreducible representation of
$\SU(2)$. We decompose $V$, $W$ as $\bigoplus V_i\otimes\rho_i$,
$\bigoplus W_i\otimes\rho_i$.
Then $a$, $b$ are maps between $V_i$ and $W_i$.
By the Clebsch-Gordan rule $\rho_i\otimes\rho_2 =
\rho_{i-1}\oplus\rho_{i+1}$, $(B_1,B_2)$ decomposes into maps
between $V_i$ and $V_{i-1}\oplus V_{i+1}$.
The dimensions $V_i$, $W_i$ are determined by $\lambda$, $\mu$, as
mentioned above. In a nutshell, the McKay quiver for $\SU(2)$ is the
double of type $A_\infty$ Dynkin graph
(figure~\ref{fig:Ainfty}). Hence we get a quiver variety of type
$A_\infty$.

\begin{figure}[htbp]
    \centering
    \begin{equation*}
        \overset{1}{\bullet}\leftrightarrows\overset{2}{\bullet}
        \leftrightarrows\overset{3}{\bullet}
        \leftrightarrows\overset{4}{\bullet}
        \leftrightarrows\cdots
    \end{equation*}
    \caption{McKay quiver for $\SU(2)$}
    \label{fig:Ainfty}
\end{figure}

A quiver variety of type $A$ can be obtained also from a framed moduli
space of $S^1$-equivariant instantons on $\RR^4$, where $S^1$ acts on
$\RR^4=\CC^2$ by $t\cdot(x,y) = (tx, t^{-1}y)$.
The reason is the same as above: (a) irreducible representations
$\rho_i$ of $S^1$ are parametrized by integers $i\in\ZZ$, i.e.,
weights, and (b) $\rho_i\otimes \CC^2 = \rho_{i-1}\oplus \rho_{i+1}$,
where $\CC^2$ is the base manifold, identified with $\rho_1\oplus
\rho_{-1}$ as an $S^1$-module. Strictly speaking, McKay quiver for
$S^1$ is slightly different from one for $\SU(2)$, and infinite in
both direction (figure~\ref{fig:Adoubleinfty}). But the quiver
varieties remain the same as $V$ is finite-dimensional.
\begin{figure}[htbp]
    \centering
        \begin{equation*}
        \cdots\leftrightarrows
        \overset{-2}{\bullet}\leftrightarrows            
        \overset{-1}{\bullet}\leftrightarrows
        \overset{0}{\bullet}\leftrightarrows\overset{1}{\bullet}
        \leftrightarrows\overset{2}{\bullet}
        \leftrightarrows\cdots
    \end{equation*}
    \caption{McKay quiver for $S^1$}
    \label{fig:Adoubleinfty}
\end{figure}

By \cite[\S5]{braverman-2007} the framed moduli space of
$S^1$-equivariant $G$-instantons on $\RR^4$ is also identified with
the intersection $\mathcal W^\mu_{G,\la}$ of a $G[[z]]$-orbit
$\Gr^\mu_G$ in the affine Grassmannian $\Gr_G = G((z))/G[[z]]$ with a
transversal slice to another $G[[z]]$-orbit $\Gr^\la_G$. Here we
regard $\la$, $\mu$ as homomorphisms $S^1\to G$, i.e., coweights of
$G$.\footnote{Here $\la$ (resp.\ $\mu$) corresponds to a homomorphism
  at $\infty$ (resp.\ $0$) of $\CC^2$. Since we follow the convention
  in \cite{Na-quiver}, this is opposite to \cite{braverman-2007}. In
  particular, $\mathcal W^\mu_{G,\la}$ is empty unless $\la\le\mu$.}
This result is true for any $G$.
\begin{NB}
    Let us use the based loop group $\Omega G$ instead of $\Gr_G$.
    Let us view an instanton as a based map $f$ from $\proj^1$ to the
    based loop group $\Omega G$ with $G=\U(r)$ \cite{Atiyah}, and
    study what $S^1$-equivariant instantons correspond to. Here a map
    $f$ is based if $f(\infty) = 1_{\Omega G}$ with the constant loop
    $1_{\Omega G}$ at the identity element in $G$.

    The rotation $S^1$-action on $\Omega G$ is given by
\begin{equation*}
    (t\cdot c)(s) = c(ts) c(t)^{-1}
\end{equation*}
for $t\in S^1$, $c\in\Omega G$, and $s$ is the variable for loops.  We
also have a homomorphism $\la\colon S^1\to G$, corresponding to the
$S^1$-module structure on $W$. We combine them to get
\begin{equation*}
    (t\ast c)(s) = \la(t) c(ts) c(t)^{-1} \la(t)^{-1}.
\end{equation*}
\begin{NB2}
    Since $(t\ast c)(1) = 1$, it is a based loop. If $c = 1_{\Omega
      G}$, i.e., $c(s) = 1$ for any $s$, $t\ast c = 1_{\Omega
      G}$. Therefore $1_{\Omega G}$ is fixed by the action.
\end{NB2}%
Then $f$ is equivariant if and only if $f(tz) = t^{-1}\ast f(z)$.

Let $g\colon \proj^1\to \Omega G$ defined by $g(z)(s) = \la(s)
(f(z)(s))$. Note $g(\infty) = \la$, where $\la$ is viewed as an
element in $\Omega G$.
Then $f$ is equivariant if and only if
$g(tz) = t^{-1}\cdot g(z)$.
\begin{NB2}
Let us give a detail:
    \begin{equation*}
        \begin{split}
            & g(tz)(s) = \la(s) (f(tz)(s)) = \la(s) t^{-1}\ast f(z)(s)
        = \la(s) \la(t)^{-1} (f(z)(t^{-1}s)) (f(z)(t^{-1}))^{-1} \la(t),
        \\
        & (t^{-1}\cdot g(z))(s) = g(z)(t^{-1}s) g(z)(t^{-1})^{-1}
        = \la(t^{-1}s) (f(z)(t^{-1}s)) (f(z)(t^{-1}))^{-1} \la(t^{-1})^{-1}.
        \end{split}
    \end{equation*}
\end{NB2}%
An equivariant map $g$ is determined by $g(1) = x\in\Gr_G$. It has the
property $t\cdot x = g(t)\to \la$ if $t\to\infty$, hence
$x\in\Gr_{G,\la}$, the transversal slice to $\Gr_G^\la$. 

On the other hand $g(0)$ is another fixed point, which is $\mu$. Then
$t\cdot x\to\mu$ as $t\to 0$. This implies $x\in\Gr_G^\mu$. Hence
$x\in \mathcal W^\mu_{G,\la}$. 

Conversely a point $x\in\mathcal W^\mu_{G,\la}$ defines an equivariant
map $g(t) = t\cdot x$ defined on $\CC^*$. It extends to $\proj^1$ with
$g(0) = \mu$, $g(\infty)=\la$.
\end{NB}%
Note that the identification of partitions with coweights in
\cite{braverman-2007} is different from one in \cite[\S10]{Na-quiver},
which will be used below.

\begin{NB}
Recall that the homology group of the resolution of $\mathcal
O_\mu\cap \mathcal S_\la$ gives a weight space in an irreducible
representation of the corresponding Lie algebra, where $\la$ gives the
highest weight and $\mu$ specifies the weight (\cite{Na-quiver}).
We regard both weights and coweights as integral vectors by expressing
them in fundamental weights and coweights. Then $\la$, $\mu$ can be
considered as coweights. The above $\la$, $\mu$ are {\it not\/} given
by this rule. We first take transposes of $\la$, $\mu$ and then apply
this rule.
\end{NB}

Thus we have identifications
\begin{equation*}
\newdimen\middlewidth
\middlewidth=3.2in
    \begin{split}
        \mathcal O_\mu\cap \mathcal S_\la \longleftrightarrow &
        \parbox[c]{\middlewidth}{\centering a framed moduli space of
          $\SU(2)$-equivariant $G$-instantons}
        \\
        &         \parbox[c]{\middlewidth}{\centering$\updownarrow$}
        \\
        & 
        \parbox[c]{\middlewidth}{ \centering a quiver variety of type $A$ }
        \\
        &         \parbox[c]{\middlewidth}{\centering$\updownarrow$}
        \\
         &
        \parbox[c]{\middlewidth}{\centering a framed moduli space of
          $S^1$-equivariant $G$-instantons}
        \longleftrightarrow
        \mathcal W^\mu_{G,\la}.
    \end{split}
\end{equation*}
The identifications $\mathcal O_\mu\cap \mathcal S_\la$ and $\mathcal
W^\mu_{G,\la}$ with quiver varieties of type $A$ were first found 
in \cite{MR1968260}.
Note however that horizontal arrows remain true for arbitrary $G$,
while vertical ones are true only for type $A$. In the top row $\la$,
$\mu$ are nilpotent orbits, while they are coweights in the bottom
row. Therefore we cannot hope a vertical relation for general $G$.
Therefore one should understand $\mathcal O_\mu\cap \mathcal S_\la
\cong \mathcal W^\mu_{G,\la}$ as a composition of the natural
horizontal identifications and accidental vertical ones.
If we apply the ADHM transform to $\SU(2)$ and $S^1$-equivariant
instantons for classical groups respectively, we will obtain different
modifications of quiver varieties. It will be discussed in
\S\S\S\ref{sec:SOSp},\ref{sec:nilp-orbits-slod},\ref{sec:aff_and_S1}.
For a finite subgroup $\Gamma\subset\SU(2)$, we can also consider
$\Gamma$-equivariant instantons in the same way. See
\subsecref{sec:SOSpALE}.

The mirror of this theory is given by $\mathcal
O_{\lambda^t}\cap\mathcal S_{\mu^t}$, where $\lambda^t$, $\mu^t$ are
transpose partitions.
(It is not clear at this moment, what is the mirror if
$\bmu^{-1}(0)^{\text{reg}}=\emptyset$, i.e., $\mu$ is not necessarily
dominant.)
This mirror symmetry can be naturally extended to the case of quiver
gauge theories of affine type A. It nicely fits with the level-rank
duality of affine Lie algebras of type A via the author's work
\cite{Na-quiver}. This was observed by de~Boer et al
\cite[\S3]{MR1454292} based on brane configurations in string theories
introduced by Hanany-Witten \cite{MR1451054}.
Further examples of the mirror symmetry will be given in
\secref{sec:more_mirror}.

\subsection{Abelian theory}\label{sec:abelian-theory}

Let us take a collection of nonzero integral vectors $u_1$, \dots,
$u_d$ in $\ZZ^n$ such that they span $\ZZ^n$. We have an exact
sequence of $\ZZ$-modules
\begin{equation}\label{eq:19}
  0 \to \ZZ^{d-n}\xrightarrow{\alpha} \ZZ^d \xrightarrow{\beta} \ZZ^n \to 0,
\end{equation}
where $\beta\colon \ZZ^d\to\ZZ^n$ is given by sending the coordinate
vector $e_i$ to $u_i$, and the kernel of $\ZZ^d\to \ZZ^n$ is
identified with $\ZZ^{d-n}$ by taking a base. We have the
corresponding exact sequence of tori:
\begin{equation}\label{eq:20}
    1 \to G = \U(1)^{d-n} \xrightarrow{\alpha} T^d = \U(1)^d 
    \xrightarrow{\beta} G_F = \U(1)^n \to 1,
\end{equation}
where maps in \eqref{eq:19} are induced homomorphisms between coweight
lattices (or equivalently fundamental groups).

Let $\bM = \HH^d$ and let $T^d$ act $\HH$-linearly on $\bM$ by
multiplication. We consider $\Hyp(\bM)\tslabar G$. It is called the
{\it abelian theory}. It is of cotangent type with $\bN = \CC^d$.

The hyper-K\"ahler quotient $\bM\tslash G$ is called a {\it toric
  hyper-K\"ahler manifold\/} and was introduced by Bielawski and
Dancer \cite{MR1792372}. Note that we have an action of $G_F$ on
$\bM\tslash G$. This group is called a flavor symmetry group, and its
importance will be explain in \secref{sec:flavor-symmetry} below.

The space $Z(\mathfrak g^*)$ is nontrivial as in quiver gauge
theories, and $\bmu^{-1}(\zeta_{\operatorname{Im}\HH})/G$ is a
hyper-K\"ahler orbifold for generic $\zeta_{\operatorname{Im}\HH}$.

\begin{NB}
    In \cite[\S6]{Cremonesi:2013lqa}, the map $\ZZ^{d-n}\to \ZZ^d$ is
    given by $S^p_i$ ($i=1,\dots,d$, $p=1,\dots, d-n$) with the
    substitution $d=N$, $n=r$. On the other hand the $(\ZZ^d)^*\to
    (\ZZ^{d-n})^*$ is given by $R^i_a$ ($i=1,\dots, d$,$a=1,\dots,n$).
\end{NB}

The abelian theory is a good example to understand the $3d$ mirror
symmetry.
We dualize the exact sequence \eqref{eq:19} to get
\begin{equation*}
    1 \to G_F^\vee \xrightarrow{\beta^\vee} (T^d)^\vee 
    \xrightarrow{\alpha^\vee} G^\vee \to 1,
\end{equation*}
where $\bullet^\vee$ denotes the dual torus, defined by
$\pi_1(\bullet)^\wedge$. ($G^\vee$ is also $\U(1)^{d-n}$, but we would
like to make our framework intrinsic.) Then we can consider another
toric hyper-K\"ahler manifold $\bM\tslash G_F^\vee$ with a
$G^\vee$-action. It was proposed in \cite[\S4]{MR1454292} that
$\Hyp(\bM)\tslabar G$ and $\Hyp(\bM)\tslabar G_F^\vee$ form a mirror dual
theories. In particular, the Coulomb branch of $\Hyp(\bM)\tslabar G$ is
$\bM\tslash G_F^\vee$.

\section{More on \texorpdfstring{$3d$}{3d} mirror symmetry}
\label{sec:more_mirror}

It is important to have many examples of $3$-dimensional mirror
symmetric pairs, as they determine the Coulomb branches as the Higgs branches of mirror theories. We give more examples in this section.

\subsection{Mirror could be a non-lagrangian theory}\label{sec:mirror-could-be}

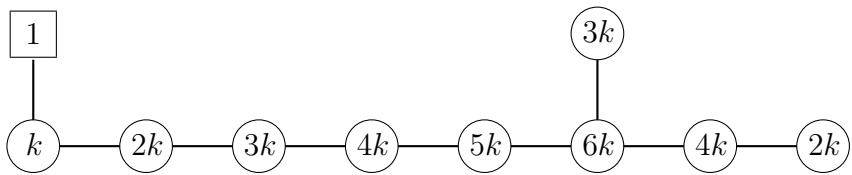
\begin{figure}[htbp]
    \centering
\setlength{\unitlength}{1mm}
\begin{picture}(112.5,24)
    \multiput(3.5,3.5)(15,0){8}{\circle{7}}
    \multiput(7,3.5)(15,0){7}{\thicklines\line(1,0){8}}
    \put(78.5,7){\thicklines\line(0,1){8}}
    \put(78.5,18.5){\circle{7}}
    \put(3.5,7){\thicklines\line(0,1){8}}
    \put(0.5,15.4){\framebox(6,6){$1$}}
    \put(2.5,2.3){$k$}
    \put(16.5,2.3){$2k$}
    \put(31.5,2.3){$3k$}
    \put(46.5,2.3){$4k$}
    \put(61.5,2.3){$5k$}
    \put(76.5,2.3){$6k$}
    \put(76.5,17.3){$3k$}
    \put(91.5,2.3){$4k$}
    \put(106.5,2.3){$2k$}
\end{picture}
\caption{$\mathcal M_C$ : $E_8$ $k$-instantons on $\RR^4$.}
    \label{fig:E8}
\end{figure}

We first remark that the mirror of the gauge theory $\Hyp(\bM)\tslabar
G$ may not be of a form $\Hyp(\bM')\tslabar G'$ for some $\bM'$ and
$G'$ in general.
For example, if we replace the diagram in Figure~\ref{fig:mirror} by
the affine Dynkin diagram of type $E_{6,7,8}^{(1)}$ as in
Figure~\ref{fig:E8}, it is expected that $\mathcal M_C$ is the framed
moduli space of $E_{6,7,8}$ $k$-instantons on $\RR^4$. This example
was found in \cite{MR1413696} for $k=1$, and in \cite{MR1454292} for
general $k$.
It is widely accepted a common belief that there are no ADHM like
description of instantons for exceptional groups. It means that the
moduli spaces cannot be given by a hyper-K\"ahler quotient
$\bM'\tslash G'$ (with finite dimensional $\bM'$, $G'$), hence the
mirror theory $B$ is not of a form $\Hyp(\bM')\tslabar G'$. In fact,
the mirror theory $B$ is known as a $3d$ Sicilian theory
\cite{Benini:2010uu}, which does not have a conventional lagrangian
description. See \subsecref{subsec:MT} below.
Nevertheless we can compute Hilbert series of instanton moduli spaces
of exceptional types by the monopole formula.\footnote{One need to
  modify the monopole formula to deal with non simply-laced
  groups. See \cite{Cremonesi:2014xha}.}
This is even more exciting, as there is only a few way to compute
them, say a conjectural blowup equation \cite{MR2199008,MR2183121}.

\subsection{Instantons on \texorpdfstring{$\RR^4/\Gamma$}{R^4/Gamma}}
\label{sec:inst-ale-space}

Let us consider a quiver gauge theory of affine type. An affine quiver
of type $ADE$ arises as the McKay quiver of a finite subgroup $\Gamma$
of $\SU(2)$. Hence the Higgs branch, the hyper-K\"ahler quotient of
$\bM$ by $G$, parametrizes $\Gamma$-equivariant $\U(\ell)$-instantons
on $\RR^4$ as in \subsecref{sec:type-A-quiver}. In fact, this is a
starting point of the work \cite{KN}, which eventually leads to the
study of quiver varieties \cite{Na-quiver}.

As we have mentioned already in \subsecref{sec:type-A-quiver}, the
mirror of a quiver gauge theory of affine type $A$ is another quiver
gauge theory again of affine type $A$. The precise recipe was given in
\cite[\S3.3]{MR1454292}.

From this example, together with Braverman-Finkelberg's proposed
double affine Grassmannian \cite{braverman-2007}, we will give an
initial step towards the determination of the mirror of the quiver
gauge theory of an arbitrary affine type as follows. It was mentioned
in a vague form in \cite[Rem.~10.13]{2014arXiv1407.0964B}.

A framed moduli space of $\Gamma$-equivariant $G$-instantons on
$\RR^4$ has discrete data, a usual instanton number, as well as
$\rho_0, \rho_\infty \colon\Gamma\to G$ homomorphisms from $\Gamma$ to
$G$ given by $\Gamma$-actions on fibers at $0$ and $\infty$. A quiver
variety, that is the Higgs branch of a quiver gauge theory of affine
type, corresponds to the case $G = \U(\ell)$. We regard $\rho_0$,
$\rho_\infty$ as $\ell$-dimensional representations of $\Gamma$. They
are given by dimension vectors $(\dim V_i)_{i\in Q_0}$, $(\dim
W_i)_{i\in Q_0}$ by
\begin{equation*}
    \rho_\infty = \bigoplus \rho_i^{\oplus \dim W_i},\qquad
    \rho_0 = \bigoplus \rho_i^{\oplus u_i} \quad
    \text{with $u_i = \dim W_i - \sum_j (2\delta_{ij} - a_{ij})\dim V_j$},
\end{equation*}
where $\{ \rho_i\}$ is the set of isomorphism classes of irreducible
representations of $\Gamma$, identified with $Q_0$ via McKay
correspondence. If some $u_i$ is negative, there is no genuine
instanton. In other words, the Higgs branch $\mathcal M_H = \bM\tslash
G$ contains no free orbits. We do not know what happens in the Coulomb
branch without this assumption.
Conversely $u_i$ determines $\dim V$ modulo the kernel of the Cartan
matrix $\mathbf C$, i.e., $\ZZ\delta$ for the (primitive) imaginary
root $\delta$.
This ambiguity is fixed by specifying the instanton number as $\sum_j
\delta_j \dim V_j$ where $\delta_j$ is the $j^{\mathrm{th}}$-entry of
$\delta$.

Let $\g$ be the complex simple Lie algebra of type $ADE$ corresponding
to $\Gamma$. Let $\g_{\aff}\begin{NB} = \widehat{\g}\oplus\CC d
\end{NB}%
$ be the associated untwisted affine Lie algebra, containing the
degree operator $d$. In \cite{Na-quiver}, an affine Lie algebra
representation was constructed by quiver varieties. Dimension vectors
give affine weights of $\g_{\aff}$ by
\begin{equation*}
   \la = \sum_i (\dim W_i) \Lambda_i, \qquad
   \mu = \sum_i (\dim W_i) \Lambda_i - (\dim V_i)\alpha_i,
\end{equation*}
where $\Lambda_i$ (resp.\ $\alpha_i$) is the $i^{\mathrm{th}}$
fundamental weight (resp.\ simple root).
The weight $\la$ is always dominant. We also have $\la\ge\mu$ by
definition. If we assume $u_i\ge 0$ for all $i$ as above, the second
weight $\mu$ is also dominant.

On the other hand, Braverman and Finkelberg \cite{braverman-2007}
associate a pair $(\lambda\ge \mu)$ of affine weights with $\rho_0$,
$\rho_\infty$ and instantons numbers as follows. They take $\Gamma =
\ZZ/\ell\ZZ$, where $\ell$ is the level of $\la$, also of $\mu$ as
$\lambda\ge\mu$. They take $G$ a simply-connected group, possibly of
type $BCFG$. Then \cite[Lemma~3.3]{braverman-2007} says a conjugacy
class of a homomorphism $\ZZ/\ell \ZZ\to G$ corresponds to a dominant
coweight $\overline\la$ of $G_\CC$ with
$\langle\overline\la,\theta\rangle\le \ell$, where $\theta$ is the
highest root of $\g$. It can be regarded as a level $\ell$ weight of
$\widehat G_\CC^\vee$, the Langlands dual of the affine Kac-Moody
group $\widehat G_\CC$. Here $\widehat G^\vee_\CC$ does not contain
the degree operator. We assign dominant weights $\overline\la$,
$\overline\mu$ of level $\ell$ to $\rho_0$, $\rho_\infty$ respectively
in this way. Then we extend them to $\la$, $\mu$ dominant weights of
the full affine Kac-Moody group $G_{\text{aff}}^\vee$ so that
\begin{equation*}
    \text{instanton number}
    = \ell \langle \la- \mu, d\rangle + \frac{(\overline\la,\overline\la)}2
    - \frac{(\overline\mu,\overline\mu)}2.
\end{equation*}
See \cite[(4.3)]{braverman-2007}. This rule only determines $\langle
\la- \mu, d\rangle$, but it is well-known that representation
theoretic information depend only on the difference $\la-\mu$.

Thus a pair of affine weights $(\la\ge\mu)$ correspond to instanton
moduli spaces in two ways, when $G$ is of type $ADE$, first in
\cite{Na-quiver}, second in \cite{braverman-2007}, as we have just
explained. Take the quiver gauge theory whose Higgs branch is the
quiver variety associated with $(\la\ge\mu)$ in the first way. Then
its Coulomb branch is expected to be the $\ZZ/\ell\ZZ$-equivariant
instanton moduli space associated with $(\la\ge\mu)$ in the second
way.

However this is not precise yet by the following reason. Since affine
weights of the Lie algebra $\g_{\aff}$ may not give weights of
$G_{\aff}^\vee$ if $G$ is simply-connected, we need to replace $G$ by
its adjoint quotient. Then \cite[Lemma~3.3]{braverman-2007} says a
homomorphism $\ZZ/\ell\ZZ\to G$ corresponds to an element in the coset
$\Lambda/W_{\mathrm{aff},\ell}$, where $\Lambda$ is the coweight
lattice of $G$,
\begin{NB}
    $\Lambda = \Hom(\CC^\times, T)$, where $T$ is a maximal torus of
    $G$.
\end{NB}%
and $W_{\mathrm{aff},\ell}$ is the semi-direct product $W\ltimes
\ell\Lambda$ of the Weyl group $W$ and $\Lambda$. Here $\ell\Lambda$
acts on $\Lambda$ naturally.
If $G$ is of {\it adjoint type}, $\Lambda$ is the weight lattice of
$G^\vee$, i.e., the weight lattice of $\g$. But
$W_{\mathrm{aff},\ell}$ is an extended affine Weyl group, i.e., the
semi-direct product of the ordinary affine Weyl group and a group
$\mathcal T$ consisting of affine Dynkin diagram automorphisms.
\begin{NB}
    It can be identified with the group of 1-dimensional
    representations of $\Gamma$ under the McKay correspondence. Here
    the group structure is given by tensor products of 1-dimensional
    representations. This can be checked by case-by-case
    analysis. More conceptual proof is explained in
    \url{http://mathoverflow.net/questions/196498}.
\end{NB}%
Then a point in the coset $\Lambda/W_{\mathrm{aff},\ell}$ does not
give an affine weight of $\widehat G_\CC^\vee$. It only gives a
$\mathcal T$-orbit.

When $G$ is of type $A_{r-1}$, this inaccuracy can be fixed: we
replace $G$ by $\U(r)$, and $(\la\ge\mu)$ by a pair of generalized
Young diagrams, in other words, dominant weights of
$\widehat{\GL}(r,\CC)$. See \cite[App.~A]{Na-branching} for a detailed
review.
\begin{NB}
    And also 2015-02-24.xoj.
\end{NB}%
If we view both Higgs and Coulomb branches as quiver varieties of
affine type $A$, the rule of the transform of dimension vectors is
given by transpose of generalized Young diagrams, as reviewed in
\cite[App.~A]{Na-branching}. It is the same as one in \cite{MR1454292}
up to a diagram automorphism.

If we take a gauge theory of {\it finite type\/} instead of affine
type, $\la$, $\mu$ are dominant weights of the finite dimensional Lie
algebra $\g$ in \cite{Na-quiver}. Then instead of
\cite{braverman-2007}, one can use just the ordinary geometric Satake
correspondence, i.e., the affine Grassmannian for $G$ of adjoint
type. In terms of instantons, we use $S^1$-equivariant $G$-instantons
on $\RR^4$. Then $\la$, $\mu$ are regarded as dominant coweights of
$G$, and correspond to $S^1$-actions on fibers at $0$ and
$\infty$. The inaccuracy disappears also in this case.
This conjectural proposal was given in
\cite[Rem.~10.7]{2014arXiv1407.0964B}, though their symplectic duality
does not make sense in general outside type $A$, as we mentioned in
Introduction.

\begin{NB}
**********************************

It is also possible that both theories $A$ and $B$ are not gauge
theories. For example, it is natural to expect that there are theories
whose Higgs branches are $G_\Gamma$-instanton moduli spaces on
$\CC^2/\Gamma'$, where $\Gamma$ and $\Gamma'$ are both finite
subgroups of $\SU(2)$. Here $G_\Gamma$ is a compact Lie group (of type
$ADE$) corresponding to $\Gamma$. The mirror should be the theory
obtained by exchanging $\Gamma$ and $\Gamma'$.
Figures~\ref{fig:mirror},~\ref{fig:E8} are example for $\Gamma=\{1\}$,
$\Gamma'=\ZZ/N\ZZ$ and $\Gamma_{E_8}$.
If both $\Gamma$ and $\Gamma'$ are exceptional, there are no ADHM type
description known, and hence both theories $A$ and $B$ are not gauge
theories.

Remark that we do not specify the fundamental group of $G_\Gamma$ nor
discrete datum classifying $G_\Gamma$-instanton moduli spaces on
$\CC^2/\Gamma'$ (like $\dim V$, $\dim W$ in a quiver gauge
theory). Therefore we do not specify theories $A$, $B$ precisely. The
statement that theories $A$ and $B$ are mirror dual must be
elaborated. It was given when $A$ is a quiver gauge theory in
\subsecref{sec:quiver-type} MODULO inaccuracy mentioned above, but the
author does not know the answer in general.
\end{NB}

\subsection{Sicilian theory}\label{subsec:MT}

There is another class of $4$-dimensional quantum field theories with
$\mathcal N = 2$ supersymmetry. They are called {\it theories of class
  $S$}. A theory $S_\Gamma(C,x_1,\rho_1,\dots,x_n,\rho_n)$ is
specified with an $ADE$ Dynkin diagram $\Gamma$, a punctured Riemann
surface $(C,x_1,\dots,x_n)$ together with a homomorphism
$\rho_i\colon\su(2)\to\g_\Gamma$ for each puncture $x_i$, where
$\g_\Gamma$ is the Lie algebra of a compact Lie group of type
$\Gamma$. It is constructed as a dimensional reduction of a
$6$-dimensional theory associated with $\Gamma$, compactified on a
Riemann surface $C$ with defects at punctures specified by
$\rho_i$. It is believed that
$S_\Gamma(C,x_1,\rho_1,\dots,x_n,\rho_n)$ does not have a lagrangian
description in general, hence is not a gauge theory studied in this
paper. See \cite{Tach-review} for a review aimed for mathematicians.

Physicists consider its Coulomb and Higgs branches. The Coulomb branch
is expected to be the moduli space of solutions of Hitchin's
self-duality equation on $C$ with boundary condition at $x_i$ given by
$\rho_i$.
On the other hand, it is asked in \cite{MR2985331} what is the
underlying complex symplectic manifold of its Higgs branch $\mathcal
M_H(S_\Gamma(C,x_1,\rho_1,\dots,x_n,\rho_n))$. The underlying complex
symplectic manifold is independent of the complex structure of
$(C,x_1,\dots,x_n)$. It is supposed to satisfy various properties
expected by physical considerations, most importantly it gives a $2d$
TQFT whose values are complex symplectic manifolds.

We can further compactify $S_\Gamma(C,x_1,\rho_1,\dots,x_n,\rho_n)$ on
$S^1_R$ and take limit $R\to 0$ to get a $3$-dimensional quantum field
theory with $\mathcal N=4$ supersymmetry. This is a $3d$ Sicilian
theory mentioned above.
It is expected that the Higgs branch is unchanged under the
compactification by $S^1_R$.

When $\Gamma$ is {\it not\/} exceptional, its mirror is supposed to be
a certain gauge theory $\Hyp(\bM)\tslabar G$
\cite{Benini:2010uu}. Therefore the Higgs branch $\mathcal
M_H(S_\Gamma(C,x_1,\rho_1,\dots,x_n,\rho_n))$ is the Coulomb branch of
a gauge theory $\mathcal M_C(\Hyp(\bM)\tslabar G)$, which we are
studying in this paper.

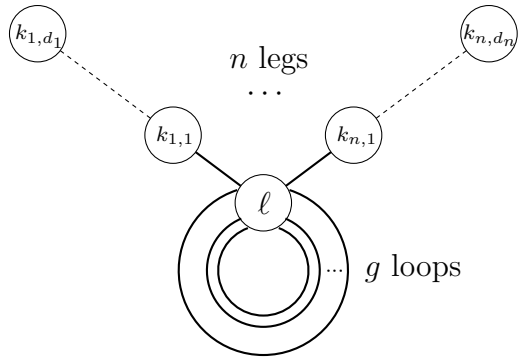
\begin{figure}[htbp]
    \centering
\setlength{\unitlength}{1.5mm}
\begin{picture}(60,35)
    \put(31,16){\circle{5}}
    \put(30.5,15){$\ell$}
    \put(23,22){\circle{5}}
    \put(39,22){\circle{5}}
    \put(21.5,21.5){$\scriptstyle k_{1,1}$}
    \put(37.5,21.5){$\scriptstyle k_{n,1}$}
    \put(29,17.5){\thicklines\line(-4,3){4}}
    \put(33,17.5){\thicklines\line(4,3){4}}
    \put(11,31){\circle{5}}
    \put(51,31){\circle{5}}
    \Dline(21,23.5)(13,29.5){0.4}
    \Dline(41,23.5)(49,29.5){0.4}
    \put(9,30.5){$\scriptstyle k_{1,d_1}$}
    \put(48.7,30.5){$\scriptstyle k_{n,d_n}$}
    \put(29.5,25){$\cdots$}
    \put(28,28){$n$ legs}
    \put(31,10){\thicklines\arc[-250,70]{4}}
    \put(31,10){\thicklines\arc[-247,67]{5}}
    \put(31,10){\thicklines\arc[-252,72]{7.5}}
    \put(36.5,9.5){$\scriptstyle\cdots$}
    \put(40,9.5){$g$ loops}
\end{picture}
\caption{Mirror of a $3d$ Sicilian theory of type $A_\ell$}
    \label{fig:mirror_Sic}
\end{figure}

Let us specify $\bM$ and $G$. First suppose $\Gamma$ is of type
$A_\ell$. If $C = S^2$, the mirror is the quiver gauge theory
associated with the star shaped quiver with $n$ legs. Entries of the
dimension vector are $\ell$ at the central vertex, and given by
$\rho_i$ on the $i^{\mathrm{th}}$ leg specified by the rule as for the
quiver construction of the nilpotent orbit $\rho_i(
\begin{smallmatrix}
    0 & 1 \\ 0 & 0
\end{smallmatrix}
)$
(see \subsecref{sec:type-A-quiver}). If $C$ has genus $g$, we add $g$
loops at the central vertex. (Figure~\ref{fig:mirror_Sic})
Note also that quivers considered in~\subsecref{sec:mirror-could-be}
are of this type.
For type $D_\ell$, we modify this quiver as in
\subsecref{sec:nilp-orbits-slod}.

The Coulomb branch $\mathcal
M_C((S_\Gamma(C,x_1,\rho_1,\dots,x_n,\rho_n))$ of a $3d$ Sicilian
theory is the Higgs branch of the gauge theory, which is the
hyper-K\"ahler quotient $\bM\tslash G$.
It is an additive version of the moduli space of homorphisms from the
fundamental group $\pi_1(C\setminus\{x_1,\dots,x_n\})$ of the
punctured Riemann surface to $\GL_{\ell+1}(\CC)$ or $\SO(2\ell,\CC)$
with prescribed conjugacy classes around punctures. (See
\cite{MR1980997}.)
There is an isomorphism between an open subset of $\bM\tslash G$ and
the actual moduli space \cite{MR2470573}, hence it is compatible
with the expectation that $\mathcal
M_C(S_\Gamma(C,x_1,\rho_1,\dots,x_n,\rho_n))$ is the Hitchin moduli
space. When we make $R\to 0$, the Hitchin moduli space is replace by
its additive version.

The monopole formula for this type of quivers is studied in \cite{Cremonesi:2014vla}.

\section{Monopole formula}\label{sec:monopole-formula}

We discuss the monopole formula in detail in this section.

\subsection{Definition}\label{sec:formula}

Let $G$ be a compact Lie group. We assume $G$ is connected hereafter
for simplicity.\footnote{The monopole formula for a disconnected group
  $\grpO(N)$ appears in \cite{Cremonesi:2014uva}.}
We choose and fix a maximal torus $T$ and a set $\Delta^+$ of positive
roots. Let $Y = Y(T)$ be the coweight lattice of $G$. Let $W$ denote
the Weyl group.

Suppose that a quaternionic representation $\bM$ (also called a {\it
  pseudoreal\/} representation) of $G$ is given. We choose an
$\HH$-base $\{ b\}$ of $\bM$ compatible with the weight space
decomposition.

We define two functions,\footnote{Following \cite{Cremonesi:2014xha},
  we change $t$ by $t^2$ from \cite{Cremonesi:2013lqa}.} one
depending on $G$ and $\bM$, another depending only on $G$, of a
coweight $\la\in Y$ by
\begin{equation}\label{eq:14}
    \begin{split}
        & \Delta(\la) \defeq - \sum_{\alpha\in\Delta^+}
        |\langle\alpha,\la\rangle| + \frac12 \sum_b |\langle \wt(b),
        \la\rangle|,
        \\
        & P_G(t;\la) \defeq \prod \frac1{1 - t^{2d_i}},
    \end{split}
\end{equation}
where $\langle\ ,\ \rangle$ is the pairing between weights and
coweights, and the product in the second formula runs over exponents
of the stabilizer $\operatorname{Stab}_G(\la)$ of $\la$.
Since we take the absolute value in the second term, $\Delta(\la)$ is
independent of the choice of $b$: it remains the same for $jb$.
\begin{NB}
    When $G$ is not connected, $P_G(t;\la)$ is defined as the Hilbert
    series of the ring of invariant functions
    $\CC[\operatorname{Lie}(\operatorname{Stab}_G(\la))]^{
      \operatorname{Stab}_G(\la)}$ on the Lie algebra of the
    stabilizer $\operatorname{Stab}_G(\la)$ of $\la$.
If $G$ is connected, the ring of invariants is a polynomial ring, and
we have
\begin{equation*}
       P_G(t;\la) =
       \prod \frac1{1 - t^{2d_i}},
\end{equation*}
where $d_i$ are {\it exponents\/} of
$\operatorname{Lie}(\operatorname{Stab}_G(\la))$.
\begin{NB2}
    Suppose $G$ is a semi-direct product of a torus and a finite group
    $\Gamma$. We take $\la = 0$. Then
    $\CC[\operatorname{Lie}(\operatorname{Stab}_G(\la))]^{
      \operatorname{Stab}_G(\la)}$ is
    $\CC[\operatorname{Lie}T]^\Gamma$. This is not a polynomial ring
    in general.
\end{NB2}%
\end{NB}%
It is well-known that $P_G(t;\la)$ is equal to the Poincar\'e
polynomial of the equivariant cohomology
$H^*_{\operatorname{Stab}_G(\la)}(\mathrm{pt})$ of a point.
\begin{NB}
    This property is certainly not true if $G$ is not connected.
\end{NB}%
Both $\Delta(\la)$ and $P_G(t;\la)$ are invariant under the Weyl group
$W$ action on $Y$.

\begin{NB}
    Even if $G$ is not connected, the Weyl group $W$ is defined as
    $N_G(T)/T$. Therefore the Weyl group of $\grpO(N)$ is larger than
    that of $\SO(N)$ for example.
\end{NB}

We assume $2\Delta(\la)\ge 1$ for any $\la\neq 0$, a `good' or `ugly'
theory in the sense of \cite{MR2610576}, hence no negative powers of
$t$ appear. (It is {\it good\/} if $2\Delta(\la) > 1$ and {\it ugly\/}
if $2\Delta(\la)\ge 1$ and not good. But we do not see any differences
of two conditions in this paper.)
Since $\Delta(\la)$ is piecewise linear, there is only
finitely many $\la$ for a given $2\Delta(\la)$. Note also that
$P_G(t;\la)$ can be expanded as a formal power series in $t$. Therefore
\begin{equation*}
    H_{G,\bM}(t) \defeq \sum_{\la\in Y/W} t^{2\Delta(\la)} P_G(t;\la).
\end{equation*}
makes sense as a formal power series in $t$.

This elementary, but combinatorially complicated expression is the
{\it monopole formula\/} for the {\it Hilbert series of the Coulomb
  branch\/} $\mathcal M_C\equiv\mathcal M_C(\Hyp(\bM)\tslabar G)$.

Recall that physicists claim that the Coulomb branch $\mathcal M_C$ is
a
\begin{NB}
conical
\end{NB}%
hyper-K\"ahler manifold with an $\SU(2)$-action rotating
hyper-K\"ahler structures $I$, $J$, $K$. One choose a complex
structure $I$, and take $\U(1)\subset\SU(2)$, fixing $I$. Then the
Hilbert series is the character of the coordinate ring $\CC[\mathcal
M_C]$ of $\mathcal M_C$ with respect to the $\U(1)$-action, endowed
with an affine scheme structure compatible with the complex structure
$I$. Thus the main claim in \cite{Cremonesi:2013lqa} is
\begin{equation*}
    H_{G,\bM}(t) = \operatorname{ch}_{\U(1)}\CC[\mathcal M_C].
\end{equation*}

\begin{Remark}
    The good or ugly condition $2\Delta(\la)\ge 1$ for any $\la\neq 0$
    means weights of $\CC[\mathcal M_C]$ are nonnegative and the
    $0$-weight space consists only on constant functions.
If this is not satisfied, it is not clear whether
\eqref{eq:11} makes sense or not.
However $\CC[\mathcal M_C]$ itself might be well-defined. It is the
case for $\Hyp(0)\tslabar\SU(2)$ for example. The only trouble is that
weight spaces might be infinite dimensional, and hence the character
$\operatorname{ch}_{\U(1)}\CC[\mathcal M_C]$ is not defined.
\end{Remark}

\subsection{Examples}\label{sec:examples}

Let us calculate $H_{G,\bM}(t)$ for the simplest example. Let
$G=\U(1)$ and $\bM = \HH = \CC \oplus\CC^*$, the vector representation
plus its dual. We identify the coweight lattice $Y({\U(1)})$ with
$\ZZ$, and denote a coweight by $m$ instead of $\la$.

There is no first term in $\Delta(m)$ as $\Delta^+ = \emptyset$. Thus
$2\Delta(m) = |m|$. This is an ugly theory. The stabilizer of $m$ is
always $\U(1)$, thus $P_G(t;m) = 1/(1-t^2)$. Therefore
\begin{equation*}
    H_{\U(1),\HH}(t) = \frac1{1-t^2} \sum_{m\in\ZZ} t^{|m|}
    \begin{NB}
    = \frac1{1-t^2}\left(\frac1{1-t} + \frac{t}{1-t}\right)
    \end{NB}
    = \frac1{(1-t)^2}.
\end{equation*}
In this case, Seiberg-Witten \cite{MR1490862} claim that the Coulomb
branch $\mathcal M_C$ is the Taub-NUT space. It is a $4$-dimensional
hyper-K\"ahler manifold with $\grpSp(1)$-action, whose underlying
complex manifold is $\CC^2$. The subgroup commuting with the complex
structure $I$ is $\U(1)$ with the multiplication action on
$\CC^2$. Therefore the Hilbert series is $1/(1-t)^2$ as expected.

If we replace $\bM$ by the direct sum of its $N$-copies, i.e., $\bM =
\HH^N$, we get
\begin{equation}\label{eq:8}
    H_{\U(1),\HH^N}(t)
    \begin{NB}
       = \frac1{1-t^2} \sum_{m\in\ZZ} t^{N|m|}
        = \frac1{1-t^2}\left(\frac1{1-t^N} + \frac{t^N}{1-t^N}\right)
    \end{NB}%
    = \frac{1+t^N}{(1-t^2)(1-t^N)}
    = \frac{1-t^{2N}}{(1-t^2)(1-t^N)^2}.
\end{equation}
It is claimed that $\mathcal M_C$ is the multi-Taub-NUT space, which
is $\CC^2/(\ZZ/N\ZZ)$ as a complex variety. It is the surface $xy=z^N$
in $\CC^3$, hence we recover the above formula if we set $\deg x =
\deg y = N$, $\deg z = 2$.
\begin{NB}
    In terms of $\CC^2/(\ZZ/N\ZZ)\ni (z_1,z_2)$, we have $x=z_1^N$,
    $y=z_2^N$, $z=z_1z_2$.
\end{NB}%

In this case, the meaning of individual terms
$t^{2\Delta(\la)}P_G(t;\la) = t^{N|m|}/(1-t^2)$ is also apparent from
the exact sequence
\begin{equation*}
    0\to\CC[\mathcal M_C] \xrightarrow{z} \CC[\mathcal M_C]
    \to \CC[\{ z= 0 \}]\to 0.
\end{equation*}
We have $(1-t^2)\operatorname{ch}_{\U(1)}\CC[\mathcal M_C] =
\operatorname{ch}_{\U(1)}\CC[\{z=0\}]$. This explains $P_G(t;\la) =
1/(1-t^2)$. Since $\{ z=0 \} = \{ xy=0\}$, we have $\CC[\{z=0\}] =
\CC[x,y]/(xy)$. Then $t^{N|m|}$ corresponds to $x^m$ for $m\ge 0$ and
$y^m$ for $m\le 0$ (and $1$ for $m=0$).

\begin{NB}
    Most of contents of this subsection is moved to
    \subsecref{sec:3d-mirror-symmetry}.

\subsection{Hyper-K\"ahler quotients and 3-dimensional mirror
  symmetry}

One can study an intrinsic meaning of the monopole formula
\eqref{eq:11} to try to {\it define\/} $\mathcal M_C$, as we will do
in this paper. On the other hand, if one {\it a priori\/} knows what
is $\mathcal M_C$, it certainly helps to understand the monopole
formula from a different direction. The $3d$ mirror symmetry provides
large examples of $\mathcal M_C$, which we explain in this subsection.
\end{NB}

\subsection{Expected properties of \texorpdfstring{$\mathcal M_C$}{MC}}\label{sec:expect-properties}

Let us give several expected properties of the Coulomb branch
$\mathcal M_C$. First of all,
\begin{enume_add}
      \item $\mathcal M_C$ contains a
    \begin{NB}
        conical         
    \end{NB}%
    hyper-K\"ahler manifold (or orbifold, more generally) with an
    $\SU(2)$-action rotating $I$, $J$, $K$, as an open dense subset.
    \label{item:hK}
\end{enume_add}

Once the Hilbert series is given, the dimension is given by the degree
of the corresponding Hilbert polynomial. We expect
\begin{enume_add}
      \refstepcounter{number}\label{item:dim}
      \item $\dim_{\mathbb H} \mathcal M_C = \dim_\RR T$. 
\end{enume_add}

It is a classical result that the fundamental group $\pi_1(G)$ of $G$
is isomorphic to the quotient of the coweight lattice $Y$ by the
coroot lattice.
\begin{NB}
    If $G$ is simply connected, the weight lattice $P$ is isomorphic
    to the dual of the coroot lattice, i.e., the lattice consisting of
    those $\lambda\in\mathfrak h^*$ such that $\langle\lambda,
    h_\alpha\rangle\in\ZZ$ for any coroot $h_\alpha$. Therefore the
    coweight lattice $Y$, which is the dual of $P$, is
    isomorphic to the coroot lattice. For general $G$, we take the
    universal cover $\tilde G$. We have an induced injective
    homomorphism $Y({\tilde G})\to Y(G)$ and the quotient is
    the fundamental group.
\end{NB}%
We can refine the Hilbert series by remembering the class of $\la$ in
$\pi_1(G)$, i.e., the coordinate ring $\CC[\mathcal M_C]$ has an
additional $\pi_1(G)$-grading. Therefore we expect
\begin{enume_add}
      \item The Pontryagin dual $\pi_1(G)^\wedge =
    \Hom(\pi_1(G),\U(1))$ acts on $\mathcal M_C$, preserving the
    hyper-K\"ahler structure.\label{item:group}
\end{enume_add}
%
In \cite{Cremonesi:2013lqa,Cremonesi:2014kwa}, an additional variable
$z$ is introduced, and \eqref{eq:11} is refined to
\begin{equation}\label{eq:15}
    H_{G,\bM}(t,z) = \sum_{\la\in Y/W} z^{J(\la)} t^{2\Delta(\la)}
    P_G(t;\la),
\end{equation}
where $J$ is the projection from $Y$ to $\pi_1(G)$. In above
examples, $G$ is a product of unitary groups, hence, $\pi_1(G) =
\ZZ^r$. Thus we expect the torus $\U(1)^r$ acts on $\mathcal M_C$. In
the above formula, $z$ is a (multi) variable for characters of the torus.

Let us check these expected properties for the abelian
case. (\ref{item:hK}) is clear as $\mathcal M_C$ is supposed to be
the hyper-K\"ahler quotient $\bM\tslash G_F^\vee$. We have
\begin{aenume}
    \setcounter{enumi}{1}
  \item $\dim_\HH \bM\tslash G_F^\vee = d-n = \dim_\RR G$.
  \item $\pi_1(G)^\wedge = (\ZZ^{d-n})^\wedge = G^\vee$ acts on
$\bM\tslash G_F^\vee$.
\end{aenume}
Therefore these two proposed properties are satisfied.

In many examples, this group action can be enlarged to a nonabelien
group action. Let us give a particular example.

Take a quiver gauge theory $\Hyp(\bM)\tslabar G$ as in
\subsecref{sec:quiver-type}. Since $G$ is a product of unitary groups,
$\pi_1(G)^\wedge$ is isomorphic to the product of $\U(1)$ for each
vertex $i\in Q_0$.
Therefore $\prod_{i\in Q_0}\U(1)$ acts on $\mathcal M_C$ from
(\ref{item:group}) above.
It is expected that a larger group containing $\prod_{i\in Q_0}\U(1)$
acts on $\mathcal M_C$ as follows.

We consider the Weyl group $W(Q)$, naturally appeared in the context
of quiver varieties \cite[\S9]{Na-quiver}: Fix $W$, or more precisely
$\dim W = (\dim W_i)\in \ZZ^{Q_0}$, but we allow $V$ to change.
We regard $(\dim W_i - \sum_j (2\delta_{ij} - a_{ij}) \dim
V_j)\in\ZZ^{Q_0}$ as a weight of the Kac-Moody Lie algebra
corresponding to $Q$.
Then we change $\dim V$ given by the usual Weyl group action on
weights.
Concretely, for each vertex $i$ without edge loops, we consider $s_i$,
which change $V$ by a new $V'$ by the following rule: A) $V'_j$ is the
same as $V_j$ if $j\neq i$. B) $\dim V'_i = \dim W_i + \sum_j a_{ij}
\dim V_j - \dim V_i$.
These $s_i$ generates the Weyl group.

Now we fix $V$ again, and introduce the subquiver $S = (S_0,S_1)$ of
$Q$ consisting of vertices $i$ such that reflections $s_i$ preserve
$\dim V$ and edges between them. We suppose $S$ is of finite type,
i.e., the underlying graph is a disjoint union of $ADE$ graphs. Let
$G_S'$ be the simply-connected\footnote{The author does not know
  whether it naturally descends to a quotient or not.} compact Lie
group corresponding to $S$. Note that $G_S'$ contains $\U(1)$'s
corresponding to vertices in $S$, as a maximal torus. We then take
other $\U(1)$'s corresponding to vertices {\it not\/} in $S$, and
define $G_S$ as the product of $G_S'$ and those $\U(1)$'s. Now
$\pi_1(G)^\wedge = \prod_{i\in Q_0} \U(1)$ is a maximal torus $T_S$ of
$G_S$.

\begin{NB}
It should be possible to see the Weyl group $W(Q)$ symmetry directly
in the formula........
\end{NB}

We also consider the group $\Gamma$ of the diagram automorphism of
preserving both $\dim V$, $\dim W$. It acts on $G_S$ by outer
automorphisms.
Then we expect
\begin{enume_add}
  \item $\Gamma\ltimes G_S$ acts on $\mathcal M_C$ compatibly with the
$\pi_1(G)^\wedge$-action, preserving the hyper-K\"ahler structure.
    \label{item:disgroup}
\end{enume_add}

In the above example in Figure~\ref{fig:mirror}, the left one contains
a type $A_{N-1}$ subgraph in the bottom. Therefore $G_S =
\SU(N)\times\U(1)$, where the extra $\U(1)$ comes from the upper
circled vertex. We also have the overall $\U(1)$ in $\SU(2)$, and
$\Gamma = \{\pm 1\}$ from diagram automorphisms. Thus their product
should act on $\mathcal M^A_C$.
This should be the same as the natural action on $\mathcal M^B_H$, the
framed moduli space of $\SU(N)$-instantons on $\RR^4$,
where $\SU(N)$ acts by the change of framing, $\U(1)\times\U(1)$ acts
on the base $\RR^4=\CC^2$ with $(t,z)\cdot (x_1,x_2) = (t z x_1, t
z^{-1} x_2)$, and the $\Gamma$-action is given by taking dual instantons.
In the example in Figure~\ref{fig:E8}, we have $E_8\times \U(1)\times
\U(1)$-action from this construction.

\begin{NB}
    In the very first example for $\mathcal M_C = \CC^2$, the refined
    monopole formula is
    \begin{equation*}
        H_{\U(1),\HH}(t,z) = \frac1{1-t^2}\sum_{m\in\ZZ} t^{|m|} z^m
        = \frac1{1-t^2}\left(\frac1{1-tz} + \frac{tz^{-1}}{1-tz^{-1}}\right)
        = \frac1{(1-tz)(1-tz^{-1})}.
    \end{equation*}
    Therefore we see $tz$ and $tz^{-1}$ on $x_1$ and $x_2$ respectively.
\end{NB}

\begin{Remark}\label{rem:parameters}
    Let us write $\pi_1(G) = Y/Y_{\mathrm{cr}}$, where $Y$ (resp.\
    $Y_{\mathrm{cr}}$) is the coweight (resp.\ coroot) lattice of $G$.
    Therefore $\pi_1(G)^\wedge$ is the kernel of the homomorphism
    $Y^\wedge\to Y_{\mathrm{cr}}^\wedge$ between Pontryagin duals of
    $Y$, $Y_{\mathrm{cr}}$.
We have $Y^\wedge = (X\otimes\RR)/X$, where $X = \Hom(Y,\ZZ)$ is the
weight lattice of $G$.
\begin{NB}
\(
     Y^\wedge = \Hom(Y, \U(1)) = \Hom(Y,\RR/\ZZ)
\)
and the exact sequence
\(
  0\to \Hom(Y,\ZZ) \to \Hom(Y,\RR) \to \Hom(Y,\RR/\ZZ) \to 0.
\)
\end{NB}%
Therefore the coweight lattice $\Hom(\U(1),\pi_1(G)^\wedge)$ is the
same as the kernel of the homomorphism $X\to
\Hom(Y_{\mathrm{cr}},\ZZ)$ given by the pairing with coroots
$Y_{\mathrm{cr}}$. It is equal to the character group of $G$.
In summary, we have a natural isomorphism
\begin{equation}
    \label{eq:36}
    \begin{NB}
    \Hom(\U(1),T_S) = 
    \end{NB}%
    \Hom(\U(1),\pi_1(G)^\wedge)
    \cong \Hom(G,\U(1))
    \cong \Hom(G_\CC,\CC^*).
\end{equation}

Thus a coweight $\chi\in \Hom(\U(1),\pi_1(G)^\wedge)$ defines a
character $G\to\U(1)$, and hence gives a stability condition and the
corresponding GIT quotient of $\bmu_\CC^{-1}(\zeta_\CC)$ by $G_\CC$ as
in \eqref{eq:37}.

Therefore an element of \eqref{eq:36} plays two roles, one on $\mathcal
M_H$, another on $\mathcal M_C$. This observation was essentially
given already in \cite{MR1413696,MR1454291}.
The former appears as a value of the hyper-K\"ahler moment map, or
{\it Fayet-Iliopoulos parameter\/} in the physics terminology.
On the other hand, when $G_S$ acts on $\mathcal M_C$, an element of
the Lie algebra of $G_S$ is called {\it mass parameter}.
\end{Remark}

\begin{NB}

    The monopole formula is a limit of a formula for a certain `index'
    of a $4$-dimensional theory, at least in special cases. The
    formula involves elliptic Macdonald operators. Let us record a
    naive generalization, at least for Macdonald polynomials.

\subsection{}

Suppose $G = \GL(k)$ and $\la = (\la_1\ge\dots\ge\la_k) =
(1^{k_1}2^{k_2}\cdots)$ is a partition. (I assume $\la_k > 0$), we
have
\begin{equation*}
    P_G(t;\la) = \prod_\alpha \frac1{(t^2;t^2)_{k_\alpha}}
    = \prod_\alpha \frac1{(1-t^2)(1-t^4)\cdots (1-t^{2k_\alpha})}.
\end{equation*}
Recall
\begin{equation*}
    c_\lambda(q,t) = \prod_{s\in\la} (1 - q^{a(s)} t^{l(s)+1}).
\end{equation*}
In the Hall-Littlewood limit $q\to 0$, we have
\begin{equation*}
    c_\lambda(0,t) = \prod_{\substack{s\in\la\\ a(s)\neq 0}} (1 - t^{l(s)+1})
    = \prod_{\alpha} (t;t)_{k_\alpha}
\end{equation*}
\end{NB}

\section{Flavor symmetry}\label{sec:flavor-symmetry}

\subsection{Line bundles over Coulomb
  branches}\label{subsec:linebundle}

Let us discuss a flavor symmetry following
\cite{Cremonesi:2014kwa}. It means that we suppose that $\bM$ is a
quaternionic representation of a larger compact Lie group $\tilde G$,
which contains the original group $G$ as a normal subgroup.
The quotient $G_F = \tilde G/G$ is called the {\it flavor symmetry
  group}. For a quiver gauge theory, we can take (at least) $G_F =
\prod_{i\in Q_0} \U(W_i)/\U(1)$ (and $\tilde G = G\times G_F$), where
$\U(1)$ is the overall scalar, which acts trivially on $\bM\tslash G$.

Note that we do not use $G_F$ to take a quotient. The gauge theory
$\Hyp(\bM)\tslabar G$ has a $G_F$-symmetric QFT in the sense of
\cite{Tach-review}. (The reader needs to remember that $\bM$ is a
representation of $\tilde G$.)
As a concrete mathematical consequence, for example, $G_F$ acts as a
symmetry group on the Higgs branch $\mathcal M_H(\Hyp(\bM)\tslabar
G)$, which is the hyper-K\"ahler quotient $\bM\tslash G$.

Let us turn to study the role of $G_F$ playing on the Coulomb branch
$\mathcal M_C(\Hyp(\bM)\tslabar G)$.
As observed in \cite{Cremonesi:2014kwa}, we can naturally put $G_F$ in
the monopole formula~\eqref{eq:11} as follows. 
Let us consider the short exact sequence of groups
\begin{equation*}
        1 \to G \xrightarrow{\alpha} \tilde G
    \xrightarrow{\beta} G_F \to 1.
\end{equation*}
(This is the same as \eqref{eq:20} in the abelian case.)
Let us fix a coweight $\la_F$ of $G_F$ and consider the inverse image
$\beta^{-1}(\la_F)$. We consider $\Delta(\la)$ and $P_G(t;\la)$ for $\la\in\beta^{-1}(\la_F)$.
Their definitions in \eqref{eq:14} remain the same. For $\Delta(\la)$,
the first term is the sum over $\Delta^+$, positive roots of $G$,
considered as positive roots of $\tilde G$.
\begin{NB}
    The root system of $\tilde G$ is the sum of root systems of $G$
    and $G_F$. Where is the reference ?
\end{NB}%
In the second term, we understand $\operatorname{wt}(b)$ as a weight
of $\tilde G$, and paired with $\la$. We do not change $P_G(t;\la)$,
we consider the stabilizer of $\la$ in $G$, {\it not\/} in $\tilde
G$. The sum is over $\beta^{-1}(\la_F)/W$, where $W$ is the Weyl group
of $G$. We get a function
\begin{equation*}
    H_{\tilde G,\bM}(t,\la_F) = \sum_{\la\in \beta^{-1}(\la_F)/W} t^{2\Delta(\la)}
    P_G(t;\la)
\end{equation*}
in $t$ together with $\la_F$.

It was found in \cite{Cremonesi:2014kwa} (and more recent one
\cite{Cremonesi:2014uva}) that this generalization turned out to be
very fruitful by two reasons:

First, let $\{ \Hyp(\bM_i\tslabar G_i)\}_{i=1,2,\dots}$ be a
collection of {\it simpler\/} gauge theories sharing the common flavor
symmetry group $G_F$. (Thus $\bM_i$ is a representation of $\tilde
G_i$, and $G_F = \tilde G_i/G_i$.)
We define a {\it complicated\/} gauge theory as $\Hyp(\bigoplus
\bM_i\tslabar \prod' G_i)$, where $\prod' G_i$ is the fiber product of
$\prod G_i$ and the diagonal $G_F$ over $\prod G_F$.
The Hilbert series of the complicated theory is written by those
$H_{\tilde G_i,\bM_i}(t;\la_F)$ ($i=1,2,3\dots)$ of simpler theories as
\begin{equation*}
    \sum_{\la_F\in Y_F/W_F} t^{-2\sum_{\alpha\in\Delta_F^+}|\langle
      \alpha,\la_F\rangle|} 
    P_{G_F}(t;\la_F) \prod_i H_{\tilde G_i,\bM_i}(t;\la_F),
\end{equation*}
where $Y_F$, $W_F$, $\Delta^+_F$ are the coweight lattice, Weyl group,
the set of positive roots of $G_F$. This is clear from the form of the
monopole formula.

Second, if $\Hyp(\bM\tslabar G)$ is a quiver gauge theory of type $A$
(or its $\grpSp/\grpO$ version), the Hilbert series are written by
Hall-Littlewood polynomials.
\begin{NB}
    It seems that a rigorous proof is missing.
\end{NB}

Combining two, one can write down the Hilbert series of Higgs branches
of $3d$ Sicilian theories in terms of Hall-Littlewood polynomials, as
an example of an application \cite{Cremonesi:2014vla}.

Since quiver varieties of type $A$ are nilpotent orbits (and their
intersection with Slodowy slices) as we mentioned in
\subsecref{sec:quiver-type}, the appearance of Hall-Littlewood polynomials
is very suggestive. They appear as dimensions of spaces of sections of
line bundles over flag varieties. See \cite{MR1223221}, where the
Euler characteristic version was found earlier in \cite{MR593631}.

For an abelian gauge theory, the dual torus of $G_F$ appears in the
quotient construction of the Coulomb branch as $\mathcal M_C =
\bM\tslash G_F^\vee$ (\subsecref{sec:abelian-theory}). Therefore a
coweight $\la_F$ of $G_F$, which is a weight of $G_F^\vee$, defines a
line bundle over the resolution
$\bmu^{-1}(\zeta_{\operatorname{Im}\HH})/G_F^\vee$ as
\begin{equation*}
    \bmu^{-1}(\zeta_{\operatorname{Im}\HH}) \times_{G_F^\vee} \CC,
\end{equation*}
where $G_F^\vee$ acts on $\CC$ by $\la_F$. It naturally has a
connection, which is integrable for any of $I$, $J$, $K$
\cite{MR1139657}. In particular, it is a holomorphic line bundle with
respect to $I$.

Based on these observations, it is natural to add the followings to
the list of expected properties:
\begin{enume_add}
      \item We have a (partial) resolution of $\mathcal M_C$ whose
    Picard group is isomorphic to the coweight lattice $Y_F$ of
    $G_F$. Moreover the character of the space of sections of a
    $\U(1)$-equivariant holomorphic line bundle $\mathcal L_{\la_F}$
    corresponding to a coweight $\la_F$ is given by the monopole
    formula $H_{\tilde G,\bM}(t,\la_F)$. (Here we replace $\mathcal
    M_C$ if necessary so that $\mathcal L_{\la_F}$ is relatively
    ample.)
\label{item:linebundle}

  \item The Weyl group $W_F$ acts on the Picard group of the partial
resolution above.
\label{item:Weyl}
\end{enume_add}

As is usual for hyper-K\"ahler manifolds, a resolution and a
deformation are related by the hyper-K\"ahler rotation. Therefore we expect
\begin{enume_add}

      \item We have a deformation of $\mathcal M_C$ parameterized by
    the Cartan subalgebra $\mathfrak h_F$ of $G_F$. The Weyl group
    $W_F$ acts on the homology group by the monodromy.
\end{enume_add}

Let us check the compatibility of the conjecture with the $3d$ mirror
symmetry.
Recall that the group $\Gamma\ltimes G_S$ acting on $\mathcal M_C^A$
(see \subsecref{sec:expect-properties}(\ref{item:disgroup})).
Since the definition of $\Gamma\ltimes G_S$ depends on the choice of
the theory $A$, let us denote it by $\Gamma^A\ltimes G_S^A$.
It is natural to expect that it is identified with $G_F^B$ (the flavor
symmetry group for the theory $B$) acting on $\mathcal M_H^B$.
This is indeed the case for the example in Figure~\ref{fig:mirror},
where $\Gamma^A\ltimes G^A_S$ are $\{\pm 1\}\ltimes
\SU(N)\times\U(1)$.
The squared $N$ gives us $\SU(N)$. The factor $\U(1)$ comes from an
{\it internal\/} symmetry of the graph. The edge loop at the circled
$k$ gives the factor $\End(\CC^k)\otimes\CC^2$ in $\bM$. Then $\U(1)$
is acting on $\CC^2$ preserving the hyper-K\"ahler
structure.\footnote{A larger group $\grpSp(1)$ acts on $\CC^2$, but
  the author does not know how to see it in the monopole formula.}
Finally $\{\pm 1\}$ is identified with the symmetry defined by
transpose of linear maps in $\bM$. (Since $\bM$ is the cotangent type,
transpose of an element in $\bN$ is in $\bN$.)

Since the mirror symmetry should be a duality, the role of $G_F^B$ in
$\mathcal M_C^B$ should be the same as the role of $\Gamma^A\ltimes
G_S^A$ playing in $\mathcal M_H^A$. The diagram automorphism group
$\Gamma^A$ induces automorphisms on $\mathcal M^A_H$. This is clear.

Let us turn to $G_S^A$. Recall that we have defined $G_S^A$ in two
steps. We first define the maximal torus of $G_S^A$ as
$\pi_1(G)^\wedge$, where the group $G$ is the one we take the
quotient. Then we consider the Weyl group invariance in the second
step. So consider $T_S^A \defeq \pi_1(G)^\wedge$ first. By
\eqref{eq:36} a coweight $\la_S\in\Hom(\U(1),T_S^A)$ defines a
character of $G$, and hence gives a (partial) resolution and a line
bundle on it for the Higgs branch $\mathcal M_H^A$ as in
\eqref{eq:37}. This is exactly the property (\ref{item:linebundle}) for
$G_F^B$ and $\mathcal M^B_C$.

Moreover, if we take $\la_S$ generic in $\Hom(\U(1),T_S)$, we expect
that the Picard group of $\bmu_\CC^{-1}(0)
\dslash\!\raisebox{-4pt}{$\scriptstyle\lambda_S$} G_\CC$ is isomorphic
to $\Hom(\U(1),T_S^A)$ and we have an action of the Weyl group of $G_S^A$,
as in (\ref{item:Weyl}).

\subsection{Abelian case}

In \cite[\S6]{Cremonesi:2013lqa}, the proposal~(\ref{item:linebundle})
was checked for the abelian theory. Let us give a different proof.

We keep the notation in \subsecref{sec:abelian-theory}. We choose and
fix a coweight $\la_F$ of $G_F$, considered also as a weight of
$G_F^\vee$. We {\it define\/} the Coulomb branch $\mathcal M_C$ by
$\bM\tslash G_F^\vee$ and its partial resolution $\tilde{\mathcal M}_C$ by 
\(
  \bmu^{-1}(\la_F,0)/G_F^\vee
= \bmu_\CC^{-1}(0)
\dslash\!\raisebox{-4pt}{$\scriptstyle\lambda_F$} (G_F^\vee)_\CC.
\)
Here $\la_F$ is considered as $(\operatorname{Lie}G_F^\vee)^*$ in the
first description as a hyper-K\"ahler quotient, and the stability
parameter in the second description as a GIT quotient. (The complex
parameter is set $0$.)
We have a relative ample line bundle $\mathcal L_{\lambda_F} =
\bmu^{-1}(\la_F,0)\times_{G_F^\vee}\CC$. Our goal is to check that the
character of the space of sections of $\mathcal L_{\lambda_F}$ is
given by $H_{\tilde G,\bM}(t,\la_F)$.

In order to have a clear picture, we consider the action of
$\pi_1(G)^\wedge = G^\vee$ on $\mathcal M_C$. We take a lift of the
action to the line bundle $\mathcal L_{\lambda_F}$. It means that we
lift $\la_F$ to a weight $\tilde\la_F$ of $(T^d)^\vee$ so that we have
the induced $G^\vee = (T^d)^\vee/G_F^\vee$-action on
$\bmu^{-1}(\la_F,0)\times_{G_F^\vee}\CC$. Then the
space of sections is a representation of $\U(1)\times G^\vee$. The
monopole formula is refined as
\begin{equation}\label{eq:21}
    H_{\tilde G,\bM}(t,\la_F) = \sum_{\la\in \ZZ^{d-n}}
    z^\la t^{2\Delta(\la)} P_G(t;\la),
\end{equation}
as in \eqref{eq:15}. Here we use the identification $\beta^{-1}(\la_F)
= \tilde\la_F + \Ima\alpha\cong \ZZ^{d-n}$, as we choose the lift
$\tilde\la_F$. We understand $z$ as a multi-variable, and $z^\la$
means $z_1^{\la_1}\cdots z_{d-n}^{\la_{d-n}}$.

Since $G$ is torus, the stabilizer of $\la$ is always $G$
itself. Therefore $P_G(t;\la) = 1/(1-t^2)^{\rank G}$. And the Weyl
group is trivial, as we have already used above.
We have 
\begin{equation*}
    2 \Delta(\la) = \sum_{i=1}^d 
    \left| (\tilde\la_F + \alpha(\la))_i \right|,
\end{equation*}
where $(\tilde\la_F + \alpha(\la))_i$ is the
$i^{\mathrm{th}}$-component of $\tilde\la_F + \alpha(\la)\in \ZZ^d$.

\begin{NB}
Let $S^p_i$ ($i=1,\dots,d$, $p=1,\dots, d-n$) denote the map
$\ZZ^{d-n}\to\ZZ^d$ in \eqref{eq:19} written in standard
coordinates. We write $\la = (m_1,\dots,m_{d-n})$ also in standard
coordinates. Then
\begin{equation*}
    2\Delta(\la) = \sum_{i=1}^d \left| \sum_{p=1}^{d-n} S_i^p m_p \right|.
\end{equation*}
\end{NB}

Let us start the proof. A key is a {\it trivial\/} observation that
the hyper-K\"ahler quotient of $\bM$ by $T^d$ is a single point at any
level of the hyper-K\"ahler moment map. It is enough to check $d=1$,
then it is obvious that the solution $(x,y)\in\CC^2$ of
\begin{equation*}
\left\{ 
  \begin{aligned}[m]
      & xy = \zeta_\CC,
\\      
      & |x|^2 - |y|^2 = \zeta_\RR
\end{aligned}
\right.
\end{equation*}
is unique up to $\U(1)$ for any $\zeta_\RR$, $\zeta_\CC$. Let us see
this in the GIT picture. Let $\zeta_\CC = 0$, as it becomes trivial
otherwise. A function on $xy=0$, which has weight $m\in\ZZ$ with
respect to the $\CC^*$-action $z\cdot(x,y) = (zx,z^{-1}y)$, is $x^m$
if $m\ge 0$, $y^{-m}$ if $m\le 0$, up to constant multiple. This
function has weight $|m|$ with respect to the dilatation action
$t\cdot (x,y) = (tx,ty)$. Therefore with respect to the
$\CC^*\times\CC^*$-action, the character of $\CC[x,y]/(xy)$ is
\begin{equation}\label{eq:22}
    \sum_{m\in\ZZ} z^m t^{|m|}.
\end{equation}

This trivial observation can be applied to our situation by
considering hyper-K\"ahler quotients of $\tilde{\mathcal M}_C$ by
$G^\vee$, which are nothing but hyper-K\"ahler quotients of $\bM$ by
$T^d$. We have the complex moment map
\begin{equation*}
    \bmu^{G^\vee}_\CC\colon \tilde{\mathcal M}_C\to
    (\operatorname{Lie} G^\vee)^* \otimes\CC \cong \mathfrak g_\CC.
\end{equation*}
We consider it as a set of functions defining the subvariety
$(\mu^{G^\vee}_\CC)^{-1}(0)$. It is a complete intersection, and the
space of sections on $(\mu^{G^\vee}_\CC)^{-1}(0)$ and that of
$\tilde{\mathcal M}_C$ differ by the factor $1/(1-t^2)^{\rank G} =
P_G(t;\la)$. Now we consider $(\mu^{G^\vee}_\CC)^{-1}(0)$ is a
quotient of the subvariety $(\mu^{T^d}_\CC)^{-1}(0)$ in $\bM$, where
$\mu^{T^d}_\CC$ is the complex moment map of the $T^d$-action on
$\bM$. The latter is given by $x_i y_i = 0$ ($i=1,\dots, d$) in the
standard coordinate system $(x_i,y_i)_{i=1}^d$ of $\bM$. A section of
$\mathcal L_{\la_F}$ with weight $\la$ with respect $G^\vee$ is a
function on $x_iy_i = 0$ with weight $\tilde\la_F+\alpha(\la)$, hence
is a monomial in either $x_i$ or $y_i$ according to the sign of
$(\tilde\la_F + \alpha(\la))_i$.
Now we deduce \eqref{eq:21} by the same argument in \eqref{eq:22}. We
replace $z^m$ by $z^\la$, $t^{|m|}$ by $t^{2\Delta(\la)}$
respectively.

As a byproduct of this proof of the monopole formula, we obtain a
linear base of the coordinate ring $\CC[\mathcal M_C]$. It is given by
monomials in components of the moment map $\bmu^{G^\vee}_\CC$ and
\begin{equation}\label{eq:23}
    \begin{NB}
      \  z^\la\defeq
    \end{NB}
    \prod_i\left(\text{$x_i^{\alpha(\lambda)_i}$ or $y_i^{-\alpha(\lambda)_i}$}\right)
\end{equation}
for each $\lambda\in\ZZ^{d-n}$. Here we take $x_i^{\alpha(\la)_i}$ if $\alpha(\la)_i\ge 0$ and $y_i^{-\alpha(\la)_i}$ otherwise.

Let us give an explicit presentation of the coordinate ring
$\CC[\mathcal M_C]$. It is an algebra over $\CC[\mathfrak g_\CC]$,
the polynomial ring over $\mathfrak g_\CC$ by the moment map $\bmu^{G^\vee}_\CC\colon \mathcal M_C\to\mathfrak g_\CC$.
Let us denote the element \eqref{eq:23} by $z^\la$. (The element gives
the corresponding character $z^\la$, and there is no fear of
confusion.) Then $\CC[\mathcal M_C]$ is $\bigoplus_{\la\in\ZZ^{d-n}}
\CC[\mathfrak g_\CC] z^\la$ with multiplication
\begin{equation*}
    z^\la z^\mu = z^{\la+\mu} \pi(\prod_i \xi_i^{d_i(\la,\mu)}),
\end{equation*}
where $\pi\colon \CC[\mathfrak t^d_\CC]\to\CC[\mathfrak g_\CC]$ is the
projection induced by the inclusion $\mathfrak g_\CC\subset\mathfrak
t^d_\CC$, $\xi_i$ is a standard linear coordinate function on
$\mathfrak t^d_\CC$, and
\begin{equation*}
    d_i(\la,\mu) =
    \begin{cases}
        0 & \text{if $\alpha(\la)_i\times\alpha(\mu)_i\ge 0$},
\\
        -\alpha(\mu)_i & \text{if $\alpha(\la)_i\ge 0$, $\alpha(\mu)_i\le 0$,
          $\alpha(\la+\mu)_i\ge 0$},
\\
        \alpha(\la)_i & \text{if $\alpha(\la)_i\ge 0$, $\alpha(\mu)_i\le 0$,
          $\alpha(\la+\mu)_i\le 0$},
\\
        -\alpha(\la)_i & \text{if $\alpha(\la)_i\le 0$, $\alpha(\mu)_i\ge 0$,
          $\alpha(\la+\mu)_i\ge 0$},
\\
        \alpha(\mu)_i & \text{if $\alpha(\la)_i\le 0$, $\alpha(\mu)_i\ge 0$,
          $\alpha(\la+\mu)_i\le 0$}.
    \end{cases}
\end{equation*}

This is because the moment map $\bmu_\CC^{G^\vee}$ is induced from
$\bmu_\CC^{T^d}$ on $\bM$, which is given by $x_i y_i$. We write
$z^\la$ by $x_i^{\alpha(\la)_i}$ or $y_i^{-\alpha(\la)_i}$, and
identify $\xi_i$ with $x_i y_i$. Then we get the above formula of
$d_i(\la,\mu)$ according to signs of $\alpha(\la)_i$, $\alpha(\mu)_i$,
$\alpha(\la+\mu)_i$.
\begin{NB}
    For example, suppose $\alpha(\la)_i\ge 0$, $\alpha(\mu)_i\le 0$,
    $\alpha(\la+\mu)_i\ge 0$. Then the $i^{\mathrm{th}}$-component of
    $z^\la z^\mu$ is
    \begin{equation*}
        x_i^{\alpha(\la)_i} y_i^{-\alpha(\mu)_i}
        = x_i^{\alpha(\la+\mu)_i} (x_i y_i)^{-\alpha(\mu)_i}.
    \end{equation*}
\end{NB}%

This formula is found by Dimofte and Hilburn \cite{DH}.

\section{Topological twist}\label{sec:topological-twist}

In this section we discuss PDE's relevant to the topologically twisted
version of our gauge theories. For $(G,\bM) = (\U(1),\HH)$, it is
nothing but the Seiberg-Witten monopole equation \cite{MR1306021}.
More general cases were discussed for example in \cite{MR1708781,MR2101297,2013arXiv1303.2971H}.
However their reductions to $2$-dimension were not discussed before,
as far as the author knows, except it is of course well known that the
dimension reduction of the Seiberg-Witten equation is the
(anti-)vortex equation.
Therefore we will discuss equations from the scratch though our
treatment is more or less a standard textbook material.

\subsection{Generalized Seiberg-Witten equations}

We consider only the case when the $4$-manifold $X$ is spin for
simplicity. See \remref{rem:pm} for a framework close to the usual Seiberg-Witten equation for a spin${}^c$ $4$-manifold under an additional assumption on $(G,\bM)$.

Recall that $\bM$ is a quaternionic representation of $G$. Therefore
$\grpSp(1)$, the group of unit quaternions, acts on $\bM$ commuting
with the $G$-action. Then $\bM$ can be made a representation of $\hat
G \defeq \operatorname{Spin}(4) \times G =
\grpSp(1)\times\grpSp(1)\times G$ in two ways, by choosing
$\grpSp(1)\times\grpSp(1)\to\grpSp(1)$ either the first or second
projection. Let us distinguish two representations, and denote them by
$\bM^+$ and $\bM^-$. We write $\operatorname{Spin}(4) =
\grpSp(1)_+\times\grpSp(1)_-$ accordingly.

\begin{NB}
The generalized Seiberg-Witten equation is basically an equation on a
pair $(A,\Phi)$ of a $G$-connection and a spinor with value in $\bM$,
but we slightly generalize the construction so that the equation makes
sense even when a $4$-manifold $X$ is not necessarily spin, as in the
ordinary Seiberg-Witten theory.
The ordinary Seiberg-Witten equation corresponds to the case $(G,\bM)
= (\U(1),\HH)$.

We assume that $G$ contains a central element acting by the
multiplication of $-1$ on $\bM$. We denote it also by $-1$ for
brevity. 
We consider $\hat G = (\operatorname{Spin}(4)\times
G)/\!\pm\!(1,1)$. We have an exact sequence $1\to G\to \hat G\to
\SO(4)\to 1$ of Lie groups.
For example, $\hat G = \operatorname{Spin}^c(4)$ for $G=\U(1)$.
We have $\operatorname{Spin}(4) = \grpSp(1)\times\grpSp(1)$, hence we
can make $\bM$ to a representation of $\hat G$ in two ways, one
through the left $\grpSp(1)$, another through the right
$\grpSp(1)$. Here $\grpSp(1)$ is considered as the group of
quaternions of unit length, and acts on $\bM$ by the $\HH$-module
structure. Let us distinguish two representations, and denote them by
$\bM^+$ and $\bM^-$. We write $\operatorname{Spin}(4) =
\grpSp(1)_+\times\grpSp(1)_-$ accordingly.
\end{NB}

We regard $\HH$ as an $\grpSp(1)_+\times\grpSp(1)_-$-module by left
and right multiplication. The quaternion multiplication gives a $\hat
G$-homomorphism $\HH\otimes\bM^-\to \bM^+$. Combining this with
$\HH\otimes\bM^+\to\bM^-$ given by $(x,m)\mapsto -\overline{x}m$, we
have a Clifford module structure (over $\RR$) on $\bM^+\oplus\bM^-$.
\begin{NB}
    Let $m\in\bM^-$, $x\in\HH$. Multiplication is given by $xm$. It is
    considered to be an element in $\bM^+$ as
    \begin{equation*}
        \begin{CD}
            (x, m) @>>> x m
            \\
            @VVV @VVV
            \\
            (g_+ x g_-^{-1}, g_- m) @>>> g_+ x g_-^{-1} g_- m = g_+xm.
        \end{CD}
    \end{equation*}
    We also consider
    \begin{equation*}
        \begin{CD}
            (x,m') @>>> - \bar{x}m'
            \\
            @VVV @VVV
            \\
            (g_+ x g_-^{-1}, g_+m') @>>>
            -\overline{g_+ x g_-^{-1}}  g_+ m' = -g_- \bar{x}m'.
        \end{CD}
    \end{equation*}
\end{NB}%

We regard the hyper-K\"ahler moment map (see
\subsecref{sec:hyper-kahl-quot} for detail), as a map
$\mu\colon\bM^+\to \mathfrak g\otimes\algsp(1)_+$ by identifying
$\algsp(1)_+$ with the space of imaginary quaternions.
\begin{NB}
    Namely $\mu_I i + \mu_J j + \mu_K k\in\mathfrak g\otimes\algsp(1)_+$.
\end{NB}%
It is equivariant under $\hat G$.
\begin{NB}
It is equivariant under $\hat G$, which act on $\mathfrak g$ by the
composition of the adjoint action and the homomorphism $\hat G\to
G/\pm 1$,
\begin{NB2}
    Note that $-1$ acts trivially on $\mathfrak g$ as it is central.
\end{NB2}%
and on $\algsp(1)_+$ through $\hat G\to\grpSp(1)_+/\!\pm\!1$.
\end{NB}

Let $X$ be an oriented Riemannian $4$-manifold. We assume $X$ is spin,
as mentioned above.
Let $P\to X$ be a principal $G$-bundle. We have the associated $\hat
G$-principal bundle, denoted by $\hat P$, by taking the fiber product
with the double cover $P_{\operatorname{Spin}(4)}$ of the orthonormal
frame bundle $P_{\SO(4)}$ of $TX$ given by the spin structure.
Let $A$ be a connection on $P$. We extend it to a connection on $\hat
P$, combining with the Levi-Civita connection on the
$P_{\operatorname{Spin}(4)}$-part. We denote the extension also by $A$
for brevity.
\begin{NB}
Let $P\to X$ be a principal $\hat G$-bundle
together with an isomorphism $P/G\cong P_{\SO(4)}$ of
$\SO(4)$-bundles, where $P_{\SO(4)}$ is the orthonormal frame bundle
of $TX$.
Let $A$ be a connection on $P$ such that the induced connection on
$P/G = P_{\SO(4)}$ is the Levi-Civita connection.
Since $\SO(4)$-part is determined, we can identify $A$ with a
connection on $P/\operatorname{Spin}(4)$, which is a $G/\!\pm\!1$-bundle.
\begin{NB2}
    We have two projections $\hat G\to G/\pm 1$ and $\hat G\to
    \SO(4)$. We add the pull-backs of a connection form
    $\omega_{P/\operatorname{Spin}(4)}$ and the Levi-Civita connection
    $\omega_{\mathrm{LC}}$ to get a connection $\omega$ on $P$. It is
    $\so(4)\oplus\g= \hat\g$-valued $1$-form. Let us check that it
    satisfies two defining properties of a connection. a) It satisfies
    $\omega(\xi^*) = \xi$ for $\xi\in\hat\g$ since
    $\omega_{P/\operatorname{Spin}(4)}$ and $\omega_{\mathrm{LC}}$
    satisfy the corresponding property for $\xi\in\g$, $\xi\in\so(4)$
    respectively. b) The equivariance under $\hat G$ also follows as
    the corresponding equivariance of
    $\omega_{P/\operatorname{Spin}(4)}$ and $\omega_{\mathrm{LC}}$.

    Conversely suppose we have a connection $\omega$ on $P$. Using the
    decomposition $\so(4)\oplus\g= \hat\g$, we define a $\g$-valued
    $1$-form $\omega_\g$ on $P$. Note that the $\so(4)$-part is the
    pull-back of the Levi-Civita connection by the assumption.

    We have $\omega_\g(\xi^*) = 0$ for $\xi\in\so(4)$, which means
    that $\omega_\g$ vanishes on tangent vectors of fiber directions
    of the projection $P\to P/\operatorname{Spin}(4)$. We have
    $\varphi_g^*\omega_\g = \omega_\g$ for
    $g\in\operatorname{Spin}(4)$, as $\operatorname{Ad}_g^{-1}(\xi) =
    \xi$ for $\xi\in\g$. Hence $\omega_\g$ descends to a connection on
    $P/\operatorname{Spin}(4)$.
\end{NB2}%
\end{NB}%
We have induced covariant derivatives $\nabla_{\!A}$ on $\hat
P\times_{\hat G}\bM^+$ and $\hat P\times_{\hat G}\bM^-$. Combining them
with Clifford multiplication, we have Dirac operators
$D_{\!A}^\pm\colon \Gamma(\hat P\times_{\hat G}\bM^\pm)\to
\Gamma(\hat P\times_{\hat G}\bM^\mp)$.

Let $\Phi$ be a section of $\hat P\times_{\hat G}\bM^+$.
Applying the hyper-K\"ahler moment map fiberwise, we regard
$\mu(\Phi)$ as a section of $\hat P\times_{\hat G}(\mathfrak
g\otimes\algsp(1)_+)$.
Note that $\hat P\times_{\hat G}\algsp(1)_+\cong
P_{\SO(4)}\times_{\SO(4)}\algsp(1)_+$ is the bundle $\Lambda^+$ of
self-dual $2$-forms.
Hence $\mu(\Phi)$ is a section of $\Lambda^+\otimes (P\times_{G}\g)$.

Now we define a generalized Seiberg-Witten equation by
\begin{equation}\label{eq:24}
    \begin{split}
        & D_{\!A}^+\Phi = 0,
        \\
        & F_A^+ = \bmu(\Phi).
    \end{split}
\end{equation}
Here $F_A^+$ is the self-dual part of the curvature of $A$. More
precisely it is of the connection on $P$, not of the extended one on
$\hat P$.
\begin{NB}
$\operatorname{Lie}\hat G =
\algsp(1)_+\oplus\algsp(1)_-\oplus\g$ valued, so we take its
$\g$-part.
\end{NB}%

\begin{Remark}\label{rem:pm}
    Assume that $G$ contains a central element acting by the
    multiplication of $-1$ on $\bM$. This is true for $(G,\bM) =
    (\U(1),\HH)$ corresponding to the usual Seiberg-Witten monopole
    equation. Then we can replace $\hat G$ by its quotient $\hat G' =
    \operatorname{Spin}(4)\times G/\pm(1,1)$. The example $G=\U(1)$
    gives $\hat G' = \operatorname{Spin}^c(4)$. Then we consider a
    principal $\hat G'$-bundle $\hat P'$ together with an isomorphism
    $\hat P'/G\cong P_{\SO(4)}$ of
    $\SO(4)=\operatorname{Spin}(4)/\!\pm 1$-bundles. The generalized
    Seiberg-Witten equation still makes sense for a connection $A$ on
    $\hat P'$ and a section $\Phi$ of $\hat P'\times_{\hat
      G'}\bM^+$. Here $A$ is supposed to induce the Levi-Civita
    connection on $P_{\SO(4)}$. In this framework, we do not need to
    assume $X$ to be spin, an existence of a lift of $P_{\SO(4)}$ to a
    $\hat G'$-bundle $\hat P'$ is sufficient. This framework is used
    in the usual Seiberg-Witten equation.
\end{Remark}

When $\bM$ is of the form $\mathbf O\otimes_\RR\HH$ for a real
representation $\mathbf O$ of $G$, we have a further
generalization. An example is $\bM = \g\otimes_\RR\HH$. The point is
$\HH$ has two $\HH$-module structures by left and right
multiplications.

We regard $\HH$ as a representation of $\operatorname{Spin}(4) =
\grpSp(1)_+\times\grpSp(1)_-$ as above. We consider $\bM$ as a
hyper-K\"ahler manifold by the left multiplication. And the moment map
is regarded as $\bmu\colon\bM\to\g\otimes\algsp(1)_+$ as above.
We have three possibilities to make $\bM=\mathbf O\otimes_\RR\HH$ as an
$\grpSp(1)_+\otimes\grpSp(1)_-$-module.
(They are possible topological twists in physics literature.)
Let $(g_+,g_-)\in\grpSp(1)_+\times\grpSp(1)_-$ and $x\in\HH$. Then the
action is either (1) $x\mapsto g_+ x$, (2) $x\mapsto g_+ x g_+^{-1}$,
or (3) $x\mapsto g_+ x g_-^{-1}$.
The case (1) is the same as the case studied above. 
In the cases (2), (3), it is better to replace $\hat G$ by
$\SO(4)\times G$ as $(-1,-1)\in\grpSp(1)_+\times\grpSp(1)_-$ acts
trivially on $\HH$.

The case (2) means that we regard $\HH$ with $S^+\otimes S^+ =
\Lambda^0\oplus \Lambda^+$.
Then $\Phi$ is a section of $(\Lambda^0\oplus\Lambda^+)\otimes
(P\times_{G}\mathbf O)$. Let us write it by $C\oplus B$. The Dirac
operator $D_{\!A}^+$ is identified with $d_{A}\oplus d_{A}^*$ via the
isomorphism $S^+\otimes S^- = \Lambda^1$, where $S^\pm$ is the spinor
bundle. Hence the generalized Seiberg-Witten equation is
\begin{equation*}
        \begin{split}
        & d_{A} C + d_{A}^* B = 0,
        \\
        & F_A^+ = \bmu(C\oplus B).
    \end{split}
\end{equation*}
This is the equation in \cite{Vafa-Witten} if $\bM=\g\otimes_\RR\HH$.

Now consider the case (3). Then $\HH$ is identified with $S^+\otimes
S^- = \Lambda^1$. Hence $\Phi$ is a section of $\Lambda^1\otimes
(P\times_{G}\mathbf O)$.  The Dirac operator is identified with
$d_A^*\oplus d_A^-$ via $S^-\otimes S^- = \Lambda^0\oplus
\Lambda^-$. Therefore the equation is
\begin{equation*}
        \begin{split}
        & d_{A}^-\Phi = 0 = d_{A}^* \Phi,
        \\
        & F_A^+ = \bmu(\Phi).
    \end{split}
\end{equation*}
This is the equation in \cite{MR2306566} if
$\bM=\g\otimes_\RR\HH$. For $\bM=\g\otimes_\RR\HH$, there is a
one-parameter family of equations parametrized by $t\in\proj^1$. See 
\cite[(3.29)]{MR2306566}.

\subsection{Roles of Higgs branch}\label{sec:roles-higgs}

Suppose $X$ is compact. We return back to the equation \eqref{eq:24}.

From the Weitzenb\"ock formula for $D^+_{\!A}$ for a solution $(A,\Phi)$ of
generalized Seiberg-Witten equations, we have
\begin{equation}\label{eq:26}
    \begin{split}
    0 & = \frac12\Delta |\Phi|^2 + |\nabla_{\!A}\Phi|^2 + \frac{s}4 |\Phi|^2
    + (F_A^+\Phi,\Phi)
\\
    & = \frac12\Delta |\Phi|^2 + |\nabla_{\!A}\Phi|^2 + \frac{s}4 |\Phi|^2
    + 2|\bmu(\Phi)|^2,
    \end{split}
\end{equation}
where $s$ is the scalar curvature.
\begin{NB}
    $((\bmu_I i + \bmu_J j + \bmu_K k)\phi,\phi) =
    2(|\bmu_I|^2+|\bmu_J|^2+|\bmu_K|^2)$.
\end{NB}%
For the ordinary Seiberg-Witten equation, we have $|\bmu(\Phi)|^2 =
|\Phi|^4/4$. Then considering a point where $|\Phi|^2$ takes a
maximum, one concludes
\(
    \sup |\Phi|^2 \le \sup_X \max(-s,0).
\)
(See e.g., \cite[Cor.~5.2.2]{MR1367507}.)

The same argument works if there exists a constant $C$ such that
$|\phi|^4\le C|\bmu(\phi)|^2$ for $\phi\in\bM$. However this is not
true in general. If there is nonzero $\phi$ such that $\bmu(\phi)=0$,
the inequality is not true. In fact, it is easy to check the
converse. If $\bmu(\phi)=0$ implies $\phi=0$, the inequality holds.
\begin{NB}
    Suppose the contrary. There exists $\phi_i$ ($i=1,2,\dots$) such
    that $|\phi_i|^4 \ge i|\bmu(\phi_i)|^2$. We may assume
    $|\phi_i|^2=1$ by multiplying a constant to $\phi_i$. Therefore
    $\bmu(\phi_i)\le 1/i$. We may further assume $\phi_i$ converges to
    $\phi_\infty$ as $i\to\infty$ with $|\phi_\infty|^2=1$. We also have
    $\bmu(\phi_\infty) = 0$, hence contradicts with the assumption.
\end{NB}%
Thus we have proved
\begin{itemize}
      \item If the Higgs branch $\mathcal M_H$ is $\{ 0\}$, $\Phi$ is
    bounded for a solution of \eqref{eq:24}.
\end{itemize}
The author learned this assertion during Witten's lecture at the
Isaac Newton Institute for Mathematical Sciences in 1996, 18 years ago.

If $\mathcal M_H\neq \{0\}$, we should study behavior of a sequence of
solutions $(A_i,\Phi_i)$ with $|\Phi_i|\to\infty$. 
We consider the normalized solution $\bar{\Phi}_i =
\Phi_i/\|\Phi_i\|_{L^2}$.
From \eqref{eq:26} we derive a bound on $\sup |\bar{\Phi}_i|$,
$\|\nabla_{\!A}\bar{\Phi}_i\|_{L^2}$,
and also
\(
   \int_X |\bmu(\bar{\Phi}_i)|^2 \le C\|\Phi_i\|_{L^2}^{-2}
\)
for a constant $C$.
\begin{NB}
\( 
   0 = \frac12 \Delta|\bar{\Phi}|^2 +
   |\nabla_A \bar{\Phi}|^2 + \frac{s}4 |\bar{\Phi}|^2 
   + 2 \|\Phi\|_{L^2}^2 |\bmu(\bar{\Phi})|^2.
\)
\end{NB}%
Thus $\bar{\Phi}_i$ converges weakly in $W^{1,2}$ and strongly in $L^p$ for any $p > 0$. Moreover $\bmu(\bar{\Phi}_i)$ converges to $0$, i.e., the limit
$\bar{\Phi}_\infty$ takes values in $\bmu^{-1}(0)$. Thus the Higgs
branch $\mathcal M_H$ naturally shows up.

Further analyses are given in e.g., \cite{2013arXiv1307.6447T} for a
special case $(G,\bM)=(\SU(2), \g\otimes\HH)$. See also \cite{2014arXiv1406.5683H} for a special case $(G,\bM)=(\U(1),\HH^N)$ and dimension $3$.

\subsection{K\"ahler surfaces}\label{sec:kahler-surfaces}

Let us consider the case $X$ is a compact K\"ahler surface, following
\cite[\S4]{MR1306021}, \cite[Chapter~7]{MR1367507}.

\begin{NB}
    Recall $\U(2)\subset\operatorname{Spin}^c(4)$ by
    \begin{equation*}
        \U(2) = \grpSp(1)\times \U(1)/\!\pm\!(1,1)\ni [g,\lambda]
        \longmapsto
        [\lambda,g,\lambda]\in
    \grpSp(1)_+\times\grpSp(1)_-\times\U(1)/\!\pm\!(1,1,1)
    = \operatorname{Spin}^c(4).
    \end{equation*}
    We understand $\operatorname{Spin}^c(4)$ as $\hat G$ with
    $G=\U(1)$. Note that $\tilde G = (\U(1)\times\grpSp(1)_-\times
    G=\U(1))/\!\pm\!(1,1,1)$ is not $\U(2)\times\U(1)$, hence the definition
    below is slightly different from the approach in \cite{MR1367507}.
\end{NB}

The bundle $\Lambda^+$ of self-dual $2$-forms decomposes as
$\Lambda^{0}\omega\oplus \Lambda^{0,2}$, where $\omega$ is the
K\"ahler form. It induces a decomposition $F_A^+ = (F_A^+)^{1,1}\oplus
F_A^{0,2}$. The moment map $\bmu(\Phi)$ also decomposes as
$\bmu_\RR(\Phi)\oplus \bmu_\CC(\Phi)$.

We replace $\hat G$ by its subgroup $\tilde G = \U(1)\times
\grpSp(1)_-\times G$. The principal bundle
$P_{\operatorname{Spin}(4)}$ has the reduction
$P_{\U(1)\times\grpSp(1)_-}$ to $\U(1)\times\grpSp(1)_-$-bundle.
We have a square root $K^{-1/2}_X$ of the anti-canonical bundle
$K_X^{-1}$, associated with the standard representation of $\U(1)$. On
the other hand, the bundle associated with the vector representation
of $\grpSp(1)_-$ is $\Lambda^{0,1}\otimes K_X^{1/2}$.
\begin{NB}
    The normalization is given so that the tensor product
    $(\Lambda^{0,1}\otimes K_X^{1/2})\otimes K_X^{-1/2} =
    \Lambda^{0,1}$.
\end{NB}%
The bundle $\hat P$ likewise has a reduction $\tilde P$ to a $\tilde
G$-bundle. It is also the fiber product of $P$ and
$P_{\U(1)\times\grpSp(1)_-}$ over $X$. The connection $A$ on $P$
extends to a connection on $\tilde P$ as before, by combined with the Levi-Civita connection.

\begin{NB}
We replace $\hat G$ by $\tilde G = (\U(1)\times
\grpSp(1)_-\times G)/\!\pm\!(1,1,1)$, which has an exact sequence
\(
   1 \to G\to \tilde G\to \U(1)\times\grpSp(1)_-/\pm(1,1) = \U(2) \to 1.
\)
We take a principal $\tilde G$-bundle $P$ with an isomorphism
$P/G\cong P_{\U(2)}$, where $P_{\U(2)}$ is the frame bundle of $\Lambda^{0,1}\otimes K_X$. This is different from the usual convention such that $P_{\U(2)}$ is the frame bundle of $\Lambda^{0,1}$,
\begin{NB2}
    It means that the standard $2$-dimensional representation gives
    $\Lambda^{0,1}$.
\end{NB2}%
but is convenient for our purpose below. 
The inclusion $\tilde G\to\hat G$ induces a $\hat G$ bundle $P'$ with
$P'/G\cong P_{\SO(4)}$, hence $P$ is considered as a reduction of a
$\hat G$-bundle in the general setting considered above.
\begin{NB2}
    $\U(2) = \U(1)\times \grpSp(1)_-/\pm$ has the negative spinor
    $S^-$ representation, which is the standard $2$-dimensional
    representation, and the positive spinor representation $S^+$,
    which is the sum of the trivial representation and the determinant
    representation through 
    \begin{equation*}
        \U(2) =
        \U(1)\times\grpSp(1)_- /\pm \ni (t,q_-) \longmapsto 
        (\begin{bmatrix}
            t & 0 \\ 0 & t^{-1}
        \end{bmatrix}, t)
        =
        \begin{bmatrix}
            t^2 & 0 \\ 0 & 1
        \end{bmatrix}
        \in
        \grpSp(1)_+\times \U(1)/\pm = \U(2).
    \end{equation*}
    The complexified tangent bundle is $\Hom(S^-, S^+)$. In our case,
    it is $S^-=\Lambda^{0,1}\otimes K_X$, $S^+=\underline{\CC}\oplus
    \det(\Lambda^{0,1}\otimes K_X) = K_X =
    (\Lambda^{0,0}\oplus\Lambda^{0,2})\otimes K_X$. Therefore
    $\Hom(S^-,S^+)$ is the same as usual.
\end{NB2}
\end{NB}

Since the multiplication of $i$ on $\bM$ commutes the action of
$\U(1)\times G
\begin{NB}
/\pm (1,1)    
\end{NB}%
$, we can view $\bM^+$ as a complex representation of $\tilde
G$. Hence $\tilde P\times_{\tilde G}\bM^+$ is a complex vector bundle.
A direct calculation shows that $\bmu_\RR(i\Phi) = \bmu_\RR(\Phi)$,
$\bmu_\CC(i\Phi) = -\bmu_\CC(\Phi)$.

We combine the complex structure $i$ with the following which is a
consequence of the Weitzenb\"ock formula:
\begin{equation}\label{eq:28}
    \int_X |D_A^+\Phi|^2 + \frac12 |F_A^+-\bmu(\Phi)|^2
    = \int_X |\nabla_A\Phi|^2 + \frac12 |F_A^+|^2 + \frac12 |\bmu(\Phi)|^2
    + \frac{s}4 |\Phi|^2.
\end{equation}
The right hand side is unchanged when we replace $\Phi$ by $i\Phi$. On
the other hand, $(A,\Phi)$ is a solution of the generalized
Seiberg-Witten equation if and only if the left hand side
vanishes. Hence $(A,\Phi)$ is a solution if and only if both $(A,\Phi)$ and $(A,i\Phi)$ are solutions, that is
\begin{equation}\label{eq:29}
    \begin{split}
        & D_A^+\Phi = 0 = D_A^+(i\Phi),
        \\
        & F_A^{0,2} = 0 = \bmu_\CC(\Phi),
        \\
        & (F_A^+)^{1,1} = \bmu_\RR(\Phi).
    \end{split}
\end{equation}
In particular, $P\times_G G_\CC$ has a structure of a holomorphic
principal bundle induced from $A$. Then $P\times_{G}\bM$ and $\tilde
P\times_{\tilde G}\bM^+ = K_X^{-1/2}\otimes (P\times_G \bM)$ are
holomorphic vector bundles.

\begin{NB}
Since above $F_A^{0,2}$ is only the $\g$-part of the curvature of $A$,
$P\times_{\tilde G}\tilde G_\CC$ may not have a structure of a
holomorphic principal bundle, but $P\times_{\tilde G}\bM^+$ is a
holomorphic vector bundle, as it involves only $\U(1)$-part in
$\U(1)\times\grpSp(1)_-/\pm (1,1)$.
\begin{NB2}
    We cannot make $P\times_{\tilde G}\bM^-$ to be a holomorphic
    vector bundle. It is not even clear what is the structure of a
    complex vector bundle.
\end{NB2}%
\end{NB}

We claim that $\Phi$ satisfies $D_A^+\Phi = 0 = D_A^+(i\Phi)$ if and
only if $\overline{\Phi}$, viewed as a section of $(\tilde
P\times_{\tilde G}\bM^+)^*
\begin{NB}
    = K_X^{1/2}\otimes (P\times_G \bM)
\end{NB}%
$, is holomorphic. It is enough to check the
assertion at each point $p$, hence we take a holomorphic coordinate
system $(z_1,z_2)$ around $p$. We write $z_1 = x_1 + iy_1$, $z_2 = x_2
+ iy_2$. We assume tangent vectors $\partial/\partial x_1$,
$\partial/\partial y_1$, $\partial/\partial x_2$, $\partial/\partial
y_2$ correspond to $1$, $i$, $j$, $k$ respectively under the
isomorphism $T_pX\cong \HH$. Then
\begin{equation*}
    \begin{split}
    D_A^+\Phi &= \sum_{\alpha=1}^2 
    \frac{\partial}{\partial x_\alpha} \cdot
    \nabla_{\!\frac{\partial}{\partial x_\alpha}}\Phi
    + 
    \frac{\partial}{\partial y_\alpha} \cdot
    \nabla_{\!\frac{\partial}{\partial y_\alpha}}\Phi
\\
&= - \nabla_{\!\frac{\partial}{\partial x_1}}\Phi
+ i\nabla_{\!\frac{\partial}{\partial y_1}}\Phi
+ j\nabla_{\!\frac{\partial}{\partial x_2}}\Phi
+ k\nabla_{\!\frac{\partial}{\partial y_2}}\Phi
\\
&= - 2\nabla_{\!\frac{\partial}{\partial z_1}}\Phi
+2\nabla_{\!\frac{\partial}{\partial \overline{z}_2}}\overline{\Phi}j.
    \end{split}
\end{equation*}
If we replace $\Phi$ by $i\Phi$, the first term is multiplied by $i$,
and the second term by $-i$. Therefore the assertion follows.

\begin{NB}
See 2014-12-13 for more detail.
\end{NB}

\begin{NB}
This is a record of an earlier attempt.

Let us proceed further.
We consider the bundle $\Lambda^{0,1}\otimes_\CC (P\times_{\tilde
  G}\bM^+)$.
In our convention, $\Lambda^{0,1}$ is the bundle associated with the representation
\(
   \HH,
\)
where the action is given by
$[\lambda,q_-,g]x = \lambda^{-1}x q_-^{-1}$ for
$[\lambda,q_-,g]\in \tilde G = \U(1)\times\grpSp(1)_-\times G/\pm$.
It is considered as a $\CC$-module by the multiplication from the
left.
\begin{NB2}
    Note that the standard representation $\lambda xq_-^{-1}$ gives
    $\Lambda^{0,1}\otimes K_X$. Then $\lambda^{-2}$ is
    $(\det\Lambda^{0,1}\otimes K_X)^{-1} = K_X^{-1}$, and
    $\lambda^{-1} xq_-^{-1}$ is $\Lambda^{0,1}$.
\end{NB2}%
Then $\Lambda^{0,1}\otimes_\CC (P\times_{\tilde G}\bM^+)$ is associated with
$\bM\otimes_\CC \HH$ where
\begin{equation*}
    [\la,q_-,g](m\otimes x) = [\la,g] m \otimes \la^{-1} xq_-^{-1}
    = gm\otimes xq_-^{-1}.
\end{equation*}
Writing $x= x_1 + x_2j$, $q_- = q_1+q_2j$, we replace $\HH$,
$\grpSp(1)_-$ with $\CC^2$, $\SU(2)$ respectively as
\begin{equation*}
    x\mapsto\begin{bmatrix} x_1 \\ x_2 \end{bmatrix}, \qquad
    q_-\mapsto \begin{bmatrix}
     \overline{q_1} & \overline{q_2} \\ -q_2 & q_1
\end{bmatrix}.
\end{equation*}
\begin{NB2}
\begin{equation*}
    x q_-^{-1} =  \overline{q_1} x_1 + \overline{q_2} x_2 + (-q_2 x_1 + q_1 x_2)j.
\end{equation*}
Further under the isomorphism
\begin{equation*}
    m\otimes
    \begin{bmatrix}
        x_1 \\ x_2
    \end{bmatrix}
    \longmapsto 
    \begin{bmatrix}
            x_1 m \\ x_2 m
    \end{bmatrix}
\end{equation*}
\begin{equation*}
    gm \otimes x q_-^{-1} = gm \otimes
    \begin{bmatrix}
        \overline{q_1} & \overline{q_2} \\ -q_2 & q_1
    \end{bmatrix}
    \begin{bmatrix}
        x_1 \\ x_2
    \end{bmatrix}
    \longmapsto 
  \begin{bmatrix}
      (\overline{q_1} x_1 + \overline{q_2} x_2) gm \\
      (- q_2 x_1 + q_1 x_2) gm
  \end{bmatrix}
  = \begin{bmatrix}
      \overline{q_1} & \overline{q_2} \\ -q_2 & q_1
  \end{bmatrix}
  \begin{bmatrix}
      x_1 m \\ x_2 m
  \end{bmatrix}.
\end{equation*}
\end{NB2}%
We further use
\begin{equation*}
    \left(\frac1{\sqrt{2}}\begin{bmatrix}
    k & - j \\
    -i & 1
\end{bmatrix}\right)^{-1}
    \begin{bmatrix}
    \overline{q_1} & \overline{q_2} \\ 
    -q_2 & q_1
\end{bmatrix}
\frac1{\sqrt{2}}\begin{bmatrix}
    k & - j \\
    -i & 1
\end{bmatrix}
=
\begin{bmatrix}
    q_1 + q_2j & 0 \\
    0 & q_1 + q_2j
\end{bmatrix}
\end{equation*}
\begin{NB2}
\begin{equation*}
    \left(\frac1{\sqrt{2}}\begin{bmatrix}
    k & - j \\
    -i & 1
\end{bmatrix}\right)^{-1}
= \frac1{\sqrt{2}}
\begin{bmatrix}
    -k & i \\ j & 1
\end{bmatrix}.
\end{equation*}
\end{NB2}%
\begin{NB2}
I have combined
    \begin{equation*}
        \left(\frac1{\sqrt{2}}\begin{bmatrix}
              1 & - j \\
              -j & 1
          \end{bmatrix}\right)^{-1}
    \begin{bmatrix}
    \overline{q_1}& \overline{q_2} \\ 
    -q_2 & q_1 
\end{bmatrix}
\frac1{\sqrt{2}}\begin{bmatrix}
    1 & - j \\
    -j & 1
\end{bmatrix}
=
\begin{bmatrix}
    \overline{q_1} - \overline{q_2}j& 0 \\
    0 & q_1 + q_2 j
\end{bmatrix}
\end{equation*}
with
\(
   -k (\overline{q_1} - \overline{q_2}j)k
   = q_1 + q_2j.
\)
\end{NB2}%
to replace $\bM\otimes_\CC\HH$ by $\bM^-\oplus\bM^-$.
We thus get a bundle isomorphism
$\Lambda^{0,1}\otimes_\CC(P\times_{\tilde G}\bM^+) \cong
P\times_{\tilde G}(\bM^-\oplus\bM^-)$. It is a direct calculation to
check that $D_A^+\oplus D_A^+ i = \DB_A$.
\end{NB}

\begin{NB}
This is a record of an earlier calculation based on
$q_- x = (q_1 + jq_2)(x_1 + jx_2) = q_1 x_1 -\overline{q_2} x_2 + 
j (q_2 x_1 + \overline{q_1} x_1)$.

\begin{equation*}
    \left(\frac1{\sqrt{2}}\begin{bmatrix}
    1 & - i \\
    -j & k
\end{bmatrix}\right)^{-1}
    \begin{bmatrix}
    q_1 & -\overline{q_2} \\ 
    q_2 & \overline{q_1}
\end{bmatrix}
\frac1{\sqrt{2}}\begin{bmatrix}
    1 & - i \\
    -j & k
\end{bmatrix}
=
\begin{bmatrix}
    q_1 + jq_2 & 0 \\
    0 & q_1 + j q_2
\end{bmatrix}
\end{equation*}
\end{NB}

\begin{NB}
    Let us take an orthonormal frame $\{ e_1, Ie_1, e_2, Ie_2\}$.
    Then
\begin{equation*}
            \begin{bmatrix}
            D_A^+ \\ D_A^+ i
        \end{bmatrix}
        = 
        \begin{bmatrix}
            \sum_{\alpha=1}^2 e_\alpha\cdot\nabla_{e_\alpha} 
            + (I e_\alpha)\cdot\nabla_{Ie_\alpha}
\\
            \sum_{\alpha=1}^2 e_\alpha\cdot i \nabla_{e_\alpha} 
            + (I e_\alpha)\cdot i \nabla_{Ie_\alpha}
        \end{bmatrix}.
    \end{equation*}
\begin{NB2}
    It is not true
    $e_\alpha \cdot i = (I e_\alpha)\cdot$, $(Ie_\alpha)\cdot i = -e_\alpha$.
    Note 
    $i 1 i = -1$, $i i i = - i$, $iji = -iij = j$, $iki = -iik = k$.
\end{NB2}

    Let $\left[
    \begin{smallmatrix}
        m_1 \\ m_2
    \end{smallmatrix}\right]\in \bM^-\oplus\bM^-$. Then as
  $\bM\otimes_\CC(\HH=\CC^2)$, it is
  \begin{equation*}
      \begin{bmatrix}
          x_1 \\ x_2
      \end{bmatrix}
      = 
      \frac1{\sqrt{2}}
      \begin{bmatrix}
          k & -j \\ -i & 1
      \end{bmatrix}
      \begin{bmatrix}
          m_1 \\ m_2
      \end{bmatrix}
      = \frac1{\sqrt{2}}
      \begin{bmatrix}
          km_1 - jm_2 \\ -i m_1 + m_2
      \end{bmatrix}.
  \end{equation*}
In fact,
\begin{equation*}
    \begin{bmatrix}
    \overline{q_1}& \overline{q_2} \\ 
    - q_2 & q_1 
\end{bmatrix}
    \begin{bmatrix}
        km_1 - jm_2 \\ -i m_1 + m_2
    \end{bmatrix}
    = 
    \begin{bmatrix}
        k (q_1+q_2j) m_1 - j (q_1+q_2j)m_2 \\
        -i(q_1+q_2j) m_1 + (q_1+q_2j) m_2
    \end{bmatrix}.
\end{equation*}
Therefore
    \begin{equation*}
        \begin{split}
            & \frac1{\sqrt{2}}
      \begin{bmatrix}
          k & -j \\ -i & 1
      \end{bmatrix}
        \begin{bmatrix}
            D_A^+ \\ D_A^+ i
        \end{bmatrix}
        = 
        \frac1{\sqrt{2}}
      \begin{bmatrix}
          k & -j \\ -i & 1
      \end{bmatrix}
        \begin{bmatrix}
            \sum_{\alpha=1}^2 e_\alpha \cdot\nabla_{e_\alpha} 
            + I e_\alpha \cdot \nabla_{Ie_\alpha}
\\
            \sum_{\alpha=1}^2 e_\alpha \cdot i \nabla_{e_\alpha} 
            + I e_\alpha\cdot i\nabla_{Ie_\alpha}
        \end{bmatrix}
\\
=\; &
        \frac1{\sqrt{2}}
        \begin{bmatrix}
            \sum_{\alpha=1}^2 -j(ie_\alpha + e_\alpha i)\cdot
            (\nabla_{e_\alpha} + i\nabla_{I e_\alpha}) \\
            - \sum_{\alpha=1}^2 e_\alpha i - (I e_\alpha) \cdot 
            (\nabla_{e_\alpha} - i\nabla_{I e_\alpha})
        \end{bmatrix}
        \end{split}
    \end{equation*}

    $\nabla^{0,1}_{e_a + i I e_a} = \nabla_{e_a} + i\nabla_{Ie_a}$.
\end{NB}

\begin{Theorem}
    A solution of the generalized Seiberg-Witten equation
    \eqref{eq:24} on a compact K\"ahler surface $X$ consists of a
    holomorphic $G_\CC$-bundle $\mathscr P$, a holomorphic section
    $\overline{\Phi}$ of $K_X^{1/2}\otimes (\mathscr
    P\times_{G_\CC}\bM)$ satisfying
    \begin{equation*}
        \begin{split}
        & \bmu_\CC(\Phi) = 0,
        \\
        & (F_A^+)^{1,1} = \bmu_\RR(\Phi)\omega.
        \end{split}
    \end{equation*}
\end{Theorem}
The equation $(F_A^+)^{1,1} = \bmu_\RR(\Phi)\omega$ corresponds to the
stability condition by the Hitchin-Kobayashi correspondence. Therefore
the moduli space of solutions has an algebro-geometric description.

\begin{Remark}
    For the original Seiberg-Witten equation for $(G,\bM) =
    (\U(1),\HH)$, $\bmu_\CC(\Phi) = 0$ can be written as
    $\alpha\overline{\beta} = 0$ where $\Phi =
    \alpha+j\beta$. Therefore we have either $\alpha=0$ or
    $\beta=0$. We do not have such a further reduction in general.
\end{Remark}

\begin{NB}
    The following is an earlier attempt. Since it is not true that
    $P\times_{\tilde G} \tilde G_\CC$ is a holomorphic bundle, the
    argument is not correct.

We further assume $\bM$ is a cotangent type, i.e., $\bM =
\bN\oplus\bN^*$.
Since the decomposition $\bM=\bN\oplus\bN^*$ is preserved by $\U(1)$,
the associated bundle $P\times_{\tilde G}\bM^+$ decomposes as
$(P\times_{\tilde G}\bN)\oplus(P\times_{\tilde G}\bN^*)$. We write
$\Phi = \alpha\oplus\beta$. We also decompose 
$D_A^- = \DB_A^*\oplus \DB_A\colon \Gamma(P\times_{\tilde G}\bM^-)
\to \Gamma(P\times_{\tilde G}\bN)\oplus\Gamma(P\times_{\tilde G}\bN^*)$.
Then we have $D_A^+ = (\DB_A, \DB_A^*)\colon
\Gamma(P\times_{\tilde G}\bN)\oplus\Gamma(P\times_{\tilde G}\bN^*)
\to\Gamma(P\times_{\tilde G}\bM^-)$.

We take the $\Gamma(P\times_{\tilde G}\bN^*)$-part of $0 = D_A^-D_A^+\Phi
= D_A^-D_A^+(\alpha\oplus\beta)$:
\begin{equation}\label{eq:27}
    0 = \DB_A\DB_A \alpha + \DB_A\DB_A^*\beta.
\end{equation}
Since $\nabla_A^*\nabla_A$ preserves the decomposition
$\Gamma(P\times_{\tilde G}\bN)\oplus\Gamma(P\times_{\tilde G}\bN^*)$,
the Weitzenb\"ock formula implies that $\DB_A\DB_A\alpha =
F_A^{0,2}\alpha$. From the equation $F_A^+ = \bmu(\Phi)$, this is equal
to $\bmu_\CC(\Phi)\alpha$. Plugging this into \eqref{eq:27}, we take
the $L^2$-inner product with $\beta$ to get
\begin{equation*}
    0 = \int_X (\bmu_\CC(\Phi)\alpha,\beta) + \| \DB_A^*\beta\|_{L^2}^2.
\end{equation*}
The first term is $\frac12 \|\bmu_\CC(\Phi)\|_{L^2}^2$, in particular,
nonnegative. Therefore we get $\bmu_\CC(\Phi) = 0$ and $\DB_A^*\beta =
0$. The first equation implies $F_A^{0,2} = 0$. We similarly get
$\DB_A\alpha = 0$.

Thus $P\times_{\tilde G}\tilde G_\CC$ has a structure of a holomorphic
principal bundle, hence $P\times_{\tilde G}\bN$, $P\times_{\tilde
  G}\bN^*$ are holomorphic vector bundles. And $\alpha$ is a
holomorphic section of the former.

The equation $\DB^*_A\beta = 0$ means $\beta$ is an anti-holomorphic
section of $P\times_{\tilde G}\bN^*$, or equivalently $\overline{\beta}$ is a holomorphic section of $K_X\otimes(P\times_{\tilde G}\bN^*)$. Thus
\begin{Theorem}
    A solution of the generalized Seiberg-Witten equation
    \eqref{eq:24} on a compact K\"ahler surface $X$ consists of a
    holomorphic $\tilde G_\CC$-bundle $\mathscr P$, a holomorphic
    section $\alpha$ of $\mathscr P\otimes_{\tilde G_\CC}\bN$ and a
    holomorphic section $\overline\beta$ of $K_X\otimes(\mathscr
    P\otimes_{\tilde G_\CC}\bN^*)$ satisfying
    \begin{equation*}
        \begin{split}
        & (F_A^+)^{1,1} = \bmu_\RR(\alpha\oplus\beta)\omega,
        \\
        & \bmu_\CC(\alpha\oplus\beta) = 0.
        \end{split}
    \end{equation*}
\end{Theorem}

The equation $(F_A^+)^{1,1} = \bmu_\RR(\alpha\oplus\beta)\omega$
corresponds to the stability condition by the Hitchin-Kobayashi
correspondence. Therefore the moduli space of solutions has an
algebro-geometric description.

For the ordinary Seiberg-Witten equation, the equation
$\bmu_\CC(\alpha\oplus\beta) = 0$ implies either $\alpha$ or $\beta$ is
$0$, as $\bmu_\CC(\alpha\oplus\beta) = \alpha\overline{\beta}$. This is
not true in general.
\end{NB}

\subsection{Dimension reduction}

Let us suppose the spin $4$-manifold $X$ is $\RR\times Y$ for a spin
$3$-manifold $Y$. The $3$-dimensional generalized Seiberg-Witten
equation is an $\RR$-invariant solution of \eqref{eq:24}.
\begin{NB}
    We take $\bar G = \operatorname{Spin}(3)\times G/\!\pm\!(1,1)$
    with an exact sequence $1\to G\to\bar G\to \SO(3)\to 1$. Choose a
    $\bar G$-bundle $P$ with an isomorphism $P/G\cong P_{\SO(3)}$,
    where $P_{\SO(3)}$ is the frame bundle of $Y$. We consider a
    connection $A$ compatible with the Levi-Civita connection, and a
    section $\Phi$ of $P\times_{\bar G}\bM$.
\end{NB}%
We regard $\mathfrak{so}(3)=\algsp(1)$ as the space of imaginary
quaternions, and we have the Clifford multiplication
$\mathfrak{so}(3)\otimes\bM\to\bM$.
\begin{NB}
 This is a $\bar G$-homomorphism.
\end{NB}%
We take a principal $G$-bundle $P\to Y$ and consider the associated
$\bar G$-bundle $\bar P$ with $\bar G \defeq
\operatorname{Spin}(3)\times G$, as in the $4$-dimensional case.
The $3$-dimensional generalized Seiberg-Witten equation is a system of
partial differential equations for a connection $A$ on $P$ and a
section $\Phi$ of $\bar P\times_{\bar G}\bM$:
\begin{equation}\label{eq:3dSW}
    \begin{split}
        & D_{\!A}\Phi = 0,
        \\
        & \ast F_A = - \bmu(\Phi).
    \end{split}
\end{equation}
Here $\bmu(\Phi)$ is a section of $\bar P\times_{\bar
  G}(\g\otimes\mathfrak{so}(3))$, and considered as a $P\times_G\g$.
\begin{NB}
$P\times_{\bar G}\g$    
\end{NB}%
valued $1$-form. The Hodge star $\ast$ send $2$-forms to $1$-forms in
$3$-dimension, the lower equation makes sense.
\begin{NB}
    The sign comes from $\rho(\ast\alpha) = -\rho(\alpha)$ where
    $\rho$ is the Clifford multiplication.
\end{NB}%

The same remark on the bound on $\Phi$ in \ref{sec:roles-higgs}
applies also in the $3$-dimensional case: We need to understand the
behavior of solutions $(A_i,\Phi_i)$ with $|\Phi_i|\to \infty$ in
order to define invariants rigorously.

For the Floer homology group, we consider the equation
\begin{equation*}
    \begin{split}
        & \frac{d}{dt}\Phi = - D_{\!A}\Phi,
        \\
        & \frac{d}{dt} A = - \ast F_A - \bmu(\Phi)
    \end{split}
\end{equation*}
on $\RR\times Y$. This is the gradient flow equation for a generalized
Chern-Simons functional given by
\begin{equation*}
    \mathcal E(A,\Phi) =
    \int_Y \operatorname{CS}(A) + \frac12(D_A\Phi,\Phi)\, d\operatorname{vol}.
\end{equation*}
Here $\operatorname{CS}$ is the (normalized) Chern-Simons functional
satisfying
\begin{equation*}
    \frac{d}{dt} \int_Y \operatorname{CS}(A)
    = \int_Y (\frac{d A_t}{dt} \wedge F_{A_t})
    = \int_Y (\frac{d A_t}{dt}, \ast F_{A_t})\, d\operatorname{vol}.
\end{equation*}
\begin{NB}
    \begin{equation*}
        \frac{d}{dt} \frac12 \int_Y (D_{A_t}\Phi_t,\Phi_t)\, d\operatorname{vol}
        = \int_Y (D_{A_t}\Phi_t,\frac{d\Phi_t}{dt})
        + \frac12 (\frac{d A_t}{dt}\cdot \Phi,\Phi)\, d\operatorname{vol}
        = \int_Y (D_{A_t}\Phi_t,\frac{d\Phi_t}{dt})
        + \langle \frac{d A_t}{dt}, \bmu(\Phi)\rangle\, d\operatorname{vol}
    \end{equation*}
\end{NB}

In the framework of a topological quantum field theory in $3d$, it is
natural to expect that one should consider the generalized
Seiberg-Witten equation \eqref{eq:3dSW} for a $3$-manifold $Y$ with
cylindrical ends. As `boundary conditions', one should look at the
translation invariant solutions of \eqref{eq:3dSW} over $Y=\RR\times C$
for a $2$-dimensional Riemann manifold $C$
\cite[\S2]{MR1734402}. Since $C$ is a K\"ahler manifold, we can apply
the argument in \subsecref{sec:kahler-surfaces} to reduce the equation
to
\begin{equation}
    \label{eq:30}
    \begin{split}
        & \text{$A$ defines a holomorphic $G_\CC$-bundle $\mathscr P$ on $C$},
        \\
        & \text{$\overline{\Phi}$ is a holomorphic section of $K_C^{1/2}\otimes
          (\mathscr P\times_{G_\CC}\bM)$},
        \\
        & \bmu_\CC(\Phi) = 0,
        \\
        & \ast F_A = \bmu_\RR(\Phi),
    \end{split}
\end{equation}
where $\ast$ is the Hodge star operator in $2$-dimension. The first
condition is automatic as $C$ is complex dimension $1$, and hence
$F^{0,2}_A$ is automatically $0$.

This is the equation for the gauge nonlinear $\sigma$-model whose
target is the Higgs branch $\bM\tslash G = \bmu^{-1}(0)/G$.

\section{Motivic Donaldson-Thomas type invariants}\label{sec:motiv-DT-invariants}

From the $(2+1)$-dimensional TQFT framework explained in
\subsecref{sec:2+1-tqft}, the Coulomb branch $\mathcal M_C$ of
$\Hyp(\bM\tslabar G)$ is expected to be related to the cohomology of
moduli spaces of solutions of \eqref{eq:30}.
This is a starting point, and must be modified as we have explained in
Introduction. We should get a reasonable answer for $(G,\bM) =
(\SU(2),0)$ and we should recover the monopole formula \eqref{eq:11}
for good or ugly theories.

Our proposal is as in \subsecref{sec:prop-defin-texorpdfs}.
Besides reproducing \eqref{eq:11}, these modifications are natural in
view of recent study of Donaldson-Thomas (DT) invariants, as we will
explain below. Recall that DT invariants were introduced as complex
analog of Casson invariants \cite{MR1634503}. As we can use
algebro-geometric techniques to handle singularities in moduli spaces,
DT invariants are easier to handle than the original Casson invariant
in a sense. Therefore it should be reasonable for us to model the
theory of DT invariants.

\begin{NB}
The following is not correct, as we take sum over bundles.

We assume $G$ is connected, as our formulation uses (rational)
equivariant cohomology groups, and hence information depending on
component groups will be lost.
\end{NB}

\subsection{Holomorphic Chern-Simons functional}

Let $C$ be a compact Riemann surface.
\begin{NB}
We consider a gauge theory
associated with a quaternionic representation $\bM$ of $G$.    
\end{NB}%
We choose and fix a spin structure, i.e., the square root $K_C^{1/2}$ of
the canonical bundle $K_C$. We also fix a ($C^\infty$) principal
$G_\CC$-bundle $P$ with a fixed reference partial connection
$\DB$. (See Remark~\ref{rem:flavor_and_degree}(b) below, for a
correct formulation.)
A {\it field\/} consists of a pair
\begin{itemize}
      \item[] $\DB + A$ : a partial connection on $P$. So $A$ is a
    $C^\infty$-section of $\Lambda^{0,1}\otimes
    (P\times_{G_\CC}\mathfrak g_\CC)$.
      \item[] $\Phi$ : a $C^\infty$-section of $K_C^{1/2}\otimes
    (P\times_{G_\CC}\bM)$.
\end{itemize}
Let $\mathcal F$ be the space of all fields.
There is a gauge symmetry, i.e., the group $\mathcal G_\CC(P)$ of all
(complex) gauge transformations of $P$ natural acts on the space
$\mathcal F$.

Our notation is slightly different from one in \eqref{eq:30}. We
replace $\overline{\Phi}$ by $\Phi$ for brevity, and $A$ is a partial connection instead of a connection.

\begin{Remarks}\label{rem:flavor_and_degree}
    (a) When we have a flavor symmetry $G_F$, we slightly change the
    setting as follows. Recall we have an exact sequence $1\to G\to
    \tilde G\to G_F\to 1$ of groups. Choose a principal
    $(G_F)_\CC$-bundle $P_F$ with a partial connection $\DB_F$, and
    also a $\tilde G$-bundle $\tilde P$ with a partial connection
    $\DB^\sim$. Then $\mathcal F$ consists of
\begin{itemize}
      \item[] $\DB^\sim + \tilde A$ : a partial connection on $\tilde
    P$. Moreover, we assume that the induced connection on $P_F$ is
    equal to $\DB_F$.
      \item[] $\Phi$ : a $C^\infty$-section of $K_C^{1/2}\otimes
    (\tilde P\times_{\tilde{G}_\CC}\bM)$.
\end{itemize}
Since the notation is cumbersome, we will restrict ourselves to the
case without flavor symmetry except we occasionally point out when a
crucial difference arises.

    (b) A $C^\infty$ principal $G_\CC$-bundle is classified by its
    `degree', an element in $\pi_1(G)$. This is compatible with what
    is explained in
    \subsecref{sec:expect-properties}(\ref{item:group}). In
    particular, we need to take sum over all degrees to define the
    Coulomb branch. Since the sum occurs at the final stage, and is
    not relevant to most of calculations, we will not mention this
    point from now.
\end{Remarks}

We define an analog of the holomorphic Chern-Simons functional by
\begin{equation}\label{eq:1}
    \CS(A,\Phi)
    = \frac12 \int_C \omega_\CC((\DB + A)\Phi\wedge\Phi),
\end{equation}
where $\omega_\CC(\ \wedge\ )$ is the tensor product of the exterior
product and the complex symplectic form $\omega_\CC$ on $\bM$. Since
$(\DB + A)\Phi$ is a $C^\infty$-section of $\Wedge^{0,1}\otimes
K_C^{1/2}\otimes(P\times_{G_\CC}\bM)$, $\omega_\CC((\DB +
A)\Phi\wedge\Phi)$ is a $C^\infty$-section of $\Wedge^{0,1}\otimes K_C
= \Wedge^{1,1}$. Its integral is well-defined.
This is invariant under the gauge symmetry $\mathcal G_\CC(P)$.

\begin{NB}
    Here is the original version:

We apply the complex moment map $\bmu_\CC$ fiberwise to $\Phi$. Then
$\bmu_\CC(\Phi)$ is a section of $K_C\otimes (P\times_{G_\CC}\mathfrak
g_\CC^*) = \Lambda^{1,0}\otimes (P\times_{G_\CC}\mathfrak
g_\CC^*)$. Now we define an analog of holomorphic Chern-Simons
functional by
\begin{equation*}
    \CS(A,\Phi)
    = \int_C \frac12 \omega_\CC(\DB \Phi\wedge\Phi) +
    \langle A\wedge\bmu_\CC(\Phi)\rangle,
\end{equation*}
where $\langle\ \wedge\ \rangle$ is the tensor product of the exterior
product and the pairing between $\mathfrak g_\CC$ and $\mathfrak
g_\CC^*$. The first term $\omega_\CC(\DB \Phi\wedge\Phi)$ is similar.
This is invariant under the gauge symmetry $\mathcal G_\CC(P)$.
\end{NB}

When $\bM$ is a cotangent type, we can slightly generalize the
construction. Let us choose $M_1$, $M_2$ be two line bundles over $C$
such that $M_1\otimes M_2 = K_C$. We modify $\Phi$ as
\begin{itemize}
      \item[] $\Phi_1$, $\Phi_2$ : $C^\infty$-sections of $M_1\otimes
    (P\times_{G_\CC} \bN)$ and $M_2\otimes (P\times_{G_\CC}\bN^*)$
    respectively.
\end{itemize}
\begin{NB}
Then $\bmu_\CC(\Phi_1,\Phi_2)$ is a section of $M_1\otimes M_2\otimes (P\times_{G_\CC}\mathfrak g_\CC^*) = \Lambda^{1,0}\otimes
(P\times_{G_\CC}\mathfrak g_\CC^*)$.
\end{NB}%
Then
\begin{equation}\label{eq:2}
    \CS(A,\Phi_1,\Phi_2) 
   = \int_C \langle (\DB+A)\Phi_1,\Phi_2\rangle.
\end{equation}
\begin{NB}
    This is the original version:
\begin{equation*}
    \begin{split}
    & \CS(A,\Phi_1,\Phi_2) 
    = \int_C \frac12 \left(\langle \DB\Phi_1,\Phi_2\rangle 
    - \langle \DB\Phi_2,\Phi_1\rangle\right)
    + \langle A\wedge\bmu_\CC(\Phi_1,\Phi_2)\rangle
\\
   =\; & \int_C \langle (\DB+A)\Phi_1,\Phi_2\rangle.
    \end{split}
\end{equation*}
\end{NB}%
It is a complex valued function on $\mathcal F$.

\begin{Remark}\label{rem:reduction}
In this expression, analogy with the holomorphic Chern-Simons
functional over a Calabi-Yau 3-fold $X$ is clear. We consider a kind
of dimension reduction from $X$ to $C$, and connection forms in the
reduced direction are changed to sections $\Phi_1$, $\Phi_2$.
\end{Remark}

Let us consider critical points of $\CS$. We take a variation in the
direction $(\dot A,\dot\Phi)$:
\begin{NB}
\begin{equation*}
    \begin{split}
    & d\CS_{(A,\Phi)}(\dot A,\dot\Phi) 
    = \int_C \omega_\CC(\DB\Phi\wedge\dot\Phi)
    + \langle \dot A\wedge\bmu_\CC(\Phi)\rangle
    + \langle A\wedge d(\bmu_\CC)_\Phi(\dot\Phi)\rangle
\\
    = \; & \int_C 
    \omega_\CC((\DB + A)\Phi\wedge \dot\Phi)
    + \langle \dot A\wedge\bmu_\CC(\Phi)\rangle.
    \end{split}
\end{equation*}
\end{NB}%
\begin{equation*}
    \begin{split}
    & d\CS_{(A,\Phi)}(\dot A,\dot\Phi) 
\\
    = \; & \frac12 \int_C \omega_\CC((\DB+A)\Phi\wedge\dot\Phi)
    + \omega_\CC((\DB+A)\dot\Phi\wedge\Phi)
    + \omega_\CC(\dot A\Phi\wedge\Phi)
\\
    = \; & \int_C 
    \omega_\CC((\DB + A)\Phi\wedge \dot\Phi)
    + \langle \dot A\wedge\bmu_\CC(\Phi)\rangle.
    \end{split}
\end{equation*}
Therefore $(A,\Phi)$ is a critical point of $\CS$ if and only if the
following two equations are satisfied:
\begin{equation}\label{eq:32}
    \begin{split}
        & (\DB+A)\Phi = 0,\\
        & \bmu_\CC(\Phi) = 0.
    \end{split}
\end{equation}
The first equation means that $\Phi$ is a holomorphic section of
$K_C^{1/2}\otimes (P\times_{G_\CC}\bM)$ when we regard $P$ as a
holomorphic principal bundle by $\DB+A$. Thus we have recovered the
second and third equations in \eqref{eq:30}.

When we have a flavor symmetry $G_F$, the partial connection $\DB_F$
on $P_F$ is fixed, and we only arrow $\tilde A$ on $\tilde P$ to vary
in the `$G$-direction'. The Euler-Lagrange equation remains the same,
if we understand $\DB+A$ involves also $\DB_F$.

\begin{Remark}\label{rem:Diaconescu}
    When $\bM$ is of an affine quiver type in
    \subsecref{sec:quiver-type}, the solution of~\eqref{eq:32} is
    analogous to quiver sheaves considered by Szendr\"oi~\cite{Sz}.
    When we further assume $\bM$ is of Jordan quiver type, it is also
    analogous to ADHM sheaves considered by Diaconescu \cite{Di}.
    In these cases, hyper-K\"ahler quotients parametrize instantons on
    $\CC^2/\Gamma$ (and its resolution and deformation) via the ADHM
    description \cite{KN}, where $\Gamma$ is a finite subgroup of
    $\SU(2)$ corresponding to the quiver. We apply the ADHM
    description fiberwise, critical points give framed sheaves on a
    compactified $\CC^2/\Gamma$-bundle over $C$.
    We do not have the corresponding $3$-fold for general $(G,\bM)$, but 
    it is philosophically helpful to keep the $3$-fold picture in mind.

    Moreover, one could probably consider analog of Gromov-Witten
    invariants, namely invariants of quasimaps by Ciocan-Fontanine,
    Kim, and Maulik, again imposing a stability condition
    \cite{CFKM}. The twist by $K_C^{1/2}$ is not included, and they
    are analog of usual Gromov-Witten invariants for stable maps from
    $\proj^1$ to $\bmu_\CC^{-1}(0)\dslash G_\CC$.
    \begin{NB}
        Suppose $\bmu_\CC^{-1}(0)\dslash G_\CC$ is a good
        quotient. Then $\bmu_\CC^{-1}(0)$ is a $G_\CC$-bundle over
        $\bmu_\CC^{-1}(0)\dslash G_\CC$. Therefore a map $\phi\colon
        \proj^1\to \bmu_\CC^{-1}(0)\dslash G_\CC$ gives a
        $G_\CC$-bundle over $\proj^1$ by pull-back. We also have a
        pull-back of the tautological homomorphism.
    \end{NB}%

    Note that there are two bigger differences between our moduli and
    one in \cite{Sz,Di,CFKM}: we assume $P$ to be a genuine $G$-bundle
    (or a vector bundle for $G = \GL(N)$), while it is just a coherent
    sheaf in \cite{Sz,Di} and the base curve $C$ is not fixed in
    \cite{CFKM}. And a stability condition in \cite{Sz,Di,CFKM} is not
    imposed here.

    As is clear from Diaconescu and his collaborators \cite{Di1,Di2},
    a change of stability conditions is worth studying, but we cannot
    impose the stability condition in order to reproduce the monopole
    formula, as is clear from the special case $(G,\bM) = (\SU(2),0)$.
\end{Remark}

If the space $Z(\mathfrak g_\CC^*) = \{ \zeta_\CC\in\mathfrak g_\CC^* \mid
\text{$\operatorname{Ad}^*_g(\zeta_\CC) = \zeta_\CC$ for any $g\in G$}\}$ is
non\-trivial, we could perturb $\CS$ by choosing a section $\tilde\zeta_\CC$ of
$K_C\otimes Z(\mathfrak g_\CC^*)$:
\begin{equation*}
    \CS_{\tilde c}(A,\Phi) = 
    \int_C \frac12 \omega_\CC((\DB +A)\Phi\wedge\Phi) -
    \langle A\wedge \tilde\zeta_\CC\rangle.
\end{equation*}
\begin{NB}
\begin{equation*}
    \CS_{\tilde c}(A,\Phi) = 
    \int_C \frac12 \omega_\CC(\DB \Phi\wedge\Phi) +
    \langle A\wedge(\bmu_\CC(\Phi)-\tilde\zeta_\CC)\rangle.
\end{equation*}
\end{NB}%
The Euler-Lagrange equation is
\begin{equation*}
    \begin{split}
        & (\DB+A)\Phi = 0,\\
        & \bmu_\CC(\Phi) = \tilde\zeta_\CC.
    \end{split}
\end{equation*}
It is not clear whether this generalization is useful or not, but we
will study the toy model case $C=\CC$ in \subsecref{sec:toy}.

\subsection{The main proposal}

Our main proposal is that the coordinate ring of the Coulomb branch
$\mathcal M_C$ of $\Hyp(M)\tslabar G$ is the equivariant cohomology
(with compact support) of the vanishing cycle associated with $\CS$:
\begin{equation}\label{eq:18}
    \CC[\mathcal M_C] \overset{\operatorname{\scriptstyle ?}}{=}
    H^{*+\dim\mathcal F-\dim\mathcal G_\CC(P)}_{c,\mathcal G_\CC(P)}
    (\text{space of solutions of \eqref{eq:32}}, 
    \varphi_{\CS}(\CC_{\mathcal F}))^*,
\end{equation}
up to certain degree shift explained later.
Since $\mathcal F$ is infinite-dimensional, it is not clear whether
the conventional definition of the vanishing cycle functor $\varphi_f$
can be applied to our situation. We expect that it is possible to
justify the definition, as the theory for usual complex Chern-Simons
functional for connections on a compact Calabi-Yau $3$-fold has been
developed Joyce and his collaborators (see
\cite{2012arXiv1211.3259B,2013arXiv1312.0090B} and the references
therein).

Since both $\dim \mathcal F$, $\dim \mathcal G_\CC(P)$ are infinite
(and its difference also as we will see below), we need to justify the
meaning of the cohomological degree in \eqref{eq:18}. This will be
done below by applying results for $\varphi_f$ proved in a finite
dimensional setting formally to our situation.
Then cohomological degree in~\eqref{eq:18} will match with the grading
on the left hand side coming from the $\U(1)$-action, appeared in the
monopole formula.
\begin{NB}
    The hidden $\algsl(2)$-symmetry comes from Lefschetz ?
\end{NB}%
We hope that this argument can be rigorously justified when the
definition of the vanishing cycle functor is settled.

However, most importantly, it is not clear at this moment how to
define multiplication in \eqref{eq:18}. In the subsequent paper
\cite{BFN}, we replace $\proj^1$ by the formal punctured disc
$D^\times = \operatorname{Spec}\CC((z))$, mimicking the affine
Grassmannian. We also replace the cohomology group by the homology
group. Then we can define multiplication by the convolution.
If $\bM = 0$, there is a natural pairing between the cohomology of the
space for $\proj^1$ (thick affine Grassmannian) and homology of the
affine Grassmannian (cf.\ \cite[Ch.6]{1995alg.geom.11007G}). 
It is not clear whether this remains true in general at this moment.
The reason which we take the dual of the cohomology group in
\eqref{eq:18} comes only from the comparison with the $\bM = 0$ case,
and hence it is not strong in our current understanding.

Next we should have a Poisson structure in \eqref{eq:18} from the
complex symplectic form of $\mathcal M_C$. In \cite{BFN}, we consider
$\CC^*$-equivariant cohomology where $\CC^*$ acts on $D^\times$ by
rotation. It will give us a noncommutative deformation of the
multiplication, and hence a Poisson structure in the non-equivariant
limit. We call it the {\it quantized Coulomb branch}.

The equivariant cohomology group \eqref{eq:18} is a module over
$H^*_{\mathcal G_\CC(P)}(\mathrm{pt})$. Therefore we should have a
morphism from $\mathcal M_C$ to $\operatorname{Spec}(H^*_{\mathcal
  G_\CC(P)}(\mathrm{pt}))$. It should factor through
$\operatorname{Spec}(H^*_G(\mathrm{pt}))$ where $\mathcal G_\CC(P)\to
G_\CC$ is given by an evaluation at a point in $C$, as $\Ker(\mathcal
G_\CC(P)\to G_\CC)$ acts freely on $\mathcal F$.
Note that $H^*_G(\mathrm{pt}) = \CC[\g_\CC]^G = \CC[\mathfrak
h_\CC]^W$ is a polynomial ring, where $\mathfrak h_\CC$ is a Cartan
subalgebra of $\g_\CC$. Since we have $\dim\mathfrak h_\CC =
\dim\mathcal M_C/2$, it is natural to expect that $\mathcal M_C$ is an
integrable system.

\subsection{Motivic universal DT invariants}

Since \eqref{eq:18} is too much ambitious to study at this moment, we
consider less ambitious one, which is the motivic universal
Donaldson-Thomas invariant defined by
\begin{equation}\label{eq:33}
    \frac{[\operatorname{crit}(\CS)]_{\mathrm{vir}}}
    {[\mathcal G_\CC(P)]_{\mathrm{vir}}}.
\end{equation}
This is expected to be obtained by taking weight polynomials with
respect to the mixed Hodge structure given by the Hodge module version
of \eqref{eq:18}. The definition of motivic DT invariants have been
worked out for degree zero by \cite{BBS}, and also by
\cite{2012arXiv1211.3259B,2013arXiv1312.0090B} and the references
therein for general cases.

Since we do not study the multiplication on \eqref{eq:18} in this
paper, our primary interest will be the Poincar\'e polynomial of
\eqref{eq:18} and its comparison with the monopole formula. For
$C=\proj^1$, we introduce a stratification on the space ($\mathcal R$
introduced below) in question such that each stratum has vanishing odd
degree cohomology groups. Therefore the Poincar\'e polynomial is equal
to the weight polynomial, hence \eqref{eq:33} looses nothing for our
purpose.

Since both $\mathcal F$, $\mathcal G_\CC(P)$ are infinite dimensional,
one still needs to justify the above definition, defined again via the
vanishing cycle functor.
Also note that $[\mathcal G_\CC(P)]_{\mathrm{vir}}$ is the virtual
motive of $\mathcal G_\CC(P)$, i.e.,
\begin{equation}\label{eq:4}
    (-\LL^{\frac12})^{-\dim\mathcal G_\CC(P)} [\mathcal G_\CC(P)],
\end{equation}
which must be interpreted appropriately.

Maybe it is worthwhile to mention one more thing need to be justified
at this stage. The moduli stack of holomorphic $G_\CC$-bundles over
$C$ is of infinite type. For example, over $\proj^1$, it can be
reduced to $T_\CC$-bundles, in other words, sums of line bundles, but
we have no control for the degrees of those line bundles.
Therefore we need to stratify $\operatorname{crit}(\CS)$ by the
Harder-Narashimhan filtration, and consider motivic invariants for
strata, and then take their sum. The last sum is infinite, so we need
to justify it. In practice, at least for $\proj^1$, the `good' or
`ugly' assumption means that we only get positive powers of $\LL$ for
each stratum, and we consider it as a formal power series in $\LL$.

\begin{Remark}
    We call the above the {\it universal\/} DT invariant, as no
    stability condition is imposed following \cite{MR2927365}. It is a
    hope that usual DT invariants and wall-crossing formula could be
    understood from the universal one as in \cite{MR2927365}. In
    \cite{MR2927365}, the authors considered representations of a
    quiver, hence arbitrary objects in an abelian category. However,
    we do not include coherent sheaves (for $G=\GL$), it is not clear
    whether this is a reasonable hope or not.
\end{Remark}

\subsection{Cutting}\label{sec:cutting}

We suppose that $\bM$ is of cotangent type, so $\bM = \bN\oplus\bN^*$.
We consider the scalar multiplication on the factor $\bN^*$. Then we have
$(t\cdot v, t\cdot w) = t(v,w)$ for $t\in\U(1)$.
\begin{NB}
If we change the condition to $(t\cdot v,t\cdot w) = t^2(v,w)$, we do
not need to assume $\bM$ is of cotangent type, as the scalar
multiplication on the total $\bM$ will work. I need to check that this
modification works or not.
\end{NB}%

We have the induced $\CC^*$-action on $\mathcal F$ given by
\begin{equation*}
    (A,\Phi) \mapsto (A, t\cdot\Phi).
\end{equation*}
Then $\CS$ is of weight $1$:
\begin{equation*}
    \CS(A,t\cdot\Phi) = t\CS(A,\Phi).
\end{equation*}
Thus the result of Behrend-Bryan-Szendr\"oi \cite{BBS} can be applied
(at least formally) to get
\begin{equation*}
    [\operatorname{crit}(\CS)]_{\mathrm{vir}}
    = - (-\LL^{\frac12})^{-\dim\mathcal F}\left(
      [\CS^{-1}(1)] - [\CS^{-1}(0)]
      \right).
\end{equation*}

We introduce the reduced space of fields by
\begin{equation*}
    \mathcal F_{\mathrm{red}} \defeq
    \{ (\DB+A, \Phi_1) \} = \mathcal F^{\CC^*}.
\end{equation*}
We have a projection $\mathcal F\to\mathcal F_{\mathrm{red}}$ induced
from $\bM\to\bN$. Note that $\CS$ is linear in $\Phi_2$ once
$(\DB+A,\Phi_1)$ is fixed. It is zero if $(\DB+A)\Phi_1 = 0$, and
nonzero otherwise. (See \eqref{eq:2}.)
Therefore the argument by Morrison
\cite{Mor} and Nagao \cite{Nagao2011} can be (at least formally) applied
to deduce
\begin{equation*}
    [\CS^{-1}(1)] - [\CS^{-1}(0)] 
    = -(-\LL^{\frac12})^{2\dim(\text{fiber})}[\mathcal R],
\end{equation*}
where $\mathcal R$ is the locus $\{ (\DB+A)\Phi_1 = 0 \}$ in $\mathcal
F_{\mathrm{red}}$. Hence
\begin{equation}
    \label{eq:3}
    [\operatorname{crit}(\CS)]_{\mathrm{vir}}
    = (-\LL^{\frac12})^{2\dim(\text{fiber}) - \dim\mathcal F}[\mathcal R].
\end{equation}

Note that $\mathcal R$ is the space of pairs of principal holomorphic
$G_\CC$-bundles $(P,\DB+A)$ together with holomorphic sections
$\Phi_1$ of the associated bundle $M_1\otimes (P\times_{G_\CC}\bN)$.

If we take $M_1 = K_C$ (and hence $M_2 = \shfO_C$) and $\bN =
\bN^* = \mathfrak g_\CC$, $\mathcal R$ is the space of Higgs bundles,
introduced first by Hitchin \cite{Hitchin}.

The fiber in \eqref{eq:3} is the space of $\Phi_2$, i.e., all
$C^\infty$ sections of $M_2\otimes (P\times_{G_\CC}\bN^*)$. This is
obviously infinite dimensional, and hence $\dim(\text{fiber})$ must be
interpreted appropriately. Observe that $\dim\mathcal F$ is also
infinite dimensional, however. Note that we also have $\dim\mathcal
G(P)$ in \eqref{eq:4}. We combine them `formally' to get
\begin{equation}\label{eq:7}
    \begin{split}
    & 2\dim(\text{fiber}) - \dim\mathcal F + \dim \mathcal G_\CC(P)\\
    = \; & \dim \{ \Phi_2 \} - \dim \{ \Phi_1 \} 
    + \dim\mathcal G_\CC(P) - \dim \{\DB+A\}.
    \end{split}
\end{equation}
Next note that
\begin{equation*}
    \begin{split}
    & \dim \{ \Phi_2 \} - \dim \{ \Phi_1 \} 
\\
    = \; &\dim C^\infty(K_C \otimes M_1^*\otimes (P\times_{G_\CC}\bN^*))
    - \dim C^\infty(M_1\otimes (P\times_{G_\CC}\bN))
\\
    =\; & \dim C^\infty(\Lambda^{0,1}\otimes M_1\otimes
    (P\times_{G_\CC}\bN))
    - \dim C^\infty(M_1\otimes (P\times_{G_\CC}\bN)),
    \end{split}
\end{equation*}
where the second equality is true because `$\dim$' is the same for the
dual space. Now this is safely replaced by a well-defined number
thanks to Riemann-Roch:
\begin{equation*}
    \begin{split}
    & - \operatorname{ind}(\text{$\DB$ on $M_1\otimes (P\times_{G_\CC}\bN)$})
\\
  =\; & - \deg (M_1\otimes (P\times_{G_\CC}\bN))
 + \rank (M_1\otimes (P\times_{G_\CC}\bN))(g-1),
    \end{split}
\end{equation*}
where $g$ is the genus of $C$. Let us write this number by
$-i(P_\bN)$.
Similarly we have
\begin{equation*}
    \begin{split}
    & \dim\mathcal G_\CC(P) - \dim \{\DB+A\}
    = \ind(\text{$\DB$ on $P\times_{G_\CC}\mathfrak g_\CC$})
\\
   = \; & 
   \begin{NB}
       \deg (P\times_{G_\CC}\mathfrak g_\CC) +        
   \end{NB}
   \dim G_\CC(1-g).
    \end{split}
\end{equation*}
Let us denote this by $i(P)$. We have
\begin{equation}
    \label{eq:5}
    \frac{[\operatorname{crit}(\CS)]_{\mathrm{vir}}}
    {[\mathcal G_\CC(P)]_{\mathrm{vir}}}
    = (-\LL^{\frac12})^{i(P)-i(P_\bN)}\frac{[\mathcal R]}{[\mathcal G_\CC(P)]}.
\end{equation}

This argument can be `lifted' to the case of the cohomology group of
vanishing cycles thanks to a recent result by Davison
\cite[Th.~A.1]{2013arXiv1311.7172D}. Applying his result formally even though the fiber is infinite dimensional, we get
\begin{equation*}
    \begin{split}
    H^{*+\dim\mathcal F-\dim\mathcal G_\CC}_{c,\mathcal G_\CC(P)}
    (\mathcal F, \varphi_{\CS}(\CC_{\mathcal F}))
    & \cong H^{*+\dim\mathcal F-\dim\mathcal G_\CC
      -2\dim(\text{fiber})}_{c,\mathcal G_\CC(P)}
    (\mathcal R,\CC)\\
    & \cong H^{*-i(P)+i(P_\bN)}_{c,\mathcal G_\CC(P)}
    (\mathcal R,\CC).
    \end{split}
\end{equation*}

\subsection{Second projection}

There is a second projection $\mathcal R\to\mathcal A(P)$, where
$\mathcal A(P)$ is the space of all (partial) connections. The fiber
at $\DB+A\in\mathcal A(P)$ is the space of holomorphic sections
$\Ker(\text{$\DB+A$ on $M_1\otimes(P\times_{G_\CC}\bN)$})$. This
depends on the isomorphism class of $\DB+A$. Let us denote the
isomorphism class by $[\mathscr P]$.  If we write $\mathscr P$, it
will be a representative of $[\mathscr P]$, considered as a
holomorphic principal $G_\CC$-bundle. Let
\begin{equation*}
    h^0(\mathscr P_\bN) \defeq 
    \dim H^0(M_1\otimes (\mathscr P\times_{G_\CC}\bN)).
\end{equation*}
Then we have
\begin{equation}\label{eq:6}
    \begin{split}
  &  (-\LL^{\frac12})^{-i(P_\bN)+i(P)}\frac{[\mathcal R]}{[\mathcal G_\CC(P)]}
\\
 = \; &\sum_{[\mathscr P]\in\mathcal A(P)/\mathcal G_\CC(P)}
   (-\LL^{\frac12})^{i(P) -i(P_\bN)+ 2h^0(\mathscr P_\bN)}
   \frac{[\mathcal G_\CC(P)\cdot \mathscr P]}{[\mathcal G_\CC(P)]}
\\
  =\; & \sum_{[\mathscr P]\in\mathcal A(P)/\mathcal G_\CC(P)}
   (-\LL^{\frac12})^{i(P) -i(P_\bN) + 2h^0(\mathscr P_\bN)}
   \frac1{[\operatorname{Aut}(\mathscr P)]}.
    \end{split}
\end{equation}
This is a well-defined object. 

Note that
\begin{equation*}
    \begin{split}
    & -i(P_\bN) + 2h^0(\mathscr P_\bN) 
    = h^0(\mathscr P_\bN) + h^1(\mathscr P_\bN)
\\
   =\; & h^0(\mathscr P_\bN) + h^0(\mathscr P_{\bN^*}) 
   = h^0(\mathscr P_\bM),
    \end{split}
\end{equation*}
where $h^1(\mathscr P_\bN) = \dim H^1(M_1\otimes(\mathscr
P\otimes_{G_\CC}\bN))$, and $\mathscr P_{\bN^*}$, $\mathscr P_{\bM}$
are vector bundles associated with representations $\bN^*$, $\bM$, and
$h^0$ denotes the dimensions of their holomorphic sections. 
In particular, this combination is always nonnegative, and makes sense
even when $\bM$ is not necessarily cotangent type.
Therefore it is reasonable to conjecture that this computation remains
true even without assuming $\bM$ is of cotangent type.
\begin{NB}
This could be
checked for a finite dimensional model (say the case studied in
\secref{sec:c=cc}).
\end{NB}

\section{Computation for \texorpdfstring{$C=\proj^1$}{C=P1}}
\label{sec:comp-P1}

\subsection{Parametrization of \texorpdfstring{$G_\CC$}{GC}-bundles}

Let us assume $C=\proj^1$. Then all holomorphic principal
$G_\CC$-bundles can be reduced to $T_\CC$-bundles, and they are unique
up to conjugation by the Weyl group. This is a result of Grothendieck
\cite{MR0087176}.
Note that line bundles are classified by its degree on
$\proj^1$. Therefore isomorphism classes of $G_\CC$-bundles are
parametrized by the coweight lattice $Y$ of $G$ modulo the Weyl group
$W$, or {\it dominant\/} coweights in other words. A coweight is
called a {\it magnetic charge\/} in the physics literature. Therefore
a coweight is denoted by $m$ in \cite{Cremonesi:2013lqa}, but we
denote a coweight by $\la$ following a standard notation in
mathematics.

Let us explain these more concretely. Suppose $G_\CC = \GL(N)$. Then a
$G_\CC$-bundle is nothing but a rank $N$ vector bundle. It decomposes
into a direct sum of line bundles
$\shfO_{\proj^1}(\la_1)\oplus\cdots\oplus\shfO_{\proj^1}(\la_N)$. As the
order of sum does not matter, we may assume $\la_1\ge\dots\ge \la_N$. This
is the dominance condition. We can make arbitrary $\la =
(\la_1,\dots,\la_N)$ to a dominance one by the action of the symmetric
group, which is the Weyl group of $\GL(N)$.

Keep in mind that rank and degree of the vector bundle are
\begin{equation*}
    \operatorname{rank} = N, \quad
    \operatorname{deg} = \sum \la_i.
\end{equation*}
The degree is called the {\it topological charge}, denoted by $J(\la)$
in \cite[(2.7)]{Cremonesi:2013lqa}. See \eqref{eq:15}.

By our formula \eqref{eq:6}, the motivic invariant is the sum over all
isomorphism classes, i.e., dominant coweights.

When we have a flavor symmetry $G_F$, we add a $(G_F)_\CC$-bundle with
a partial connection $\DB_F$ as additional data. Its isomorphism class
is also parametrized by a dominant coweight $\la_F$. It is a {\it
  background\/} flux, i.e., it enters in the formula, but it is fixed,
and we do not sum over $\la_F$.

Let us note also that the calculation below is true in the level of
Poincar\'e polynomials, not only as motives. Considering isomorphism
classes of $G_\CC$-bundles, we have a stratification of the space
$\mathcal R$. Each stratum is a vector bundle over a $\mathcal
G_\CC(P)$-orbit, and hence has no odd cohomology groups. Therefore the
short exact sequences associated with the stratification splits, hence
the cohomology group is isomorphic (as a graded vector space) with the
sum of cohomology groups of strata.

\subsection{Calculation of contributions from matters}

Let $B(\bN)$ be a base of $\bN$ compatible with weight space
decomposition. For $b\in B(\bN)$, let $\wt(b)$ be its weight. 
Hence
\begin{equation*}
    \bN = \bigoplus_b \CC b.
\end{equation*}
We denote the pairing between weights and coweights by $\langle\ ,\
\rangle$.
When a principal bundle $\mathscr P$ is reduced to a $T_\CC$-bundle,
the associated vector bundle decomposes as
\begin{equation*}
    \mathscr P_\bN = \bigoplus_b  \mathscr P\times_{T_\CC}\CC b,
\end{equation*}
a sum of line bundles. If $\la$ is the coweight for $\mathscr P$, the
degree of the line bundle is $\langle\la, \wt(b)\rangle + \deg M_1$.
Then
\begin{equation*}
    h^0(\mathscr P_\bN) + h^1(\mathscr P_\bN)
    = \sum_{b} |\langle \la, \wt(b)\rangle + \deg M_1 + 1 |.
\end{equation*}
This is because
\begin{equation*}
    \dim H^0(\shfO_{\proj^1}(n)) + \dim H^1(\shfO_{\proj^1}(n))
    = | n + 1 |.
\end{equation*}
For our standard choice $M_1 = M_2 = \shfO_{\proj^1}(-1)$, we have
\begin{equation*}
    \sum_{b} |\langle \la, \wt(b)\rangle|,
\end{equation*}
which coincides with the second term in $\Delta(\la)$ in \eqref{eq:14}
if we replace $-\LL^{\frac12}$ by $t$. (Recall $\Delta(\la)$ appears
as $t^{2\Delta(\la)}$ in \eqref{eq:11}).

The calculation remains the same for $\bM$ not necessarily cotangent
type.

\subsection{Calculation of automorphism groups (\texorpdfstring{$G_\CC=\GL(N,\CC)$}{GC=GL(N,C)}
  case)}

Let us first suppose that $G_\CC$ is $\GL(N,\CC)$.
\begin{NB}
    Or more generally products of general linear groups as for quiver
    gauge theory $\Hyp(\bM)\tslabar G$.
\end{NB}%
Then $\mathscr P$ is a usual vector bundle of rank $N$.
\begin{NB}
or a
collection of vector bundles.
\end{NB}%
We write $\la = (\la_1\ge \la_2\ge\dots\ge
\la_N)$, a non-increasing sequence of integers. We also write $\la =
(\cdots (-1)^{k_{-1}} 0^{k_0} 1^{k_1} \cdots )$, which is a
modification of the usual notation for partitions, i.e., $-1$ appears
$k_{-1}$ times, $0$ appears $k_{0}$ times, etc in the sequence $\la$.

In this notation, we have
\begin{NB}
    \begin{equation*}
        \begin{split}
    i(P) &= \sum_{\alpha,\beta} k_\alpha k_\beta
    \chi(\shfO_{\proj^1}(\alpha), \shfO_{\proj^1}(\beta))
\\
     &= \sum_{\alpha,\beta} k_\alpha k_\beta (\beta - \alpha + 1)
     = \sum_{\alpha,\beta} k_\alpha k_\beta,
\\        
        \end{split}
    \end{equation*}
\end{NB}%
\begin{equation*}
    \begin{split}
    \dim \End(\mathscr P) & = \sum_{\alpha,\beta} k_\alpha k_\beta
    \dim \Hom(\shfO_{\proj^1}(\alpha), \shfO_{\proj^1}(\beta))
\\
   & =
   \sum_{\beta\ge\alpha}k_\alpha k_\beta (\beta - \alpha + 1).
    \end{split}
\end{equation*}
For $\operatorname{Aut}(\mathscr P)$, we replace the block diagonal
entries $\alpha = \beta$ by $\GL(k_\alpha)$. Therefore
\begin{equation}\label{eq:16}
    [\operatorname{Aut}(\mathscr P)] = 
    [\End(\mathscr P)] \prod_\alpha   \frac{[\GL(k_\alpha)]}{[\End(\CC^{k_\alpha})]}.
\end{equation}
The ratio $[\GL(k)]/[\End(\CC^k)]$ is easy to compute and well-known:
\begin{equation*}
    \begin{split}
    & [\GL(k)]/[\End(\CC^k)] 
    = (\LL^k - 1)(\LL^k - \LL) \cdots (\LL^k - \LL^{k-1}) \LL^{-k^2}
\\
    = \; & (-1)^k (1-\LL)(1-\LL^2)\cdots (1-\LL^k)
    \LL^{-\frac{k(k+1)}2}.
    \end{split}
\end{equation*}
Thus
\begin{equation*}
    \frac{(-\LL^{\frac12})^{i(P)}}{[\operatorname{Aut}(\mathscr P)]}
    = (-1)^{\sum k_\alpha} (-\LL^{\frac12})^{d}
    \prod_\alpha \frac1{(1-\LL)(1-\LL^2)\cdots (1-\LL^{k_\alpha})},
\end{equation*}
where
\begin{equation*}
    \begin{split}
    & d = i(P) - 2 \dim \End(\mathscr P) + \sum k_\alpha(k_\alpha+1)
\\
  =\; & -2 \sum_{\beta\ge\alpha} k_\alpha k_\beta(\beta-\alpha)
  - 2\sum_{\beta\ge\alpha} k_\alpha k_\beta + 
  \sum_{\alpha,\beta} k_\alpha k_\beta + \sum k_\alpha(k_\alpha + 1)
\\
  =\; & -2 \sum_{\beta\ge\alpha} k_\alpha k_\beta(\beta-\alpha) + \sum k_\alpha.
    \end{split}
\end{equation*}
The first term is
\begin{equation*}
    -2 \sum_{i< j} (\la_i - \la_j),
\end{equation*}
which is equal to the first term in $\Delta(\la)$ in \eqref{eq:14},
again if we replace $-\LL^{\frac12}$ by $t$.

The term
\begin{equation*}
    \prod_\alpha \frac1{(1-\LL)(1-\LL^2)\cdots (1-\LL^{k_\alpha})}
\end{equation*}
is $P_{\U(N)}(t;\la)$ if we put $\LL = t^2$.

Therefore up to the factor $(-1)^{\sum k_\alpha}(-\LL^{\frac12})^{\sum
  k_\alpha}$, it coincides with \eqref{eq:11}. Note that $\sum
k_\alpha$ is $N = \operatorname{rank} G$, and hence is
independent of $\la$.

Recall that the rank of the group is $\dim_\HH \mathcal M_C$ (\subsecref{sec:expect-properties}(\ref{item:dim})). Therefore
\begin{equation*}
    H_{G,\bM}(t) = (\LL^{\frac12})^{-\dim_\HH\mathcal M_C}
    \frac{[\operatorname{crit}(\CS)]_{\mathrm{vir}}}
    {[\mathcal G_\CC(P)]_{\mathrm{vir}}}.
\end{equation*}
The virtual motive is shifted already by $-\dim/2$, hence
$H_{G,\bM}(t)$ is shifted by $-\dim$ in total. This shift is
unavoidable, as the monopole formula starts with $1$ which corresponds
to constant functions on $\mathcal M_C$. This convention is not taken
in the cohomology side.
\begin{NB}
    This might suggest that $H_{G,\bM}(t)$ should be `Poincar\'e
    dual', but I cannot make anything rigorous. And I do not
    understand the sign either.
\end{NB}

\subsection{Calculation of automorphism groups (general case)}

Let us turn to a general case. The relevant calculation was done in
\cite[\S7]{MR2329314}. We reproduce it with shortcuts of a few
arguments for the sake of the reader.

As in \cite[Def.~2.4]{MR2329314}, we add the {\it torsor\/} relation
to our motivic ring: $[\mathscr P] = [X][G]$ for a principal $G$
bundle $\mathscr P$ over a variety $X$.

Let $P$, $U$, $L$ be the parabolic subgroup, its unipotent radical and
its Levi quotient associated with the dominant coweight $\la$. ($L =
\operatorname{Stab}_G(\la)$ in the previous notation.)
\begin{NB}
    That is $P = \{ g\in G \mid\text{$\lim_{t\to
        0}\operatorname{Ad}(\la(t))(g)$ exists} \}$ and $L =
    G^{\la(\CC^*)}$.
\end{NB}%
Since the underlying $C^\infty$ principal $G_\CC$-bundle $P$ will not
occur any more except through the formula $i(P)$ (which is $\dim
G_\CC$ in our case), there is no fear of confusion. Let $\mathfrak u$
be the Lie algebra of $U$, and $\mathscr P_{\mathfrak u}$ be the
associated vector bundle. Then generalizing the computation for
$G=\GL(N,\CC)$, we have

\begin{Lemma}[\protect{\cite[Prop.~5.2]{MR688263}}]\label{lem:Ramanathan}
We have $\operatorname{Aut}(\mathscr P) = L\ltimes H^0(\proj^1, \mathscr P_{\mathfrak u})$.
\end{Lemma}

The proof is not difficult. It could be an exercise for the reader.

Since $\mathscr P$ is now a $T$-bundle, we have
\begin{equation*}
    \mathscr P_{\mathfrak u} = \bigoplus_{\substack{\alpha\in\Delta^+ \\
      \langle\la,\alpha\rangle > 0}}
    \mathscr P_{\mathfrak g_\alpha},
\end{equation*}
where $\mathscr P_{\mathfrak g_\alpha}$ is the line bundle associated
with the root subspace $\mathfrak g_\alpha$. Its degree is
$\langle\la,\alpha\rangle$. Hence
\begin{equation*}
    \dim H^0(\proj^1,\mathscr P_{\mathfrak u})
    = \sum_{\substack{\alpha\in\Delta^+\\ \langle\la,\alpha\rangle > 0}}
    \left(\langle\la,\alpha\rangle + 1\right).
\end{equation*}
Therefore
\begin{equation*}
    \frac{[H^0(\proj^1,\mathscr P_{\mathfrak u})]}{[U]}
    = \mathbb L^{2\langle\la,\rho\rangle},
\end{equation*}
where $\rho$ is the half sum of positive roots. Note
$2\langle\la,\rho\rangle = \sum_{\alpha\in\Delta^+} \langle
\la,\alpha\rangle$. Since we take $\la$ dominant, this is the first
term (up to sign) of $\Delta(\la)$ in \eqref{eq:14}.

\begin{NB}
    Note that $\dim G[[z]] \lambda = 2\langle \lambda,\rho\rangle$,
    the $G[[z]]$-orbit in the affine Grassmannian.
\end{NB}

We substitute this into \lemref{lem:Ramanathan}. Since $LU=P$,
\begin{equation*}
    [\operatorname{Aut}(\mathscr P)]
    = \mathbb L^{2\langle\la,\rho\rangle} [P]
    = \mathbb L^{2\langle\la,\rho\rangle} \frac{[G]}{[G/P]},
\end{equation*}
where $G/P$ is the partial flag variety. (We implicit used the torsor
relation. See \cite[p.637]{MR2329314} for detail.)

The partial flag variety $G/P$ has the Bruhat decomposition. Therefore
we only need to compute ordinary cohomology. To connect with the ring
of invariant polynomials, we use an isomorphism of equivariant
cohomology groups:
\begin{equation*}
    H^*_G(G/P) \cong H^*_L(\mathrm{pt}).
\end{equation*}
The spectral sequence relating equivariant and ordinary cohomology
groups collapses for $G/P$ (as there is no odd degree
cohomology). Hence $H^*_G(G/P) = H^*(G/P) \otimes
H^*_G(\mathrm{pt})$. Thus we get
\begin{equation*}
    [G/P] = \frac{[H^*_L(\mathrm{pt})]}{[H^*_G(\mathrm{pt})]}
    = \frac{P_G(-\mathbb L^{\frac12};\lambda)}{P_G(-\mathbb L^{\frac12};0)}.
\end{equation*}
\begin{NB}
    Let us check for $G=\SL(2)$, $P=B$. Then $P_G(t;\lambda) =
    1/(1-t^2)$, $P_G(t;0) = 1/(1-t^4)$. Therefore
    \begin{equation*}
        \frac{P_G(-\mathbb L^{\frac12};\lambda)}
        {P_G(-\mathbb L^{\frac12};0)} = \frac{1-\mathbb L^2}{1-\mathbb L}
        = 1 + \mathbb L.
    \end{equation*}
\end{NB}%

From the special case $P=B$ of this computation, we have
\begin{equation*}
    [B] 
    \begin{NB}
        = \frac{[G]}{[G/B]} =
        \frac{[G][H^*_G(\mathrm{pt})]}{[H^*_T(\mathrm{pt})]}
    \end{NB}%
    = [G] P_G(-\mathbb L^{\frac12};0) (1-\mathbb L)^{\rank G},
\end{equation*}
as $[H^*_T(\mathrm{pt})] = (1-\mathbb L)^{-\rank G}$. On the other
hand, $B$ is the product of $T$ and its unipotent radical. Hence $[B]
= (\mathbb L-1)^{\rank G} [\mathbb L]^{\frac12(\dim G_\CC - \rank G)}$. We
get
\begin{equation*}
    [G] P_G(-\mathbb L^{\frac12};0) =
    (-1)^{\rank G}[\mathbb L]^{\frac12(\dim G_\CC-\rank G)}.
\end{equation*}

Now we combine all these computation to get
\begin{equation*}
    \frac{(-\mathbb L^{\frac12})^{i(P)}}{[\operatorname{Aut}(\mathscr P)]}
    \begin{NB}
    = (-\mathbb L^{\frac12})^{\dim G_\CC} \mathbb L^{-2\langle\la,\rho\rangle}
    \frac{P_G(-\mathbb L^{\frac12};\la)}{[G] P_G(-\mathbb L^{\frac12};0)}
    \end{NB}
    = (-1)^{\rank G}(-\mathbb L^{\frac12})^{-4\langle\la,\rho\rangle + \rank G}
    P_G(-\mathbb L^{\frac12};\la).
\end{equation*}
This coincides with the contribution of terms not involving $\bM$ in
the monopole formula up to the factor
\(
    (-1)^{\rank G}(-\mathbb L^{\frac12})^{\rank G}.
\)

\section{Computation for \texorpdfstring{$C=\CC$}{C=CC}}\label{sec:c=cc}

\subsection{Motivic universal DT invariant for \texorpdfstring{$\mathfrak g_\CC\times\bM$}{gC x M}}

Since we assume $C$ is compact, we cannot apply the argument in
\secref{sec:motiv-DT-invariants} to $C=\CC$. However we further
consider the dimension reduction in the $\CC$-direction to a point as
in \remref{rem:reduction}. In practice, it means that we replace
$\DB+A$ by an element $\xi$ in the Lie algebra $\mathfrak g_\CC$. We
thus arrive at
\begin{equation*}
    \CS(\xi,\Phi) = \frac12 \omega_\CC(\xi\Phi,\Phi)
    = \langle\xi, \bmu_\CC(\Phi)\rangle.
\end{equation*}
This is a function on a finite dimensional space $\mathcal F =
\mathfrak g_\CC\times\bM$, and all computation in
\secref{sec:motiv-DT-invariants} become rigorous. We repeat our
assertions in this setting:
\begin{enumerate}
      \item 
$(\xi,\Phi)$ is a critical point of $\CS$ if and only if
\begin{equation*}
    \xi \Phi = 0, \quad \bmu_\CC(\Phi) = 0.
\end{equation*}

\item The motivic universal DT invariant is defined by
\begin{equation*}
    \frac{[\operatorname{crit}(\CS)]_{\mathrm{vir}}}{[G_\CC]_{\mathrm{vir}}}.
\end{equation*}

  \item Assuming $\bM$ is of cotangent type (i.e., $\bM =
\bN\oplus\bN^*$), we use a $\CC^*$-action given by
$(\xi,\Phi_1,\Phi_2)\mapsto (\xi,\Phi_1,t\Phi_2)$ to get
\begin{equation*}
  \frac{[\operatorname{crit}(\CS)]_{\mathrm{vir}}}{[G_\CC]_{\mathrm{vir}}}
  = 
  \frac{[\mathcal R]}{[G_\CC]},
\end{equation*}
where $\mathcal R$ is the locus $\xi\Phi_1 = 0$ in $\mathcal F_{\mathrm{red}}
= \{ (\xi,\Phi_1)\in\mathfrak g_\CC\times\bN\}$.

Note that \eqref{eq:7} vanishes in this setting.

\item We further have
\begin{equation}\label{eq:10}
  \frac{[\operatorname{crit}(\CS)]_{\mathrm{vir}}}{[G_\CC]_{\mathrm{vir}}}
  = \frac{[\mathcal R]}{[G_\CC]}
  = \sum_{[\xi]\in \mathfrak g_\CC/G_\CC} \frac
  {\LL^{\dim(\text{$\Ker\xi$ on $\bN$})}}{[\operatorname{Stab}_{G_\CC}(\xi)]}.
\end{equation}
\end{enumerate}
All these are rigorous now.

\subsection{Another \texorpdfstring{$\CC^*$}{C^*}-action}\label{sec:toy}

It is interesting to note that we have another $\CC^*$-action
$(\xi,\Phi)\mapsto (t\xi,\Phi)$, which makes sense without the
cotangent type condition. It is of weight $1$, even when we perturb
$\bmu_\CC$ by $\zeta_\CC\in Z(\mathfrak g_\CC^*)$.
The cutting for this action was used by Mozgovoy
\cite{2011arXiv1107.6044M} for $(G,\bM)$ of quiver type.

Then the application of results on the cutting implies as in \subsecref{sec:cutting}
\begin{equation}\label{eq:9}
   \frac{[\operatorname{crit}(\CS)]_{\mathrm{vir}}}{[G_\CC]_{\mathrm{vir}}}
   = (-\LL^{\frac12})^{-(\dim \bM - 2\dim\mathfrak g_\CC)}
  \frac{[\bmu^{-1}_\CC(\zeta_\CC)]}{[G_\CC]}.
\end{equation}
Note that $\dim \bM - 2\dim \mathfrak g_\CC$ is the expected dimension
of $\bmu^{-1}_\CC(\zeta_\CC)/G_\CC$.

Let us again suppose $\bM$ is of cotangent type and consider the
restriction of the projection $\bmu_\CC^{-1}(\zeta_\CC)\to \bN^*$ given
by $(\Phi_1,\Phi_2)\mapsto \Phi_2$. When $\bM$ is of quiver type, it
was studied by Crawley-Boevey and his collaborators
\cite{CH,CB,CBVB}. The argument can be modified to our setting as
follows.

Let us fix $\Phi_2\in\bN^*$ and consider an exact sequence
\begin{equation*}
    0 \to \operatorname{Lie}(\operatorname{Stab}_{G_\CC}(\Phi_2))
    \to \mathfrak g_\CC \to \bN^*,
\end{equation*}
where the second arrow is the inclusion and the third arrow is the
action of $G_\CC$ on $\bN^*$ : $\xi\mapsto \xi\Phi_2$. We consider the
transpose
\begin{equation*}
    \bN \to \mathfrak g_\CC^* \xrightarrow{t}
    \operatorname{Lie}(\operatorname{Stab}_{G_\CC}(\Phi_2))^* \to 0.
\end{equation*}
From the definition of the moment map, the first arrow is given by
$\Phi_1\mapsto \bmu_\CC(\Phi_1,\Phi_2)$. Let $K_{\Phi_2}$ denote the kernel
of $\bN\to\mathfrak g_\CC^*$.

If we have a solution of $\bmu_\CC(\cdot,\Phi_2) = \zeta_\CC$ then
$\zeta_\CC$ is in the image of the first map. Hence we must have
$t(\zeta_\CC) = 0$ by the exact sequence. Moreover, if $t(\zeta_\CC) =
0$, the space of solutions is an affine space modeled by
$K_{\Phi_2}$. Therefore we are led to introduce the following
condition:

\begin{Definition}
    We say $\Phi_2$ is {\it $\zeta_\CC$-indecomposable\/} if
    $t(\zeta_\CC) = 0$, in other words,
    \begin{equation*}
        \langle \xi, \zeta_\CC\rangle = 0 \quad
        \text{for any $\xi\in
          \operatorname{Lie}(\operatorname{Stab}_{G_\CC}(\Phi_2))$}.
    \end{equation*}
\end{Definition}

If $\bM$ is of quiver type (with $W=0$) and $\zeta_\CC$ is generic,
this is equivalent to that $\Phi_2$ is indecomposable in the usual
sense. See the proof of \cite[Prop.~2.2.1]{CBVB}. On the other hand,
this imposes nothing if $\zeta_\CC = 0$.

This condition is invariant under $G_\CC$-action. Let
$\bN^*_{\text{$\zeta_\CC$-ind}}$ be the constructible subset of
$\bN^*$ consisting of $\zeta_\CC$-indecomposable $\Phi_2$.

Let us go back to \eqref{eq:9}. We have
\begin{equation}\label{eq:34}
  \begin{split}
    (-\LL^{\frac12})^{-(\dim \bM-2\dim\mathfrak g_\CC)}
  \frac{[\bmu^{-1}_\CC(\zeta_\CC)]}{[G_\CC]}
  &= \sum_{[\Phi_2]\in\bN^*_{\text{$\zeta_\CC$-ind}}/G_\CC}
  \frac{\LL^{\dim K_{\Phi_2}- \dim \bN +\dim \mathfrak g_\CC}}
  {[\operatorname{Stab}_{G_\CC}(\Phi_2)]}
\\
  & = 
  \sum_{[\Phi_2]\in\bN^*_{\text{$\zeta_\CC$-ind}}/G_\CC}
   \frac{[\operatorname{Lie}(\operatorname{Stab}_{G_\CC}(\Phi_2))]}
   {[\operatorname{Stab}_{G_\CC}(\Phi_2)]},
  \end{split}
\end{equation}
where we have used the exact sequence to compute the alternating sum
of the dimension.

In the situation of \cite{CBVB}, we replace $G_\CC$ by its quotient
$G_\CC/\CC^*$ as explained at the end of \subsecref{sec:quiver-type}.
If $\zeta_\CC$ is generic and hence $\Phi_2$ is indecomposable in the
usual sense, $\operatorname{Stab}_{G_\CC/\CC^*}(\Phi_2)$ and its Lie
algebra is isomorphic by the exponential.
Then each term in the summand is $1$.
Hence the sum is the motif of the space of indecomposable
modules. This is one of crucial steps in their proof of Kac's
conjecture for indivisible dimension vectors. See
\cite[Prop.~2.2.1]{CBVB}.

\begin{NB}
    It should be possible to write down the motif of the space of
    indecomposable modules by a combinatorial expression, like the
    number of rational points.
\end{NB}

Suppose $\zeta_\CC = 0$, hence $\Phi_2$ is an arbitrary element. When
$(G,\bM)$ is of quiver type with $W=0$, the right hand side of
\eqref{eq:34} was computed in \cite{2011arXiv1107.6044M}. Let us
introduce variables $x_i$ for each $i\in Q_0$ and denote $\prod_i
x_i^{v_i}$ by $x^{\mathbf v}$ for $\mathbf v = (v_i)_{i\in
  Q_0}$. Summing up to the motivic Donaldson-Thomas invariants for all
dimension vectors, the generating function is given by
\begin{equation}\label{eq:35}
    \sum_{\mathbf v}
    \frac{[\operatorname{crit}(\CS)]_{\mathrm{vir}}}{[G_\CC]_{\mathrm{vir}}}
    x^{\mathbf v} = 
    \exp\left(\sum_{d=1}^\infty
      \frac{\sum_{\mathbf v}a_{\mathbf v}(\LL^d) x^{d\mathbf v}}{d(1-\LL^{-d})}
      \right),
\end{equation}
where $a_{\mathbf v}(q)$ is the Kac polynomial counting the number of
absolute indecomposable representations of dimension vector $\mathbf
v$ of the finite field $\mathbb F_q$. See
\cite[Th.~1.1]{2011arXiv1107.6044M}.

Note that Kac polynomials have combinatorial expressions very similar
to the monopole formula due to Kac-Stanley (see \cite[p.90]{MR718127}
and also \cite{MR1752774} for a detail of the proof).

\begin{NB}
*********************************************************************

\begin{equation*}
    b_\lambda(t) = \prod_{i\ge 1}\varphi_{m_i(\lambda)}(t),
\end{equation*}
where $m_i(\lambda)$ is the number of times $i$ occurs as a part of
$\lambda$, and $\varphi_r(t) = (1-t)(1-t^2)\cdots (1-t^r) = (t;t)_r$.
\end{NB}

\begin{NB}
\subsection{Calculation of stabilizers}

Let us return back to the case $\zeta_\CC = 0$ and the equation
\eqref{eq:10}.

\begin{center}
    To be finished.
\end{center}
\end{NB}

\appendix

\section{Instantons for classical groups}\label{sec:classical}

We give further examples of hyper-K\"ahler quotients arising as
instanton moduli spaces for $\SO/\grpSp$ groups on $\RR^4$ with
various equivariant structures.

\subsection{The case of $\RR^4$}\label{sec:SOSp}

It is well-known that the ADHM description of $\SU(N)$-instantons on
$\RR^4$ can be modified for $\SO/\grpSp$ groups.

For $\SO(N)$ $k$-instantons, we take the vector representation $W$ of
$\SO(N)$, the vector representation $V$ of $\grpSp(k)$ and set
\begin{equation}\label{eq:SOSp}
    \begin{split}
    \bM = \{ (B_1,B_2,&a,b)\in
    \Hom(V,V)^{\oplus 2}\oplus \Hom(W,V)\oplus\Hom(V,W) 
\\
&
\left|
\begin{aligned}[m]
    &(B_\alpha v, v') = (v, B_\alpha v')  \quad
    \text{for $v,v'\in V$, $\alpha=1,2$},
\\
    &(aw, v) = (w, bv) \quad \text{for $v\in V$, $w\in W$}
\end{aligned}\right\},
    \end{split}
\end{equation}
where $(\ ,\ )$ is either the symplectic form or orthogonal form on
$V$ or $W$ respectively. Then the framed instanton moduli space is the
hyper-K\"ahler quotient $\bmu^{-1}(0)/\grpSp(k)$.

For $\grpSp(N)$ $k$-instantons, we replace $W$ by the vector
representation of $\grpSp(N)$, $V$ by the vector representation of
$\grpO(k)$. It should be warned that the group is not
$\SO(k)$.
\begin{NB}
Since we assume $G$ is connected in the monopole formula, it is not
clear how to compute the Hilbert series of the Coulomb branch for $\Hyp(\bM)\tslabar\grpO(k)$.
\end{NB}%

\subsection{Nilpotent orbits and Slodowy slices for classical groups}
\label{sec:nilp-orbits-slod}

Let us generalize discussions in \subsecref{sec:type-A-quiver} to
classical groups. This straightforward generalization was mentioned
\cite[Remark~8.5(4)]{Na-quiver} for the $\SU(2)$-equivariant case, but
not explicitly written down before.

Suppose $(B_1,B_2,a,b)$ in \eqref{eq:SOSp} corresponds to an
$\SU(2)$-equivariant instanton. Then $V$, $W$ are representations of
$\SU(2)$, and $(B_1,B_2)$, $a$, $b$ are $\SU(2)$-linear. Here we mean
the pair $(B_1,B_2)$ is $\SU(2)$-equivariant, when it is considered as
a homomorphism in $\Hom(V,V\otimes \rho_2)$, where $\rho_2$ is the
vector representation of $\SU(2)$.

\begin{figure}[htbp]
    \centering
\setlength{\unitlength}{1mm}
\begin{picture}(70,28)
    \multiput(5,23)(20,0){3}{\circle{10}}
    \multiput(0,0)(20,0){3}{\framebox(10,8){}}
    \multiput(10,23)(20,0){3}{\thicklines\line(1,0){10}}
    \multiput(5,18)(20,0){3}{\thicklines\line(0,-1){10}}
    \put(1,22){$\scriptstyle\grpSp(\frac{v_1}2\!)$}
    \put(21,22){$\scriptstyle\grpO(v_2)$}
    \put(41,22){$\scriptstyle\grpSp(\frac{v_3}2\!)$}
    \put(63,22){$\cdots$}
    \put(0.5,3){$\scriptstyle\grpO(w_1)$}
    \put(20.5,3){$\scriptstyle\grpSp(\frac{w_2}2\!)$}
    \put(40.5,3){$\scriptstyle\grpO(w_3)$}
    \put(63,3){$\cdots$}
\end{picture}
\caption{Intersection of a nilpotent orbit and a Slodowy slice}
    \label{fig:SOSp}
\end{figure}
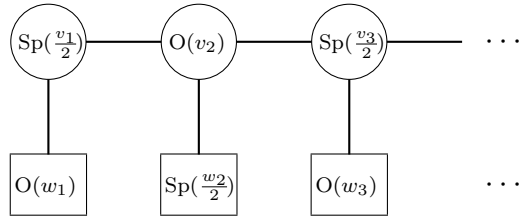

Let us decompose $V$, $W$ as $\bigoplus V_i\otimes\rho_i$, $\bigoplus
W_i\otimes\rho_i$ as in \subsecref{sec:type-A-quiver}.
Since $\rho_i$ has a symplectic or orthogonal form according to the
parity of $i$, we have either symmetric or orthogonal forms on $V_i$,
$W_i$. For example, for $\SO(N)$-instantons, linear maps $B_1$, $B_2$,
$a$, $b$ can be put into a graph (see Figure~\ref{fig:SOSp}) as
before. 
Groups $\grpSp(\frac{v_1}2)$, $\grpO(w_1)$, etc in circles or
squares indicate, we have a symmetric or an orthogonal form on $V_1$,
$W_1$, etc. (Here $v_i = \dim V_i$, $w_i = \dim W_i$.)
The space $\bM$ consists of linear maps, both directions for
each edges, satisfying the transpose condition like in \eqref{eq:SOSp}.
And the group $G$, by which we take the hyper-K\"ahler quotient, is
the product of groups in circles.
It is not relevant here whether we should put either
$\grpO(w_1)$,\dots or $\SO(w_1)$,\dots, as we only need orthogonal
forms.
However, it is matter that for $\grpO(v_2)$,\dots, as $G$ is the
product of groups in circles.
As a special case $w_2=w_3=\cdots=0$, we recover Kraft-Procesi's
description of classical nilpotent orbits \cite{MR694606}.
This special case also appeared later in \cite{MR1382722}.

A nilpotent orbit $\shfO_\la$ in $\SO$ corresponds to an even
partition $\la$. It corresponds to that $\dim W_{\mathrm{even}}$ is even, and is compatible with that $W_{\mathrm{even}}$ has a symplectic form.
Similarly $\shfO_\mu$ corresponds to an even partition as $\dim
V_{\mathrm{odd}}$ is even.

\begin{NB}
Let us recall the rule which associates $\dim V_i$, $\dim W_i$ with
two nilpotent orbits $\mathcal O_\lambda$, $\mathcal O_\mu$ given in
\cite[\S8]{Na-quiver}.

We define a partition $\lambda$ by
\(
    (1^{w_1} 2^{w_2} 3^{w_3} \cdots).
\)
We have a corresponding nilpotent matrix $N_\lambda$ with Jordan cells
consisting of $w_1$-times size $1$, $w_2$-times size $2$, and so on.
Since $W_{\mathrm{even}}$ have symplectic forms,
$w_{\mathrm{even}}=\dim W_{\mathrm{even}}$ are all even.
This is the well-known restriction on the partition $\lambda$
corresponding to a nilpotent orbit $\mathcal O_\lambda$ in $\SO$.
For nilpotent orbits in $\grpSp$, the restriction is that $w_i$ is
even for odd $i$.

Another partition $\mu$ is defined by 
\(
   (1^{u_1} 2^{u_2} 3^{u_3} \cdots),
\)
where
\begin{equation*}
    u_i = w_i + v_{i-1} + v_{i+1} - 2v_i.
\end{equation*}
(Here we set $v_i$ or $w_i=0$ if we do not have corresponding vector
spaces in Figure~\ref{fig:SOSp}.)
\begin{NB2}
    Therefore $u_{n+1} = v_{n}$ for the last $n$ with $v_n \neq 0$.
\end{NB2}%
Since $v_{\mathrm{odd}}$ are even, we find that $u_{\mathrm{even}}$
are all even. Thus $\mu$ corresponds to a nilpotent orbit $\mathcal
O_\mu$ in $\SO$.
\end{NB}

\begin{NB}
    Special orbits, to be written.
\end{NB}

\subsection{Affine Grassmannian and \texorpdfstring{$S^1$}{S1}-equivariant instantons}\label{sec:aff_and_S1}

The discussion in the previous subsection can be modified to the case
of $S^1$-equivariant instantons on $\RR^4$. We have a different quiver
as the dual of $\rho_i$ is $\rho_{-i}$ for irreducible representations
of $S^1$.

For an $\SO(r)$-instanton, the corresponding quiver is
Figure~\ref{fig:SOaffineGr}.
\begin{figure}[htbp]
    \centering
\setlength{\unitlength}{1mm}
\begin{picture}(70,28)
    \multiput(5,23)(20,0){3}{\circle{10}}
    \multiput(0,0)(20,0){3}{\framebox(10,8){}}
    \multiput(10,23)(20,0){3}{\thicklines\line(1,0){10}}
    \multiput(5,18)(20,0){3}{\thicklines\line(0,-1){10}}
    \put(1,22){$\scriptstyle\grpSp(\frac{v_0}2\!)$}
    \put(21,22){$\scriptstyle\U(v_1)$}
    \put(41,22){$\scriptstyle\U(v_2)$}
    \put(63,22){$\cdots$}
    \put(0.5,3){$\scriptstyle\grpO(w_0)$}
    \put(20.5,3){$\scriptstyle\U(w_1)$}
    \put(40.5,3){$\scriptstyle\U(w_2)$}
    \put(63,3){$\cdots$}
\end{picture}
\caption{$\mathcal W^\mu_{G,\la}$ for $G=\SO(r)$}
    \label{fig:SOaffineGr}
\end{figure}
From the isomorphisms $V\cong V^*$, $W\cong W^*$ given by the
symplectic and orthogonal forms, we have $V_i\cong V_{-i}^*$,
$W_i\cong W_{-i}^*$. Therefore we do not need to consider $V_i$, $W_i$
for $i<0$.
The moment map takes value in
$\RR^3\otimes(\algsp(v_0/2)\oplus\mathfrak{u}(v_1)\oplus\mathfrak{u}(v_2)\oplus\cdots)$. The
$\mathfrak{u}(v_1)\oplus\mathfrak{u}(v_2)\oplus\cdots$-component is as
for ordinary quiver gauge theories.
The $\algsp(v_0/2)$-component is given by
\begin{equation*}
    B_{0,1} B_{1,0} - B_{1,0}^\vee B_{0,1}^\vee + i_0 i_0^\vee,
\end{equation*}
where $B_{0,1}\colon V_1\to V_0$, $B_{1,0}\colon V_0\to V_1$,
$i_0\colon W_0\to V_0$. And $\vee$ is defined by
\(
   (B_{1,0} v_0, v_{-1}) = (v_0, B_{1,0}^\vee v_{-1}),
\)
by using the symplectic pairing $(\ ,\ )$ between $V_1$ and $V_{-1}$,
$V_0$ and itself, etc.
\begin{NB}
    We have
    \begin{equation*}
        \begin{split}
            & (B_{0,1} B_{1,0} v_0, v'_0) = (B_{1,0} v_0, B_{0,1}^\vee v'_0)
            = (v_0, B_{1,0}^\vee B_{0,1}^\vee v'_0),
\\
& (i_0 i_0^\vee v_0, v'_0) = (i_0^\vee v_0, i_0^\vee v'_0)
= \ve_W(i_0^\vee v'_0, i_0^\vee v_0) = \ve_W(i_0 i_0^\vee v'_0, v_0)
= \ve_V\ve_W (v_0, i_0 i_0^\vee v'_0),
        \end{split}
    \end{equation*}
    where $\ve_W = \pm 1$ according to the group at $w_0$ is
    $\SO(w_0)$ or $\grpSp(w_0/2)$. Similarly $\ve_V$ is $\mp 1$
    according to the group at $v_0$ is $\grpSp(v_0/2)$ or
    $\SO(v_0)$. Therefore $\ve_V \ve_W = -1$.
\end{NB}%
Thus $(G,\bM)$ for $\SU(2)$-equivariant and $S^1$-equivariant
instantons are different for classical groups.

\subsection{The case of \texorpdfstring{$\RR^4/\Gamma$}{R^4/Gamma}}\label{sec:SOSpALE}

Let $\Gamma$ be a finite subgroup of $\SU(2)$. Consider a
$\Gamma$-equivariant $G$-instanton on $\RR^4$. The case $G=\U(\ell)$ was explained in \subsecref{sec:inst-ale-space}. So we explain how it can be modified to the case when $G$ is $\SO/\grpSp$.

As above, the isomorphisms $V\cong V^*$, $W\cong W^*$ induces
$V_i\cong V^*_{i^*}$, $W_i\cong W^*_{i^*}$, where $i^*$ is determined
by the McKay correspondence as $\rho_i^*\cong\rho_{i^*}$. It fixes the
trivial representation $\rho_0$, and hence induces a diagram
involution on the finite Dynkin diagram. It is the same diagram
involution given by the longest element $w_0$ of the Weyl group as
$-w_0(\alpha_i) = \alpha_{i^*}$, where $\alpha_i$ is the simple root
corresponding to the vertex $i$. In the labeling in \cite[Ch.~4]{Kac},
it is given by $i^* = \ell-i+1$ for type $A_\ell$, $1^* = 5$, $2^* =
4$, $3^*=3$, $4^* = 2$, $5^* = 1$, $6^* = 6$ respectively.
For type $D_\ell$ with odd $\ell$, it is given by
\begin{equation*}
    i^* =
    \begin{cases}
        \ell -1 & \text{if $i=\ell$}, \\
        \ell & \text{if $i=\ell-1$},\\
        i & \text{otherwise},
    \end{cases}
\end{equation*}
\begin{NB}
    For $D_\ell$ with odd $\ell$, it exchanges two `tails'. For $E_6$,
    it is a reflection at the center.
\end{NB}%
It is the identity for other types.

When $i^* = i$, we determine whether $\rho_i\cong\rho_i^*$ is given by
a symplectic or orthogonal form as follows: 
The trivial representation, assigned to $i=0$, is orthogonal.
For type $A_{\ell}$ with odd $\ell$ with $i= {(\ell+1)/2}$, it is
orthogonal. For other types, $\rho_{i_0}$ for the vertex $i_0$
adjacent to the vertex $i=0$ in the affine Dynkin diagram is the
representation given by the inclusion $\Gamma\subset\SU(2)$. Therefore
$\rho_{i_0}\cong \rho_{i_0}^*$ is symplectic. If $i$ is adjacent to
$i_0$ and $i^* = i$, then $\rho_i\cong\rho_i^*$ is orthogonal. If $j$
is adjacent to such an $i$ with $j^*=j$, then $\rho_j\cong\rho_j^*$ is
symplectic, and so on.
From this rule, we determine either $V_i\cong V_i^*$, $W_i\cong W_i^*$
is symplectic or orthogonal.

We impose constraint on linear maps $B$, $a$, $b$ between $V_i$,
$W_i$'s induced from \eqref{eq:SOSp} either as in
\subsecref{sec:nilp-orbits-slod} or \subsecref{sec:aff_and_S1}.

Examples discussed in \secref{sec:classical} are not of cotangent type
in general. We also remark that $G$ is not connected if it contains
$\grpO(N)$ as a factor. A criterion on the complete intersection
property in \subsecref{sec:compl-inters} is not known in general. See
\cite{Choy} for special cases.

\begin{NB}
\section{Monopole moduli spaces and homology of affine Grassmannian}

The Coulomb branch $\mathcal M_C$ of the gauge theory $G=\SU(k)$, $\bM
= 0$ is identified with $M^0_k$, the moduli space of centered
monopoles by a physical argument (\cite{MR1490862} for $k=2$,
\cite{MR1443803} for arbitrary $k$). The theory is neither good nor
ugly, but our main proposal *** produces this answer. We explain it in
this section.

\subsection{Homology of affine Grassmannian}

When $\bM=0$, our proposal *** states $\mathcal M_C$ is the spectrum of
the equivariant Borel-Moore homology group
$H^{G_\CC[[z]]}_*(\Gr_{G_\CC})$ of the affine Grassmannian
$\Gr_{G_\CC} = G_\CC((z))/G_\CC[[z]]$ of $G_\CC$, where $G_\CC = \SL(k)$.
This homology group has an algebra structure given by the
convolution. It is commutative and its spectrum is described in
\cite[Th.~2.12]{MR2135527}, \cite[Th.~3]{MR2422266} for general
$G_\CC$.
\begin{NB2}
    The result is true for $G_\CC = \GL(k)$ ?
\end{NB2}%
Let us review this result.

Let $\check G_\CC$ be the Langlands dual group and $\check{\mathfrak
g}_\CC$ its Lie algebra.
Let $\rho\colon\algsl(2)\to\check{\mathfrak g}_\CC$ be a Lie algebra
homomorphism such that 
$y = \rho(
\begin{smallmatrix}
    0 & 0 \\ 1 & 0
\end{smallmatrix}
)$
is a principal nilpotent element in $\check{\mathfrak g}_\CC$.
Let $\check N$ be the unipotent subgroup of $\check G_\CC$ whose Lie algebra
$\check{\mathfrak n}$ is the sum of negative eigenspaces of $h = \rho(
\begin{smallmatrix}
    -1 & 0 \\ 0 & 1
\end{smallmatrix}
).$
\begin{NB2}
    We have $[h,y] = 2y$, $[h,x]=-2x$.
\end{NB2}%
If we regard $y$ as an element of $\check{\mathfrak n}^*$ via an invariant
pairing on $\check{\mathfrak g}_\CC$, it is stabilized by $\check N$.
\begin{NB2}
    $(y, [n_1,n_2]) = 0$ for $n_1$, $n_2\in\mathfrak n$ as the
    $(-1)$-eigenspace of $h$ is $0$.
\end{NB2}%

The cotangent bundle $T^* \check G_\CC = \check
G_\CC\times\check{\mathfrak g}_\CC$ is a holomorphic symplectic
manifold with a $\check G_\CC\times \check G_\CC$-action by left and
right multiplication. Here we identify $\check{\mathfrak g}^*_\CC$
with $\check{\mathfrak g}_\CC$ by the invariant inner product.
We have a moment map $\bmu_\CC\colon \check G_\CC\times\check{\mathfrak
  g}_\CC\to \check{\mathfrak g}_\CC^*\oplus\check{\mathfrak g}_\CC^*$
given by $\bmu_\CC(g,\xi) = (\xi,-\operatorname{Ad}(g^{-1})\xi)$ for
$g\in \check G_\CC$, $\xi\in\check{\mathfrak g}_\CC$.
Let $\overline\bmu_\CC\colon \check G_\CC\times\check{\mathfrak
  g}_\CC\to \check{\mathfrak n}^*\oplus\check{\mathfrak n}^*$ be the
moment map for the $\check N\times \check N$-action, that is the
composite of $\bmu_\CC$ and the natural projection $\check{\mathfrak
  g}_\CC^*\oplus\check{\mathfrak g}_\CC^*\to\check{\mathfrak
  n}^*\oplus\check{\mathfrak n}^*$. Now \cite[Th.~2.12]{MR2135527},
\cite[Th.~3]{MR2422266} state
\begin{equation}
    \label{eq:25}
    \operatorname{Spec}(H^{G_\CC[[z]]}_*(\Gr_{G_\CC})) \cong
    \overline\bmu_\CC^{-1}(y,y)/\check N\times\check N.
\end{equation}

\subsection{\texorpdfstring{$\SU(2)$}{SU(2)}-monopoles}

Let $M_k$ be the framed moduli space of charge $k$ $\SU(2)$-monopoles
on $\RR^3$ (see \cite{MR934202}). By \cite{MR709461}, it is naturally
bijective to the moduli space of solutions of Nahm's equations
\begin{equation*}
    \nabla_t T_\alpha + \frac12 \sum \varepsilon_{\alpha\beta\gamma}[T_\beta,T_\gamma] = 0 \qquad (\alpha,\beta,\gamma=1,2,3)
\end{equation*}
on the interval $(-1,1)$ such that $T_\alpha$ has at most simple poles
at $t=\pm 1$ and its residue $\operatorname{res}_{t=\pm 1}T_\alpha$
gives an irreducible $k$-dimensional representation $\rho$ of
$\SU(2)$. 
They are even isomorphic as hyper-K\"ahler manifolds \cite{MR1215288}.
We view $M_k$ as the moduli space of solutions of Nahm's
equation hereafter.

By \cite{MR1438643} $M_k$ is a submanifold of $T^* \check G_\CC$
($\check G_\CC=\GL(k)$) defined as follows. 
Let $\rho\colon\algsl(2)\to\check{\mathfrak g}_\CC$ as above. We
define Kostant-Slodowy slice $S(\rho) = y + Z(x)$, where $x = \rho(
\begin{smallmatrix}
    0 & 1 \\ 0 & 0
\end{smallmatrix}
)$,
$y = \rho(
\begin{smallmatrix}
    0 & 0 \\ 1 & 0
\end{smallmatrix}
)$
and $Z(x)$ is the centralizer of $x$ in $\check{\mathfrak g}_\CC$.
Now \cite[Cor.~4.1]{MR1438643} states $M_k =
\bmu^{-1}_\CC(S(\rho)\times S(\rho))$.

Let us further rewrite $M_k$ as a holomorphic symplectic quotient of
$T^*\check G_\CC$. The result is implicit in
\cite[\S3]{MR1438643}. The point is that $S(\rho)$ is
$\nu^{-1}(y)/\check N$, where $\check N$ is the unipotent subgroup of $\check
G_\CC$ as above, and $\nu\colon\check{\mathfrak g}^*_\CC\to\mathfrak
n^*$ is the natural projection.
\begin{NB2}
   Note $\nu^{-1}(y) = y + \mathfrak b$.
\end{NB2}%
Since $\overline\bmu_\CC = (\nu\times\nu)\circ \bmu_\CC$, we have
\(
    M_k \cong \overline\bmu_\CC^{-1}(y,y)/\check N\times\check N.
\)
Combining with \eqref{eq:25}, we get
\begin{equation*}
    M_k \cong \operatorname{Spec}(H^{\GL(k)[[z]]}_*(\Gr_{\GL(k)})).
\end{equation*}

\subsection{Centered monopoles}
Let us next consider the case of centered monopoles.
We first review the definition of centered monopoles.
Let $\tilde M_k$ be the space of solutions, and $\mathcal G_k^0$ be
the group of gauge transformations, i.e., the space of maps
$\gamma\colon [-1,1]\to \U(k)$ such that $\gamma(\pm 1) = 1$. We thus
have $M_k = \tilde M_k/\mathcal G_k^0$. We introduce a larger group
$\mathcal G_k$ consisting of maps $\gamma$, for which we require
$\gamma(-1) = 1$ and $\gamma(1)$ commutes with the pole $\alpha$ of
$T_\alpha$ for $\alpha=1,2,3$. Since $\operatorname{res}T_\alpha$
gives an irreducible representation, $\gamma(1)$ must be
scalar. Moreover, the action of $\mathcal G_k$ is still free by the
irreducibility of $(\nabla_t, T_\alpha)$. We have a free action of
$\mathcal G_k/\mathcal G^0_k = \U(1)$ on $M_k$.

We also have an $\RR^3$-action on $\mathcal N_k$ given by
$T_\alpha\mapsto T_\alpha + ix_\alpha$ ($x_\alpha\in\RR$,
$\alpha=1,2,3$). The quotient
\begin{equation*}
    M^0_k = (M_k/\U(1))/\RR^3
\end{equation*}
is called the moduli space of centered monopoles. In \cite{MR934202},
this space was introduced in terms of monopoles, one can check that it
is equivalent to the above definition. The detail is left as an
exercise for the reader.
\begin{NB2}
    See ``2014-11-13 Nahm's equation and cyclic covering.xoj''.
\end{NB2}%

Choose a trivialization of the vector bundle, and write $\nabla = d +
T_0$. By the gauge transformation
\begin{equation*}
    \gamma(t) \defeq \exp\left(\frac1k
      \int_{-1}^s \tr T_0 
      \right) \in \mathcal G_k,
\end{equation*}
we can make $\tr T_0 = 0$. The Nahm equation implies that
$\frac{d}{dt} \tr T_\alpha = 0$, hence we can make $\tr T_\alpha = 0$
by the $\RR^3$-action. Therefore
\begin{equation*}
    M^0_k = \{ (T_0,T_\alpha) \colon (-1,1) \to \su(k) \}/\mathcal G_{\SU(k)},
\end{equation*}
where $(d+T_0,T_\alpha)$ satisfies Nahm's equation and the condition
of the pole. The group $\mathcal G_{\SU(k)}$ is the subgroup of
$\mathcal G_k$ preserving the condition $\tr T_0 = 0$. Therefore it
consists of maps $\gamma\colon [-1,1]\to \SU(k)$ with $\gamma(-1) = 1$,
$\gamma(1) \in \ZZ_k = \U(1)\cap\SU(k)$.

Since $T_0$, $T_\alpha$ are $\su(k)$-valued, we can replace $\mathcal
G_{\SU(k)}$ by the space of maps $\gamma\colon [-1,1]\to
\SU(k)/\ZZ_k$. Moreover as $[-1,1]$ is contractible, such $\gamma$
lifts uniquely to $\SU(k)$ when we set $\gamma(-1) = 1$. Therefore the
group is unchanged under this replacement. Thus
\(
  \mathcal G_{\SU(k)} = \mathcal G^0_{\SU(k)/\ZZ_k}, 
\)
where $\mathcal G^0_{\SU(k)/\ZZ_k}$ is the space of maps
$\gamma\colon [-1,1]\to \SU(k)/\ZZ_k$ such that $\gamma(\pm 1) = 1$.
\begin{NB2}
    We can represent the gauge group $\mathcal G^0_{G}$ for a compact
    Lie group $G$ with $\operatorname{Lie} G = \su(k)$, as
    \begin{equation*}
        \mathcal G^0_G = \left\{ \gamma\in\mathcal G^0_{\SU(k)/\ZZ_k} \middle|
          [\gamma] \in \pi_1(G) \right\},
    \end{equation*}
    where $[\gamma]$ is the class in $\pi_1(\SU(k)/\ZZ_k) = \ZZ_k$,
    which is well-defined as $\gamma(\pm 1) = 1$, and we regard
    $\pi_1(G)$ as a subgroup of $\pi_1(\SU(k)/\ZZ_k)$ via the covering
    $G\to \SU(k)/\ZZ_k$.
\end{NB2}%
Since \cite[Cor.~4.1 and \S3]{MR1438643} is a result for an arbitrary
compact Lie group, we can apply it to get
\(
    M^0_k \cong \overline\bmu^{-1}_\CC(y, y)/\check N\times\check N
\)
as above, where $\check G_\CC = \SL(k)/\ZZ_k$ now. The Langlands dual
group of $\SL(k)/\ZZ_k$ is $\SL(k)$, hence
\begin{equation*}
    M^0_k \cong \operatorname{Spec}(H^{\SL(k)[[z]]}_*(\Gr_{\SL(k)})),
\end{equation*}
as expected.
\end{NB}

\section{Hilbert series of symmetric products}\label{sec:app}

We compute
\begin{equation*}
    H_k(t,z) = \sum_{\la=(\la_1\ge\cdots\ge \la_k)}
    z^{\sum \la_i} t^{\sum_i N|\la_i|} P_{\U(k)}(t,\la).
\end{equation*}
We formally set $H_0(t,z) = 1$ and consider the generating function
over all $k$. Our goal is to show that
\begin{equation}\label{eq:13}
    \sum_{k=0}^\infty H_k(t,z) \Lambda^k
    = \exp\left(
      \sum_{d=1}^\infty \frac{\Lambda^d}d
      H_1(t^d,z^d)
      \right).
\end{equation}
The first term $H_1(t,z) = (1-t^{2N})/{(1-t^2)(1-t^Nz)(1-t^Nz^{-1})}$
is the Hilbert series of $\CC^2/(\ZZ/N\ZZ)$, and the above is its
plethystic exponential. Then it is the Hilbert series of symmetric
powers.

Let us first note that there is a one-to-one correspondence between
dominant coweights $\la = (\la_1\ge\cdots\ge \la_k)$ and
$(k_0,k_1,\cdots; m)\in\ZZ_{\ge 0}^\infty\times\ZZ$ such that
\begin{aenume}
      \item $k_0\neq 0$.
      \item $k_0 + k_1 + \cdots = k$. In particular, there are only
    finitely many nonzero $k_\alpha$.
\end{aenume}
The correspondence is given by $m=\la_k$ and $k_\alpha = \# \{ i \mid
\la_i - \la_k = \alpha\}$. This is an extension of usual two types of
presentations of partitions.

We write $(t;t)_k = (1-t)(1-t^2)\cdots (1-t^k)$ as usual. Under the
bijection we have
\begin{equation*}
    \begin{split}
    & \sum \la_i = \sum_\alpha k_\alpha (m+\alpha),\quad
    \sum |\la_i| = \sum_\alpha k_\alpha | m+\alpha|,
\\
   & P_{\U(k)}(t,\la) = \prod_{\alpha=0}^\infty \frac1{(t^2;t^2)_{k_\alpha}}.
    \end{split}
\end{equation*}
Hence
\begin{equation*}
    \begin{split}
    & \sum_{k=0}^\infty H_k(t,z)\Lambda^k = 1 
    + \sum_{m\in\ZZ} \sum_{\substack{k_0=1\\ k_1,k_2,\dots = 0}}^\infty
    \prod_{\alpha=0}^\infty
    \frac{t^{N|m+\alpha|k_\alpha} z^{(m+\alpha)k_\alpha}\Lambda^{k_\alpha}}
    {(t^2;t^2)_{k_\alpha}}
\\
   =\; &
   1 
    + \sum_{m\in\ZZ} \prod_{\alpha=0}^\infty
    \sum_{\substack{\text{$k_\alpha= 0$ for $\alpha\neq0$}\\ 
        \text{$k_\alpha=1$ for $\alpha=0$}
      }}^\infty
    \frac{t^{N|m+\alpha|k_\alpha} z^{(m+\alpha)k_\alpha}\Lambda^{k_\alpha}}
    {(t^2;t^2)_{k_\alpha}}.
    \end{split}
\end{equation*}
By the $q$-binomial theorem (see e.g., \cite[Ch.~I, \S2, Ex.~4]{Mac})
we have
\begin{equation*}
    \sum_{k_\alpha=0}^\infty
        \frac{t^{N|m+\alpha|k_\alpha} z^{(m+\alpha)k_\alpha}\Lambda^{k_\alpha}}
        {(t^2;t^2)_{k_\alpha}}
        = \frac1{(t^{N|m+\alpha|} z^{m+\alpha}\Lambda;t^2)_\infty},
\end{equation*}
where $(a;t)_\infty = (1-a)(1-at)\cdots$. For $\alpha=0$, we subtract
the first term $k_0 = 0$, which is $1$. Hence this is equal to
\begin{equation}\label{eq:12}
    1 + \sum_{m\in\ZZ} \left( \prod_{\alpha=0}^\infty
      \frac1{(t^{N|m+\alpha|} z^{m+\alpha}\Lambda;t^2)_\infty}
      - \prod_{\alpha=1}^\infty
      \frac1{(t^{N|m+\alpha|} z^{m+\alpha}\Lambda;t^2)_\infty}\right).
\end{equation}

We separate the sum to two parts $m\ge 0$ and $m < 0$. First suppose
$m\ge 0$. Note that $|m+\alpha| = m+\alpha$ in this case. We change
$m+\alpha$ to $\alpha$ and get
\begin{equation*}
    \sum_{m=0}^\infty 
    \left(\prod_{\alpha=m}^\infty
      \frac1{(t^{N\alpha} z^{\alpha}\Lambda;t^2)_\infty}
      - \prod_{\alpha=m+1}^\infty
      \frac1{(t^{N\alpha} z^{\alpha}\Lambda;t^2)_\infty}\right).
\end{equation*}
We cannot take the sum $\sum_{m=0}^\infty$ separately as each sum
diverges. But it is possible once we subtract $1$ from each
term. Hence we get
\begin{equation*}
    \begin{split}
        & \sum_{m=0}^\infty \left(\prod_{\alpha=m}^\infty
          \frac1{(t^{N\alpha} z^{\alpha}\Lambda;t^2)_\infty} - 1
        \right) - \sum_{m=0}^\infty \left( \prod_{\alpha=m+1}^\infty
          \frac1{(t^{N\alpha} z^{\alpha}\Lambda;t^2)_\infty} -
          1\right)
        \\
        =\; & \prod_{\alpha=0}^\infty \frac1{(t^{N\alpha}
          z^{\alpha}\Lambda;t^2)_\infty} - 1.
    \end{split}
\end{equation*}
The last $-1$ cancels with $1$ in \eqref{eq:12}.

Let us turn to $m<0$. We have
\begin{equation*}
    \begin{split}
    & \sum_{m<0} \left( \prod_{\alpha=0}^\infty
      \frac1{(t^{N|m+\alpha|} z^{m+\alpha}\Lambda;t^2)_\infty}
      - \prod_{\alpha=1}^\infty
      \frac1{(t^{N|m+\alpha|} z^{m+\alpha}\Lambda;t^2)_\infty}\right)
\\
    =\; &
    \prod_{\alpha=0}^\infty \frac1{
      (t^{N\alpha} z^\alpha \Lambda;t^2)_\infty}
    \sum_{m=1}^\infty
    \left(\prod_{\alpha=1}^m
      \frac1{(t^{N\alpha} z^{-\alpha}\Lambda; t^2)_\infty}
      - \prod_{\alpha=1}^{m-1}
      \frac1{(t^{N\alpha} z^{-\alpha}\Lambda; t^2)_\infty} \right).
    \end{split}
\end{equation*}
We use the same trick as above. We subtract $1$ from each term in the
sum $\sum_{m=1}^\infty$ and separate the sum. We get
\begin{equation*}
    \prod_{\alpha=0}^\infty \frac1{
      (t^{N\alpha} z^\alpha \Lambda;t^2)_\infty}
    \left(\prod_{\alpha=1}^\infty
      \frac1{(t^{N\alpha} z^{-\alpha} \Lambda;t^2)_\infty} - 1 \right).
\end{equation*}

Adding $1$ and terms with $m\ge 0$, $m < 0$, we get
\begin{equation*}
    \prod_{\alpha=0}^\infty \frac1{
      (t^{N\alpha} z^\alpha \Lambda;t^2)_\infty}
    \prod_{\alpha=1}^\infty
      \frac1{(t^{N\alpha} z^{-\alpha} \Lambda;t^2)_\infty}.
\end{equation*}
Now it is straightforward to check \eqref{eq:13}.

\begin{center}
    \bf Note Added in Proof
\end{center}

After this paper was posted to ArXiv, several important progresses
have been made. Bullimore et al.\ write a paper
\cite{2015arXiv150304817B}, where it is argued from a physical
intuition that $\CC[\mathcal M_C]$ is embedded into a localization of
another $\CC[\mathcal M_C]$ of the abelian gauge theory where the
gauge group $G$ is replaced by its maximal torus $T$. This embedding
is rigorously given in the definition of \cite{BFN}.
At the same time, \cite{2015arXiv150304817B} discusses Coulomb
branches of quiver gauge theories of type $ADE$ when $\mu$ is
\emph{not} necessarily dominant. It clarifies a problem raised in
\subsecref{sec:type-A-quiver}.

The sequel \cite{BFN}, announced here, is written, and it is shown
that the definition there reproduces many examples in quiver gauge
theories \cite{blowup}. Further examples have been studied in later
papers by the author with collaborators.

\bibliographystyle{myamsalpha}
\bibliography{nakajima,mybib,coulomb}

\end{document}